\documentclass[final,5p,times,twocolumn]{elsarticle}

\usepackage{amssymb}
\usepackage[utf8]{inputenc}
\usepackage[T1]{fontenc}
\usepackage{lmodern}
\usepackage{booktabs}
\usepackage{textcomp}
\usepackage{graphicx}
\usepackage{subfigure}
\usepackage{xspace}
\usepackage{pgfplots}
\pgfplotsset{compat=1.17}
\usepackage{nicefrac}
\usepackage[greek,english]{babel}
\usepackage{textgreek}
\usepackage{hyperref}

\newcommand{\csh}{C\hbox{-}S\hbox{-}H\xspace}
\newcommand{\alite}{{alite}\xspace}
 
\journal{Cement and Concrete Research}

\begin{document}

\begin{frontmatter}

\title{Argon Broad Ion beam sectioning and high resolution scanning electron microscopy imaging of hydrated alite}

\author{Florian Kleiner}
\author{Christian Matthes}
\author{Christiane Rößler}
\address{F. A. Finger-Institute for Building Materials Science, Bauhaus-University Weimar, 99423 Weimar, Germany}

\begin{abstract}
Scanning electron microscopy (SEM) imaging is able to visualize micro- to nano-structures of cement and concrete.
A prerequisite is that the sample preparation preserves the native structure of the specimen.
In this study, argon Broad Ion Beam (BIB) sectioning  is compared to state-of-the-art sample preparation (resin embedding, polishing) for hydrated alite. 
Additionally, it is investigated if during BIB, sample cooling is beneficial to avoid deterioration of cement hydrates.
The aim is to quantitatively measure pore size distributions in hardened \alite pastes.

Therefore, not only optimized sample preparation but also optimized imaging conditions are investigated.
Finally, it is demonstrated that by image analysis pores down to a diameter of 5\,nm in hydrated \alite pastes can be quantitatively analysed.
\end{abstract}


\end{frontmatter}

\section{Introduction}

The preparation of smooth, flawless surfaces is an important prerequisite for high resolution scanning electron microscopy (SEM) and image analysis.
Good quality images of cross sections are important for an easy image analysis and can be used as reference data to validate other techniques like mercury intrusion porosimetry or to calibrate results of proton nuclear magnetic resonance spectroscopy \cite{Naber.2020}.
Furthermore, a very clean surface allows the collection of electron backscatter diffraction patterns, which provide valuable information about the crystalinity and the crystal orientation of minerals \cite{Roler.2017}.

Investigations of cement pastes and concrete samples have been performed on specimens which were vacuum impregnated and embedded in low viscosity resins, ground and polished since the 1980s till today \cite{Scrivener.1984,Kjellsen.1990,Kjellsen.2003,Scrivener.2004,Hu.2016}.

A clean and smooth surface of the sample without introducing artefacts like scratches is nearly impossible using  abrasive diamond suspensions or polishing pads. 
This is especially the case, if a specimen contains minerals of different hardness \cite{Ji.2014} or is embedded in resin \cite{Desbois.2010}.
During mechanical polishing, the softer material (e.g. resin) is more easily removed and spread over the surface.
These effects result in a bad feature resolution of the image.
For material with low or unconnected porosity, resin impregnation is often not completely possible.
As a result, subsequent sample preparation and imaging may be hindered or impossible.

Scrivener et al. described that it is possible to get a feature resolution of calcium silicate hydrate (\csh) phases, ettringite and pores down to about 100\,-\,200\,nm using conventional polishing techniques \cite{Scrivener.2004}.
Therefore, especially phases with nano-scaled structures such as ettringite or \csh can not be differentiated by these techniques mainly because of damages due to mechanical preparation.

\begin{figure}[htbp]
    \centering
    \subfigure[Captured using an Everhart-Thornley detector in BSE mode providing very few details.]{\label{fig:artefacts_a}\includegraphics[width=0.49\linewidth ]{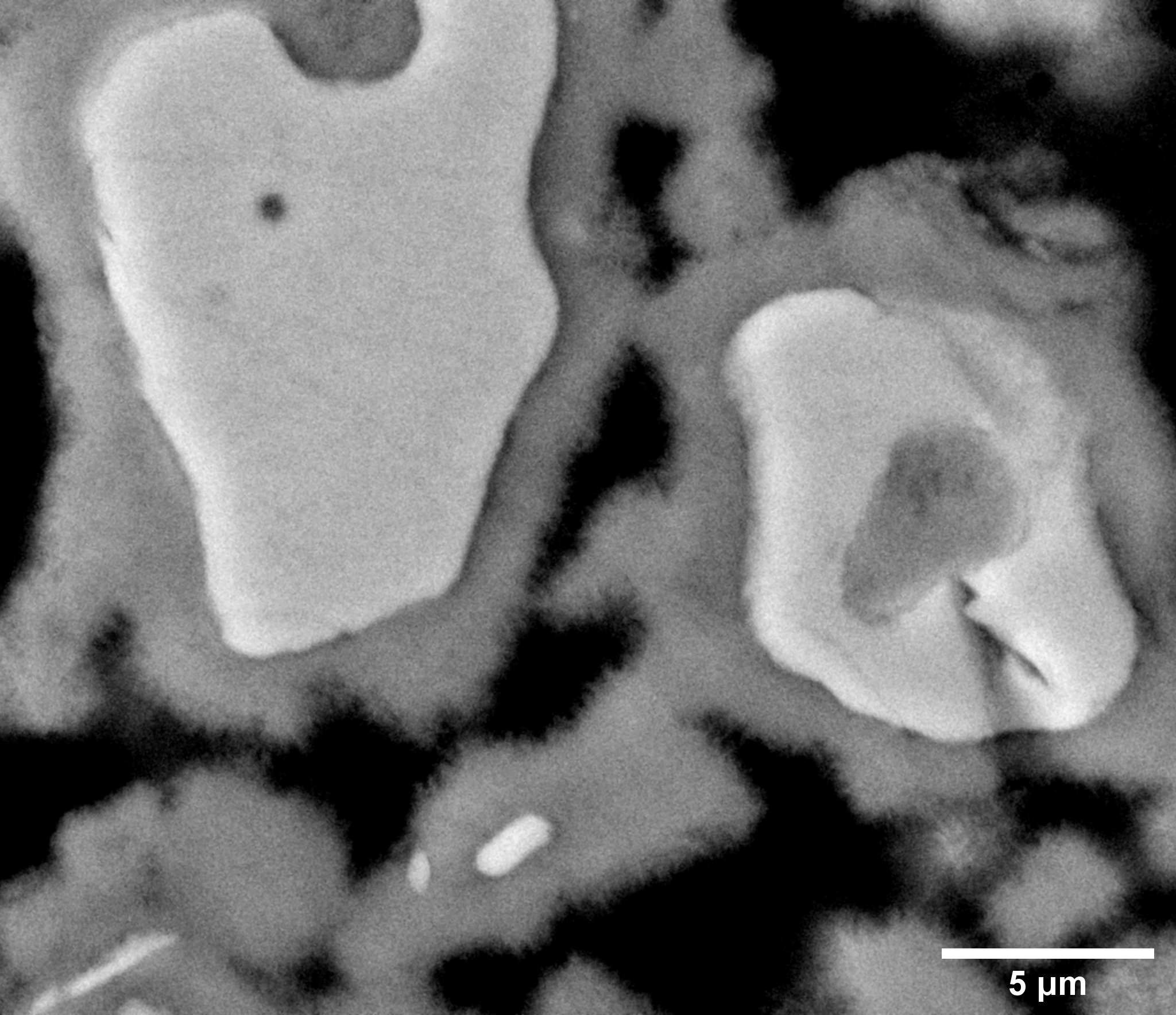}}\hspace{0.005\linewidth}
    \subfigure[Captured using a modern diode BSE detector providing finer details.]{\label{fig:artefacts_b}\includegraphics[width=0.49\linewidth ]{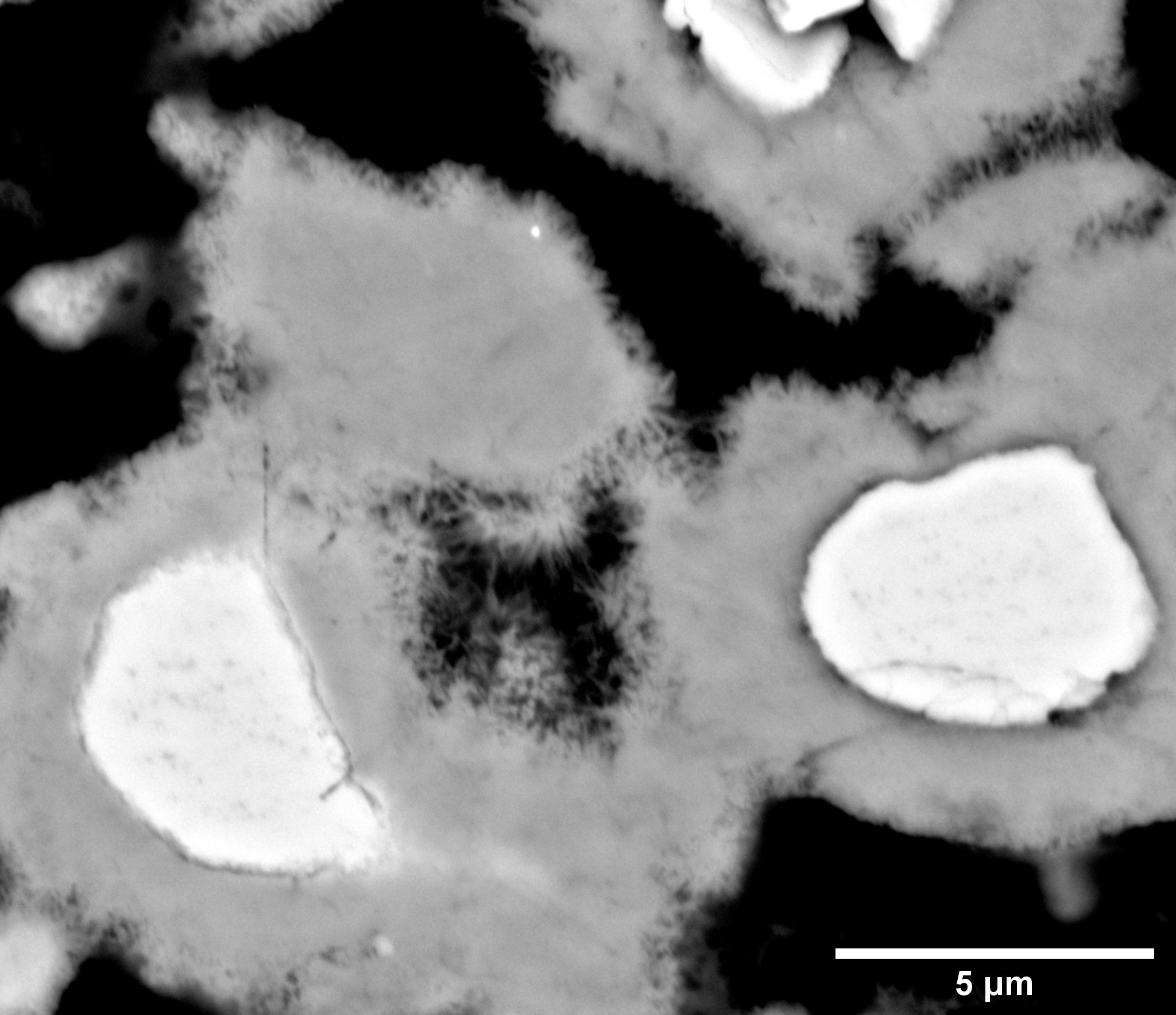}}
    \caption{High voltage (15\,kV, 0.8\,nA) BSE images by two different detectors of a 28\,d hydrated \alite specimen, mechanically polished, carbon coated.}
    \label{fig:artefacts}
\end{figure}

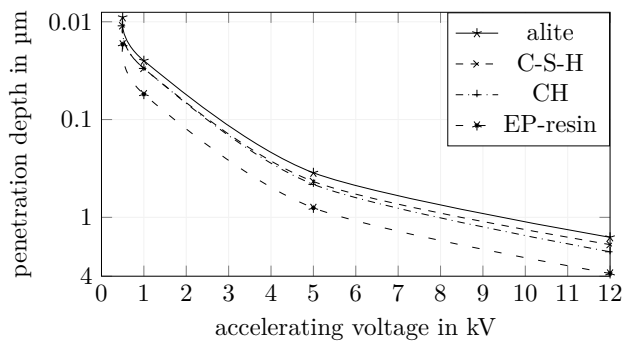
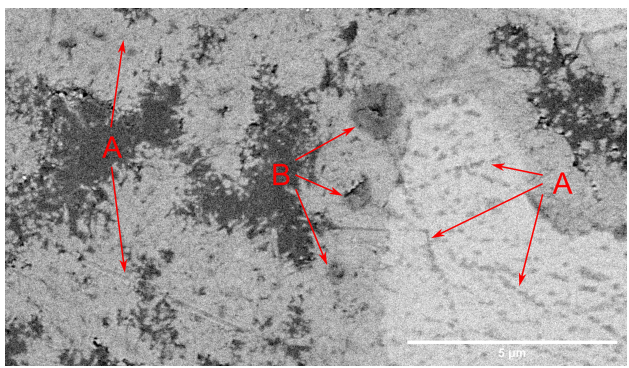
\begin{figure}[!htb]
    \centering
    \subfigure[Calculated penetration depths in different materials, modelled using the \textsc{Electron Flight Simulator} \cite{SmallWorld.2004}]{
   	 \label{fig:penetration_depth_a}
    \begin{tikzpicture}
        \begin{semilogyaxis}[
        	log ticks with fixed point,
            legend style={at={(1,1)},anchor=north east},
            xmin=0, xmax=12,
            ymin=0.008, ymax=4,
            width=0.99\linewidth,
            height=0.6\linewidth,
            grid=major,
            xlabel={accelerating voltage in kV},
            ylabel={penetration depth in \textmu m},
            xtick={0,1,...,12},
            ytick={0,0.01,0.1,1,4},
            y dir=reverse,
            grid style={line width=.1pt, draw=gray!10},
            ]
            \addplot[smooth,color=black, mark=star] table[x=Voltage,y=C3S,col sep=semicolon] {figure02a.csv};
            \addlegendentry{\alite};
            \addplot[smooth,color=black, dashed, mark=x] table[x=Voltage,y=CSH,col sep=semicolon] {figure02a.csv};
            \addlegendentry{\csh};
            \addplot[smooth,color=black,dashdotted, mark=+]  table[x=Voltage,y=CH,col sep=semicolon] {figure02a.csv};
            \addlegendentry{CH};
            \addplot[smooth,color=black, loosely dashed, mark=10-pointed star]  table[x=Voltage,y=Epoxy,col sep=semicolon] {figure02a.csv};
            \addlegendentry{EP-resin};
        \end{semilogyaxis}
    \end{tikzpicture}
   }
    \subfigure[A low voltage (1.5\,kV, 0.8\,nA) BSE image showing the surface damage (A: scratches, B: oil residue) due to mechanical polishing.]{\label{fig:penetration_depth_b}\includegraphics[width=0.99\linewidth]{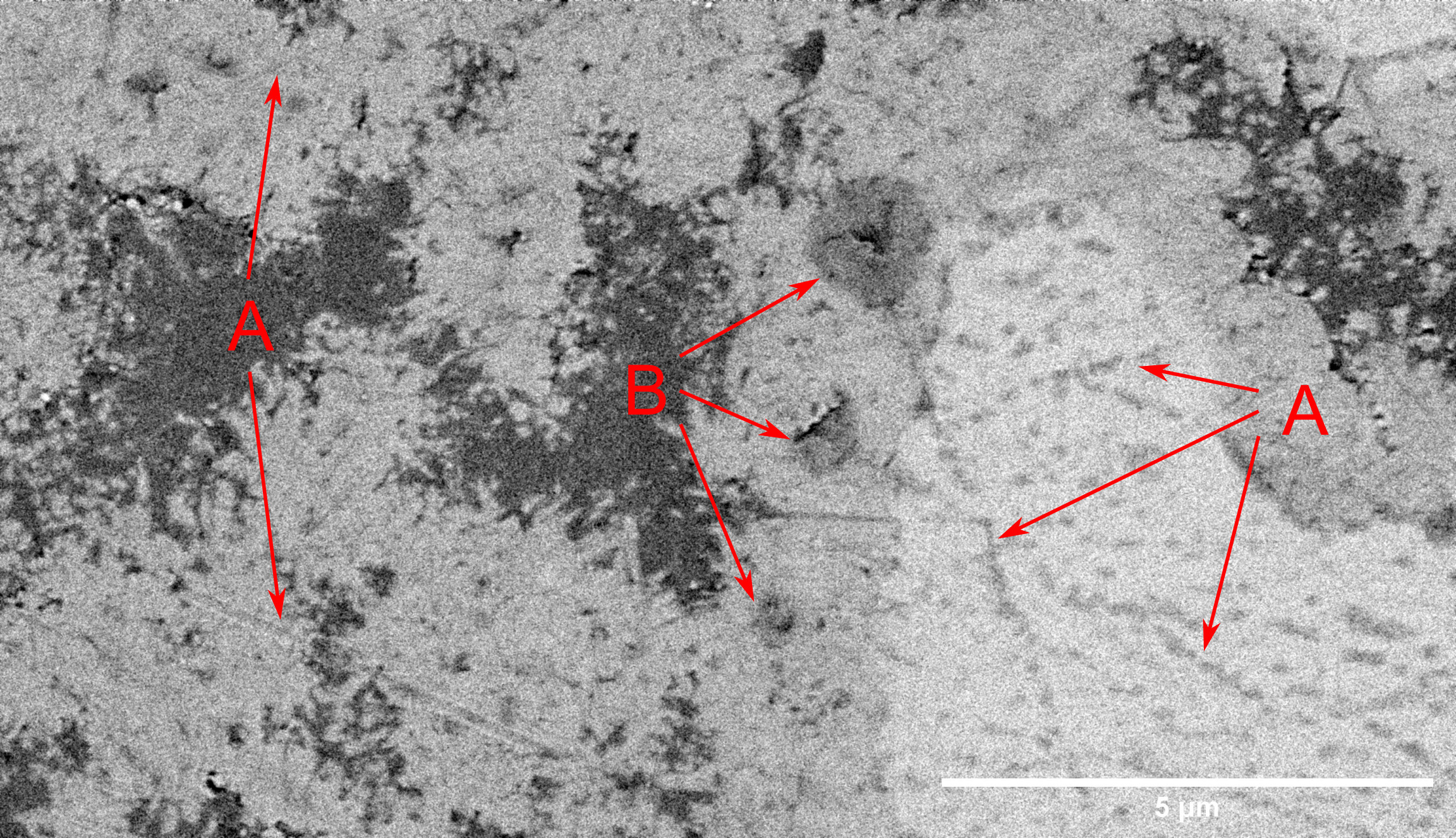}}
    \caption{Visualization of the low penetration depth of the electron beam and sample-preparation defects that become visible at low acceleration voltage.}
    \label{fig:penetration_depth}
\end{figure}

In the past, backscatter electron (BSE) detectors only gave good image results at high voltages (>\,10\,kV), which further reduced the effective resolution (Figure \ref{fig:artefacts_a}).
Modern diode BSE detectors deliver a improved resolution (Figure \ref{fig:artefacts_b}).
Since the SEM detectors available today are able to work at low acceleration voltages, a conductive coating of samples is no longer a requirement in SEM imaging.
The penetration depth of the electron beam at 1\,kV for the investigated materials is relatively low (Figure \ref{fig:penetration_depth_a}).
In combination with more sensitive BSE detectors the obtainable imaging resolution is improved \cite{Bell.2013} (chapter 4.6), but the preparation defects become very visible (Figure \ref{fig:penetration_depth_b}).

\begin{figure}[!htb]
    \centering
    \includegraphics[width=\linewidth]{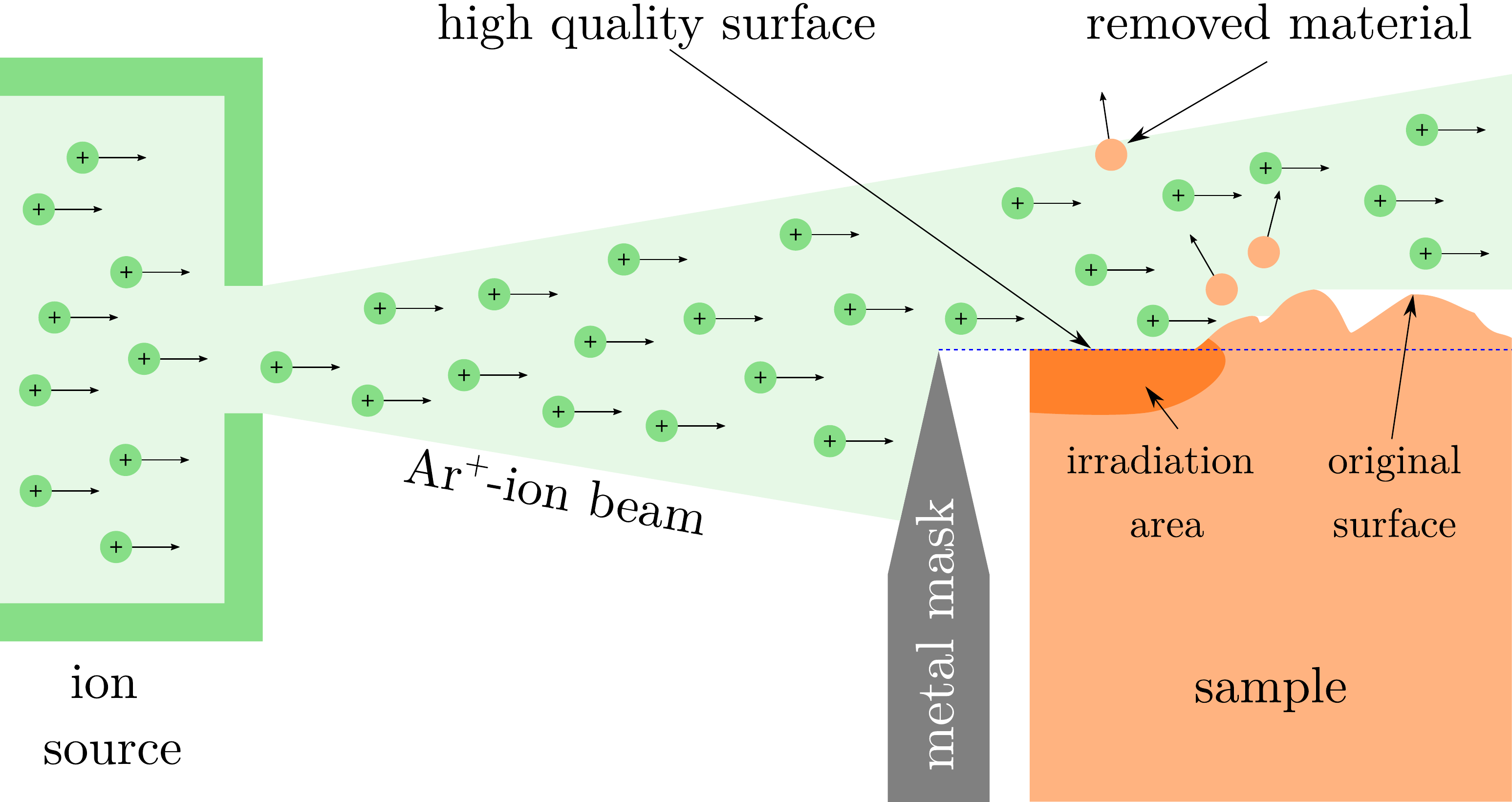}
    \caption{Principle of Ar$^+$-broad ion beam milling, side view.}
    \label{fig:bib_scheme}
\end{figure}

This study presents argon broad ion beam (BIB) sectioning \cite{Harper.1983,Grunewald.1987} as an alternative method to prepare cementitious specimen surfaces for SEM.
In a BIB, accelerated ions are shot under a grazing angle of incidence on the specimen surface. 
A sharp tungsten blade makes it possible to erode the material in a controlled manner and to create a flat and smooth surface (Figure \ref{fig:bib_scheme}).

Specimens prepared using argon BIB have a polished surface with limited damage to the nano-scale features of interest and pores without debris \cite{Desbois.2009,Desbois.2010}.
Furthermore, the necessity of resin intrusion is eliminated \cite{Desbois.2009}, while it is still possible \cite{Roler.2017} and in some cases feasible to do so.

\csh contains structural water and thus is known to undergo structural changes when exposed to high temperatures \cite{TajueloRodriguez.2017}. 
Argon BIB sectioning normally causes an increase in sample temperature \cite{Kim.2003}.
Preparing \csh specimens for transmission electron microscopy (TEM) using a cooling stage was already proposed by Groves et al. \cite{Groves.1986} and Richardson et al. \cite{Richardson.1993}.
This proposal was strengthened by Viguier and Mortensen who calculated a theoretical temperature rise up to 673\,K \cite{Viguier.2001} in TEM lamellae and Park et al. measured up to 603\,K in a thin TEM lamella \cite{Park.2007}.
Nevertheless, Groves et al. \cite{Groves.1986} were proposing, that there is no significant difference between the microstructure of hydrated \alite prepared using normal and low-temperature conditions.
It should be noted, that the respective images published by Groves et al. \cite{Groves.1986} show extensive damage in the pore structure of the inner \csh induced by the electron beam, which could mask the damages induced by the ion beam.
Furthermore, for thin TEM lamellae the temperature increase due to ion beam polishing is expected to be different than for the approximately 5\,mm thick bulk material that is BIB sectioned for large area SEM imaging.
Thus, it could still be beneficial to use a cooling stage to prevent structural damage of the cement hydrates during BIB sectioning \cite{Roler.2006}.

Desbois et al. presented the advantages of this method to visualize complex structures of e.g. clay stone \cite{Desbois.2009} and sandstone \cite{Desbois.2010} using BIB and cryo-BIB in combination with (cryo) focused ion beam (FIB).
In contrast to FIB, BIB allows to prepare large sample areas in the range of a few square millimetres \cite{Hauffe.2003,Desbois.2013} instead of typically 100\,-\,200\,\textmu m$^2$ using FIB.
In some cases cryo-BIB is even used as a tool for tomography \cite{Desbois.2013,Mac.2018}, while the slice thickness still is relatively large and inconsistent (min. 350\,nm\,$\pm$\,20\,\% ) compared to FIB (5\,-\,10\,nm) and the effort required is very high \cite{Desbois.2013}.

It is well known, that the electron beam tends to damage phases like \csh using TEM and SEM at high magnifications, due to thermal stress \cite{Richardson.1993, Roler.2006}.
Therefore, high resolution imaging of hydrated \alite samples using SEM is only possible at low voltage and low beam current in order to avoid beam damage on hydrated phases.

The present study will demonstrate and discuss advantages and limitations of BIB sectioning for high resolution imaging and quantitative image analysis of hardened \alite pastes.
Typical irradiation artefacts, caused by the electron beam and the ion beam on this material are shown and their impact on image analysis is discussed.
Additionally, possible solutions for their avoidance or correction methods are proposed.

\section{Materials and methods}
\subsection{Experimental}

For all investigations, commercially available alite (M3 polymorph, Vustah, Czech Republic) was used. 
Its chemical composition, as shown in Table \ref{tab:composition}, was determined by X-ray fluorescence spectroscopy (XRF).

\begin{table}[ht]
	\centering
	\caption{Chemical composition of alite measured by XRF.}
	\begin{tabular}{cc}
		\toprule
			Oxide & Alite [wt.-\%]\\
		\midrule
CaO 				& 70.3 \\
SiO$_2$			& 26.5 \\
MgO				& 2.02 \\
Al$_2$O$_3$	& 0.34 \\
P$_2$O$_5$ & 0.13\\
Fe$_2$O$_3$	& 0,07 \\
Na$_2$O		& 0.04 \\
K$_2$O			& 0.03 \\
TiO$_2$			& 0.03 \\
Mn$_2$O$_3$ & 0.02 \\
LOI & 0.46\\
		\midrule
		$\sum$ & 99.8 \\
		\bottomrule
	\end{tabular}
	\label{tab:composition}
\end{table}

The specimens were prepared in a nitrogen filled glovebox to prevent carbonation. 
Alite with a mean particle size of 10\,\textmu m was mixed with water (\nicefrac{$w$}{$s$} = 0.5) and filled into sample moulds (7\,$\times$\,7\,$\times$\,50\,mm).
The specimens were cured for 1,  7 and 28 days in the same glovebox.
After this time, the \alite reaction was stopped by immersing the prisms in ethanol (98\,\%) for 20 minutes.
The specimens were dried at 40\,$^\circ$C and then saw cut using a target surfacing system (EM TXP, Leica).
Some specimens were polished using sandpaper / diamond polishing paste from 1.0 to 0.5\,\textmu m in the same device.
Since \alite is sensitive to water, the preparation process was done using ethanol as rinsing solution.
Afterwards, the sample was dried in a vacuum chamber.
To obtain images of fractured surfaces of the 24\,h hydrated specimen, the prisms were cracked using pliers until they were small enough to fit on a SEM-sample holder.

Furthermore, some of the samples shown in this study were embedded in low viscosity epoxy-resin (resin R1370, hardener R1376, Agar Scientific) by immersing a 7\,$\times$\,7\,$\times$\,3\,mm sample-section in resin, followed by degassing in two 30 minute cycles.
Subsequently, the specimen was removed from the still uncured resin and placed face-down on a glass surface and cured at 40\,$^\circ$C in an oven.

Using standard mechanical polishing, sections were polished in incremental steps with diamond oil slurries of particle sizes 15, 3, 1, and 0.25\,\textmu m.
Carbon coating of approx. 8\,-\,10\,nm thickness was applied on selected samples (SCD 500, Baltec).

The dried and polished 7 and 28\,d \alite samples were transferred into a triple ion beam milling system (EM TIC 3X, Leica).
Argon BIB sectioning is carried out under vacuum (10 to 40\,mbar) conditions.
The three argon ion beams intersect at a sharp edged tungsten mask forming a milling sector of approximately of 100\,$^\circ$ (Figure \ref{fig:bib_scheme}).
The mask and the specimen surface are arranged in a way that only 20\,-\,100\,\textmu m of the sample are exposed to the ion beam above the mask.

\begin{figure}[htbp]
    \centering
    \includegraphics[width=0.7\linewidth]{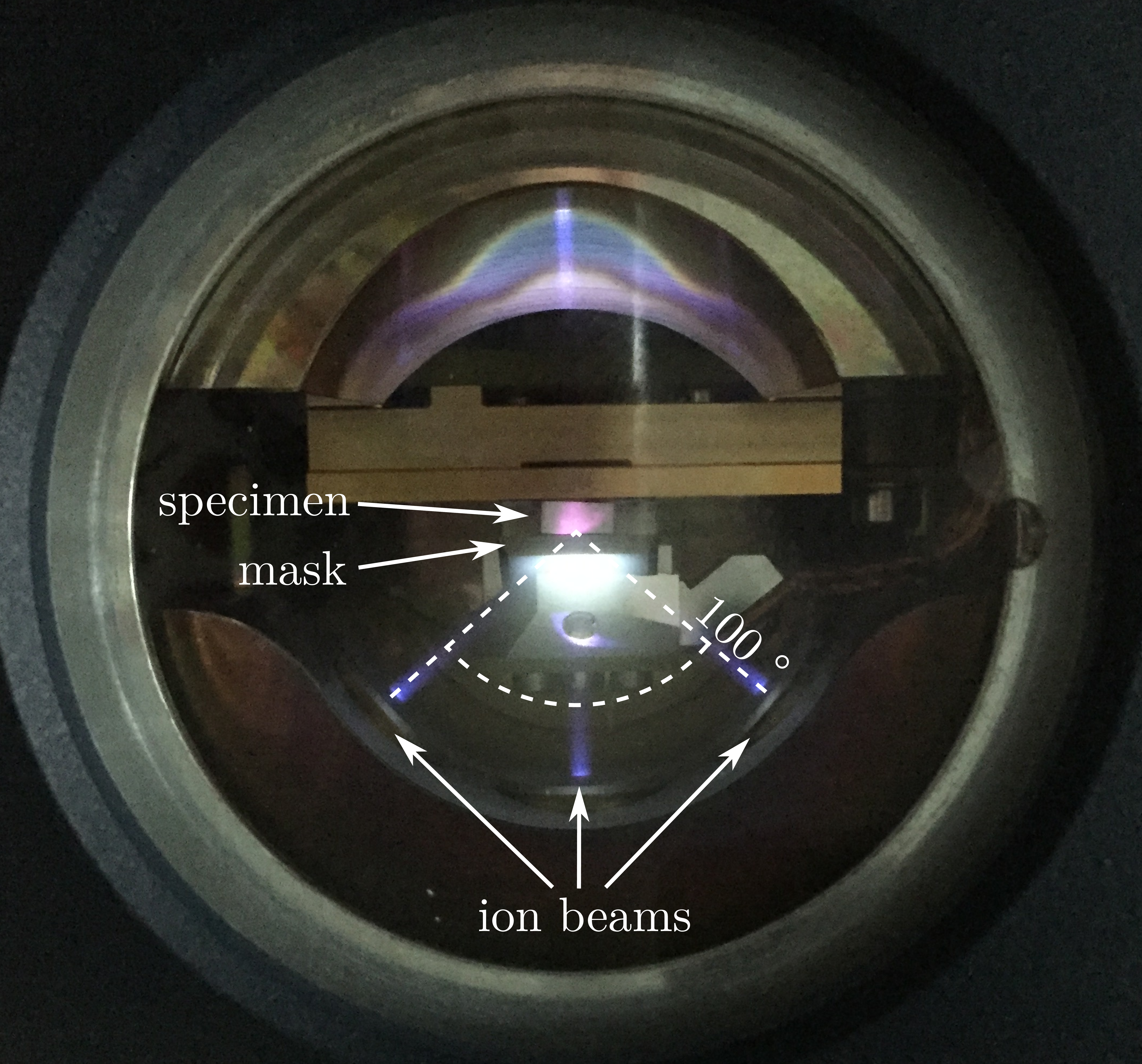}
    \caption{Milling of a specimen. The ion beams appear as violet-blueish rays. }
    \label{fig:bib_device}
\end{figure}

\begin{figure}[htbp]
    \centering
    \includegraphics[width=\linewidth]{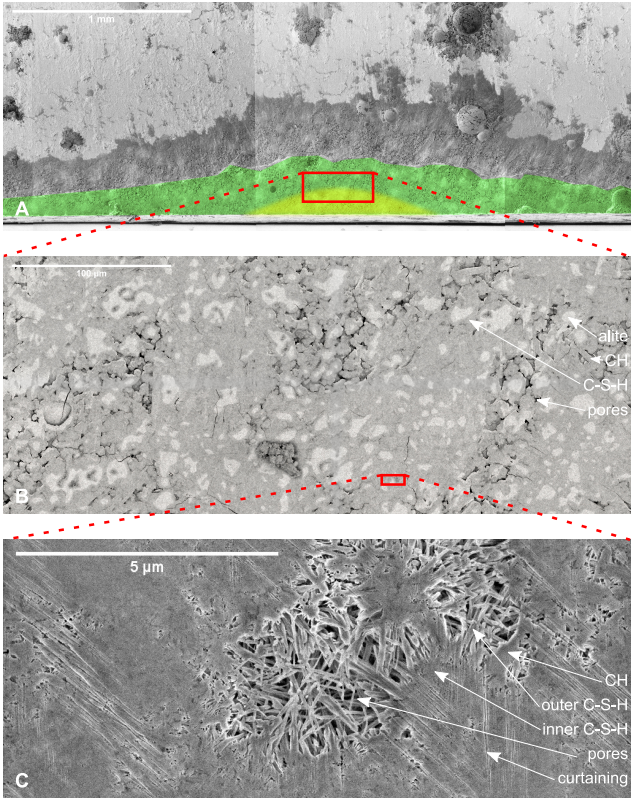}
    \caption{Surface of 28 days hydrated \alite (7.0$ \times $2.5\,mm) prepared using cryo-BIB (same specimen orientation as shown in Figure \ref{fig:bib_scheme}. The area used for SEM imaging (approx. 0.99\,mm$^2$) is tinted green. Images A and B are large area, high resolution BSE images composed of multiple individual images. The lower image C shows a SE image with typical morphology of fibrous outer \csh phases and dense inner \csh. Furthermore, some stripes caused by the BIB process (curtaining) are visibile.}
    \label{fig:zoom_area}
\end{figure}

The BIB system used, allows to cool the specimen using liquid nitrogen to -140\,$^\circ$C during the milling process.
The process gas was argon with a purity of 99.999\,\%.
The examined samples were initially prepared using an acceleration voltage of 6\,kV and 2\,mA gun current for 3 hours for the bulk milling.
Afterwards, the parameters were changed to 3\,kV, 1\,mA for another 3 hours as a polishing step. 
The same parameters were used without cooling to compare both preparation methods.
An area of up to one mm$^2$ can be analysed afterwards as shown in Figure \ref{fig:zoom_area}.

The uncoated, as-prepared surface of all samples was characterized using an ultra high resolution SEM (Helios G4 UX DualBeam, thermoFischer Scientific and Nova Nano SEM 230, FEI).
The images shown are acquired using the through-the-lens detector (TLD) in secondary electron detection (SE) or in BSE mode.
For the high resolution imaging using the TLD-SE (high vacuum), the SEM-immersion mode was used.
In some cases these settings were combined with a stage bias and a monochromator.
These settings allow the observation of pores structures of hydrated \alite at nano-scale resolution.

\subsection{Semi automatic segmentation}
To segment a statistically relevant amount of pores using the acquired images, a segmentation by hand, using arbitrary thresholds is not suitable.
Therefore, the image analysis was performed using a custom-written python script \cite{Kleiner.2021}.
The half-automated process automatically determines the scale and denoises the SE images of hydrated alite cross sections using non-local-means filtering \cite{Buades.2011}.

To relate the segmented pores to a phase like \csh, it is necessary to segment these phases.
In many cases, the gray values of the solid phases like \csh, CH and \alite are very similar or even overlapping.
Therefore, the script offers a method for manual segmentation, to mark areas that should be excluded from the analysis (eg. other phases, large pores or image and preparation artefacts).

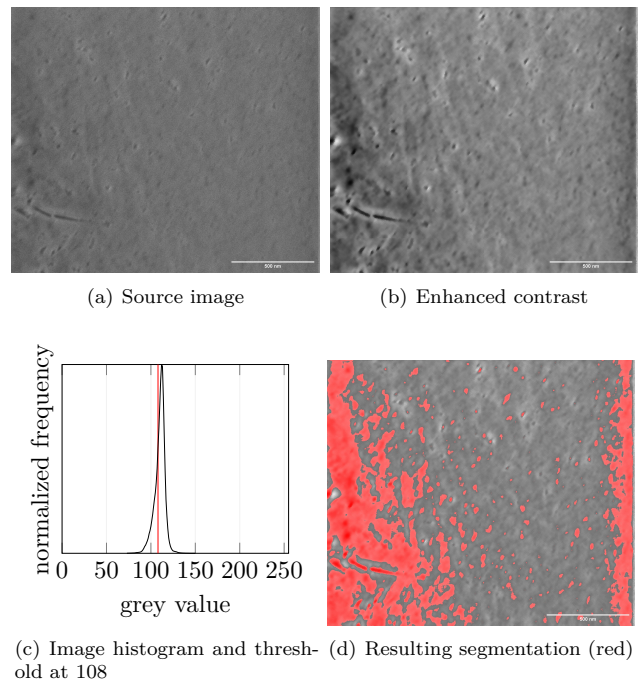
\begin{figure}[htbp]
    \centering
    \subfigure[Source image]{\label{fig:fixed_threshold_a}\includegraphics[width=0.49\linewidth]{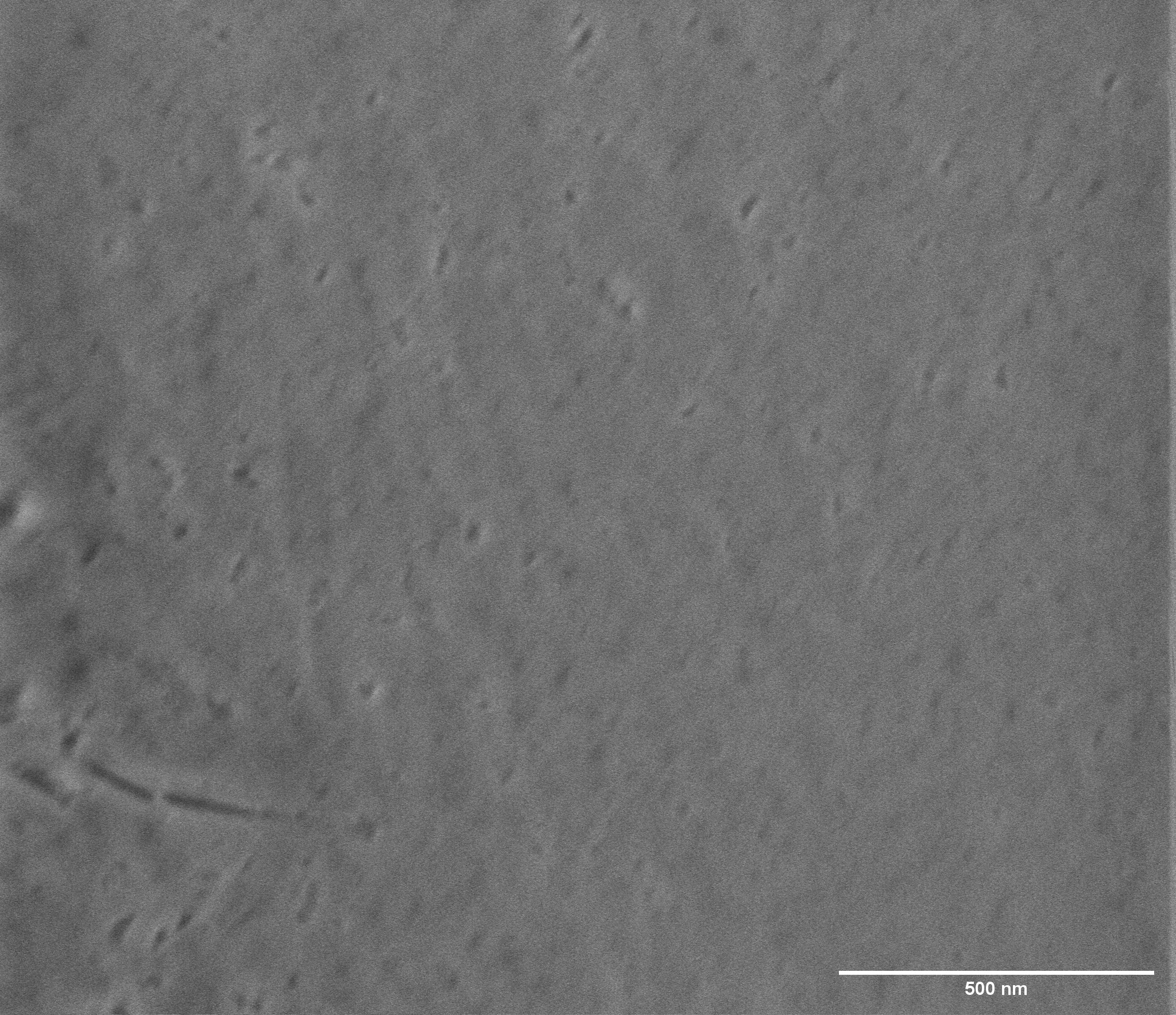}}\hspace{0.005\linewidth}
    \subfigure[Enhanced contrast]{\label{fig:fixed_threshold_b}\includegraphics[width=0.49\linewidth]{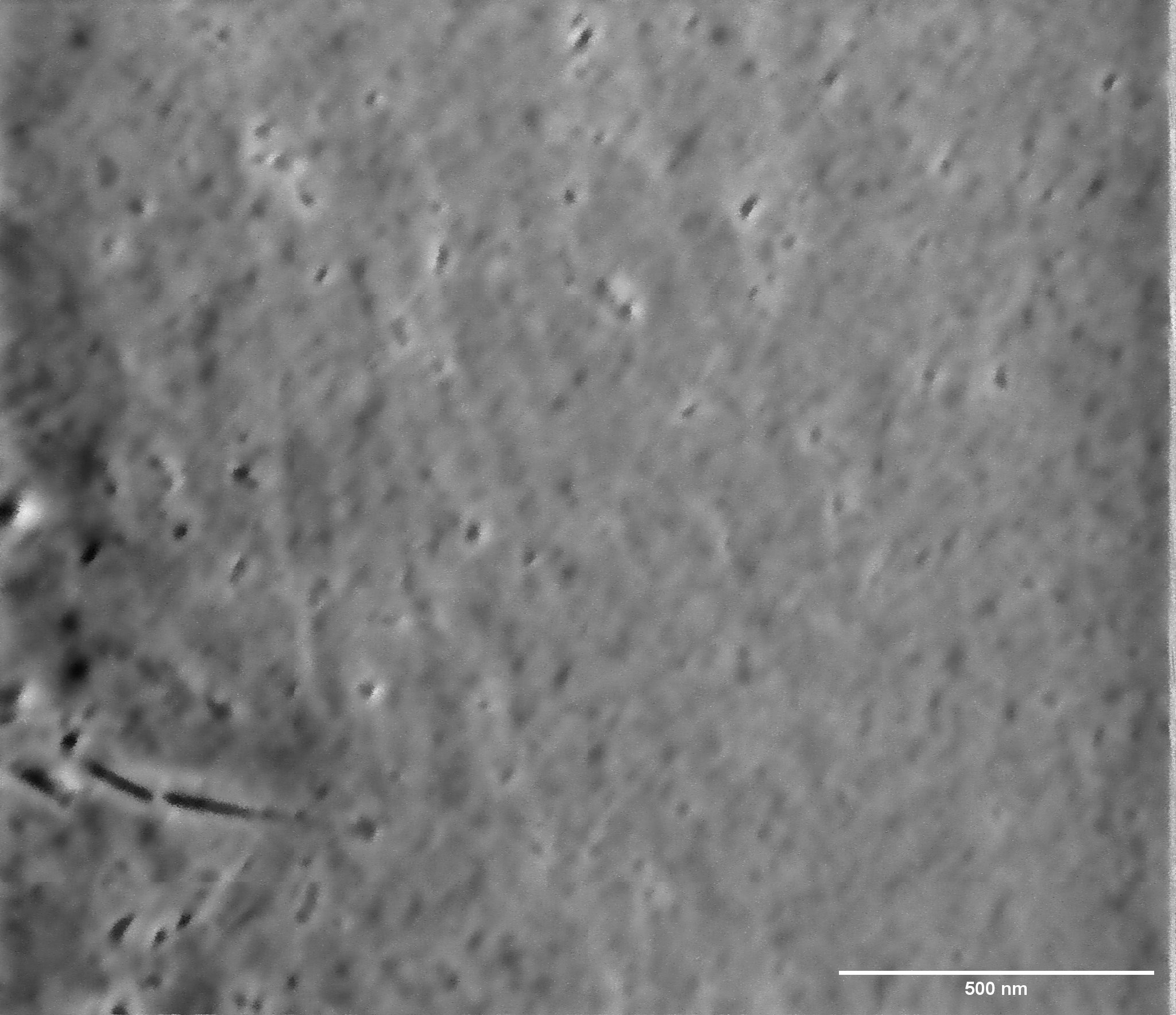}}
    \subfigure[Image histogram and threshold at 108]{\label{fig:fixed_threshold_c}
    \begin{tikzpicture}
        \begin{axis}[
            legend style={at={(1,1)},anchor=north east, font=\scriptsize},
            xmin=0, xmax=255,
            ymin=0, ymax=1,
            width=0.54\linewidth,
            height=0.48\linewidth,
            grid=major,
            xlabel={grey value},
            ylabel style={align=center},
            ylabel={normalized frequency},
            xtick={0,50,...,250},
            ytick=\empty,
            grid style={line width=.1pt, draw=gray!10},
            ]
            \draw[red] ({axis cs:108,0}|-{rel axis cs:0,1}) -- ({axis cs:108,0}|-{rel axis cs:0,0});
            \addplot[color=black] table[x=value,y=normed,col sep=comma] {Histogram_C3S28dBIB6KV6h_016_cut.csv};
        \end{axis}
    \end{tikzpicture}
   }
    \subfigure[Resulting segmentation (red)]{\label{fig:fixed_threshold_d}\includegraphics[width=0.49\linewidth]{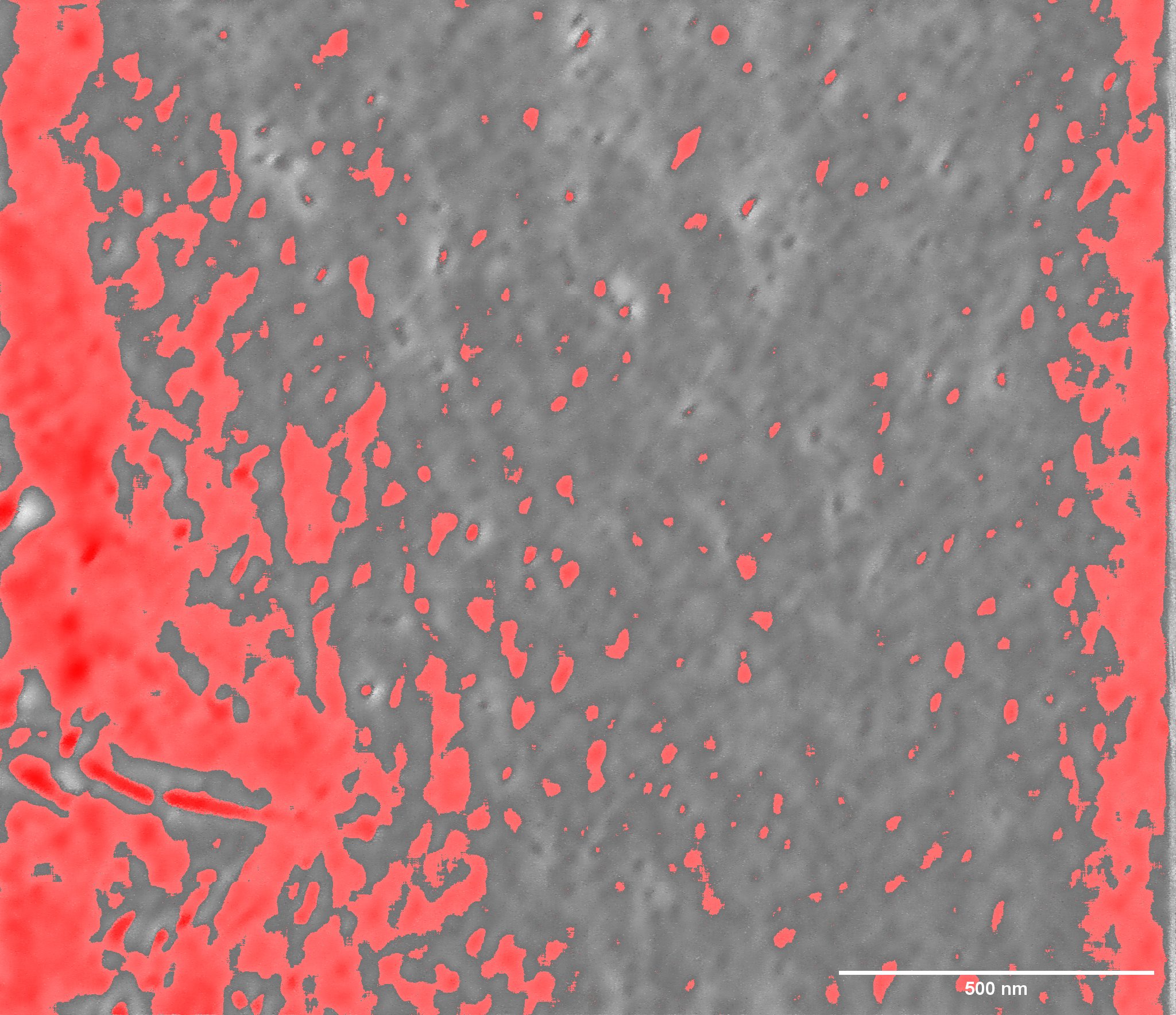}}
    \caption{Illustration of the usage of a global threshold value on the image of dense inner \csh of 28\,d hydrated alite prepared using BIB at 20\,$^\circ$C. TLD-SE, 2\,kV, 89\,pA}
    
    \label{fig:fixed_threshold}
\end{figure}

Overflow based algorithms for pore segmentation as proposed by Wong et al. \cite{Wong.2006} result in a global threshold value for an image. 
They require fairly distinct brightness differences between the object of interest (pores) and the surrounding areas (\csh).
Furthermore it is assumed, that pores and background have no overlapping gray values within the image.
As shown in Figure \ref{fig:fixed_threshold_a} this does not apply for the TLD-SE images of dense inner \csh. This becomes more obvious in Figure \ref{fig:fixed_threshold_b} with an enhanced contrast and also in the narrow histogram in Figure \ref{fig:fixed_threshold_c}.
Applying a global threshold on such images will result in a significant over-segmentation or no segmentation at all, if other phases or large pores are included in the image.
Figure \ref{fig:fixed_threshold_d} shows the segmentation results based on threshold values determined by algorithm proposed by Shanbhag \cite{Shanbhag.1994}.
 
Therefore, the pores within the alite are segmented using the local thresholding algorithm proposed by Phansalkar et al. \cite{Phansalkar.10.02.201112.02.2011}.
It is derived from the Sauvola thresholding method \cite{Sauvola.2000}, but modified to deal especially with low contrast images as shown in Figure \ref{fig:fixed_threshold_a}.
The algorithm calculates a threshold in a given window radius (e.g. 20\,px) to segment the pores instead of using a single thresholding value for the whole image.
Phansalkar thresholding is implemented in the python script but also easily accessible in the the Auto Local Threshold function of Fiji (ImageJ) \cite{Schindelin.2012}, delivering results comparable to the python script using the parameters $k$ = 0.15 and $r$ = 0.3.
After the application of morphological operators to denoise the binary images, a pore size distribution (PSD) refereed to the surrounding phase or to the whole image can be calculated.
Since a single image very often contains very few pores, a PSD delivers a very noisy result due to a low pore count. 
Therefore, also the chord-length density function (CDF) \cite{torquato.2002} (chapter 2.5) is calculated in $x$- and $y$- directions for every pixel row and column to improve the data density.
The CDF is derived from the measurement of the chord lengths between the intersection of two-phase interfaces.
In this case the chords within a pore are measured and plotted in a histogram.
For all calculations the pores are considered to be circular.

\section{Results and discussion}

\begin{figure}[htbp]
    \centering
    \subfigure[24\,h hydrated \alite]{\label{fig:fractured_surfaces_a}\includegraphics[width=0.49\linewidth]{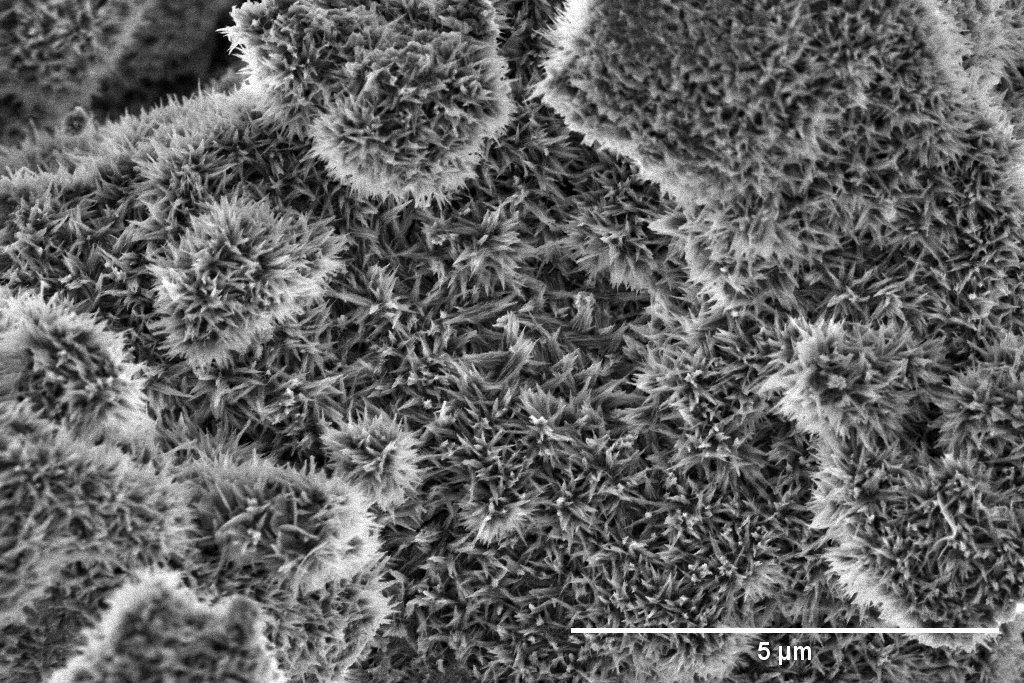}}\hspace{0.005\linewidth}
    \subfigure[28\,d hydrated \alite]{\label{fig:fractured_surfaces_b}\includegraphics[width=0.49\linewidth]{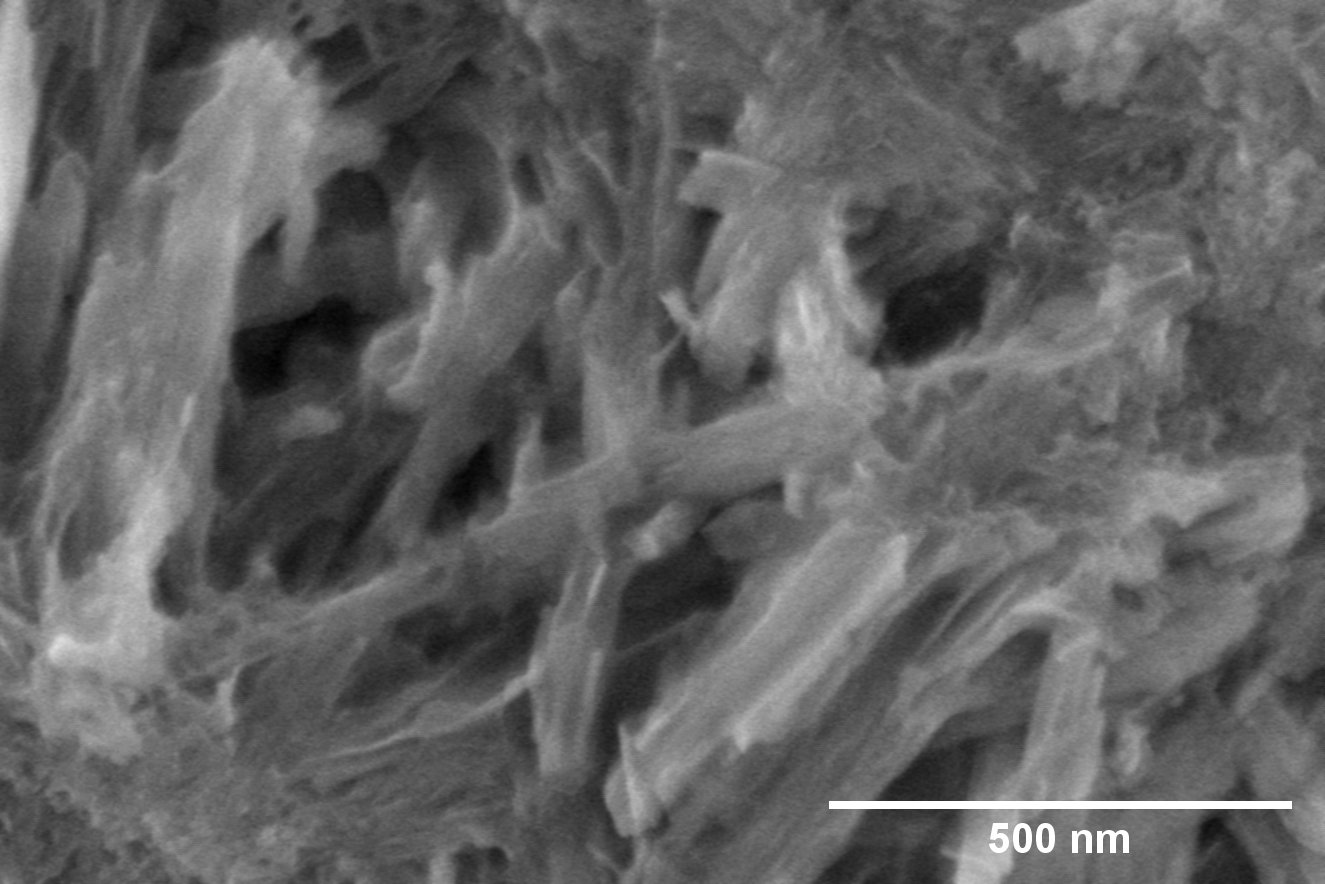}}
    \caption{Fractured surfaces of hydrated \alite specimens. Acquisition parameters: TLD-SE, left: 2\,kV, 7\,pA, right: 4\,kV, 1.25\,nA, stage bias 2\,kV, immersion mode, high vacuum.}
    \label{fig:fractured_surfaces}
\end{figure}

The goal of this investigation was to achieve high resolution images of hydrated alite with as little influence of the preparation steps. 
Figure \ref{fig:fractured_surfaces} shows such images using fractured surfaces of a 28\,d hydrated \alite as an example.

The \csh-needles are clearly visible and due to the low magnification in Figure \ref{fig:fractured_surfaces_a}, no significant deformation of the microstructure due to the high vacuum can be seen. 
This is supported by similar images acquired using a low vacuum SEM by Stark et al. \cite{Stark.2001}. 
According to findings by Gajewicz et al. \cite{Gajewicz.2016}, drying at 60\,$^\circ$C causes changes in the water saturation and the size of gel pores of the \csh. 
Therefore, it may be possible that there are changes within the microstructure at the nano-scale.
Nevertheless, Figure \ref{fig:fractured_surfaces_b} shows no immediate signs of nano- and micro-cracks due to vacuum drying and SEM imaging.
Therefore it is assumed that these images represent a close-to native view of the \csh microstructure.

\begin{figure}[!ht]
    \centering
    \subfigure[\csh needles (2\,kV, 22\,pA)]{\label{fig:reference_images_a}\includegraphics[width=0.49\linewidth]{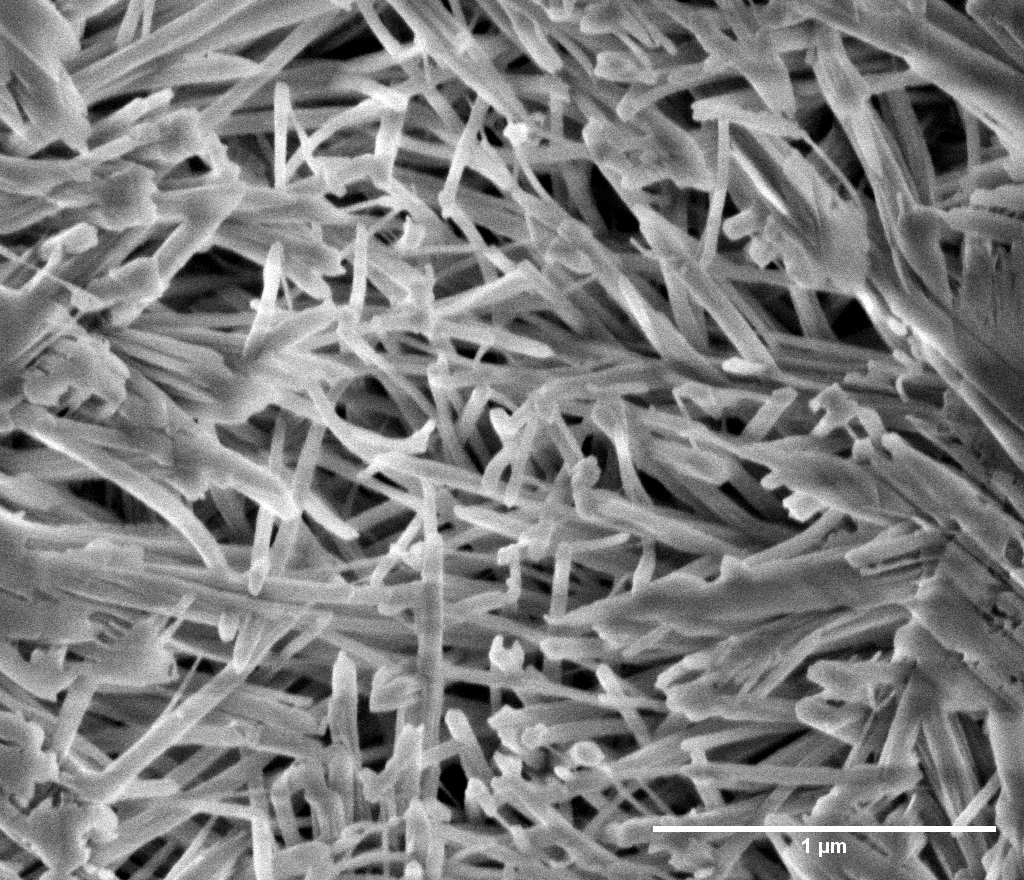}}\hspace{0.005\linewidth}
    \subfigure[ A: Outer \csh, B: inner \csh, C: unhydrated \alite (1.5\,kV, 6.8\,pA)]{\label{fig:reference_images_b}\includegraphics[width=0.49\linewidth]{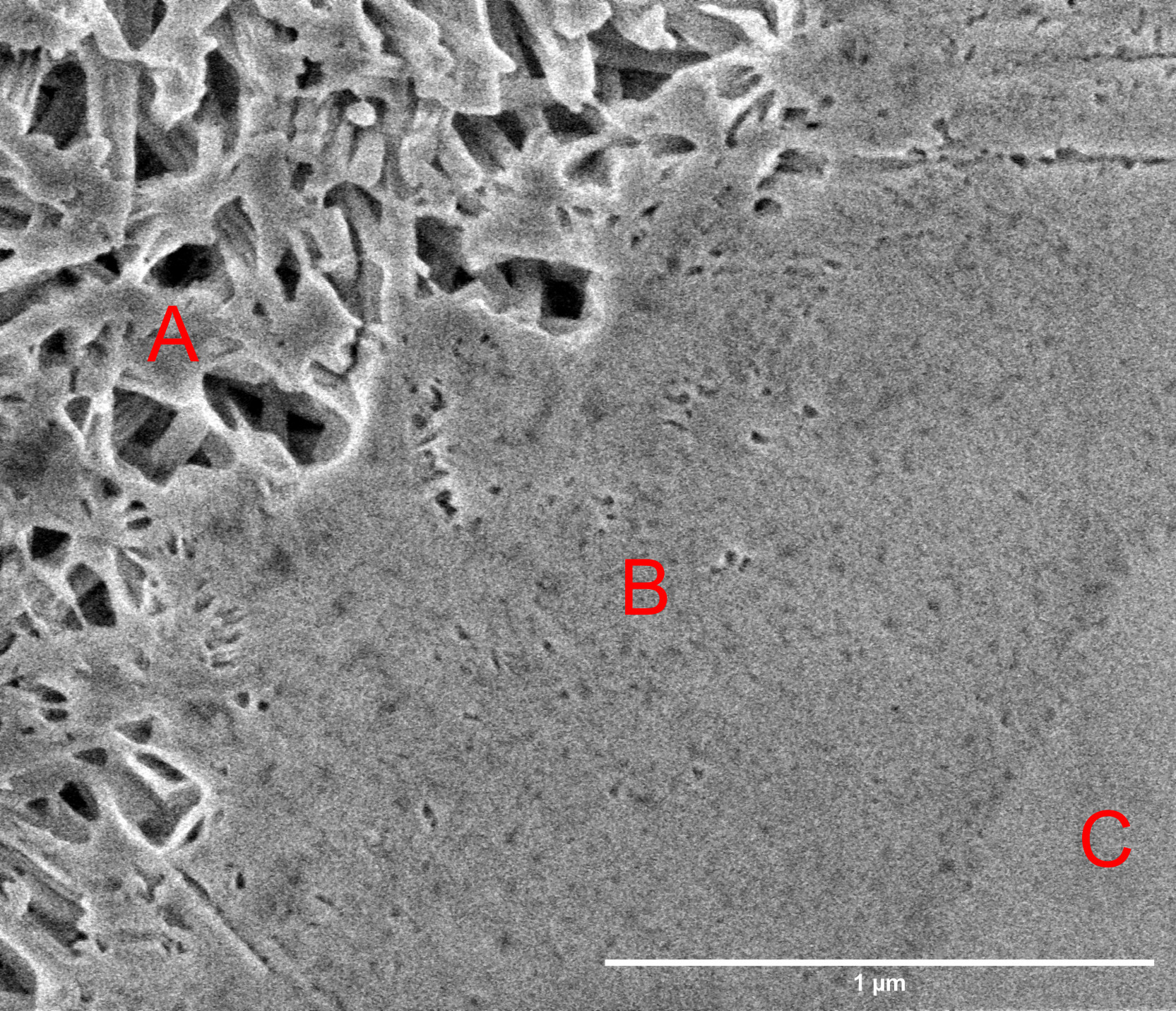}}
    \caption{SE images of 28\,d hydrated \alite without visible preparation artefacts after argon BIB sectioning.}
    \label{fig:reference_images}
\end{figure}

However, it is not possible to process any pore analysis on an undefined rough surface.
Therefore, the proposed preparation method using BIB is applied to these specimens.
While utilizing BIB can produce good results as shown in Figure \ref{fig:reference_images}, some artefacts inherent to this method will be discussed in the following sections.

\subsection{Effect of low voltage SEM on \csh}

\begin{figure}[htbp]
    \centering
    \subfigure[Initial images before final focus]{\includegraphics[width=0.49\linewidth]{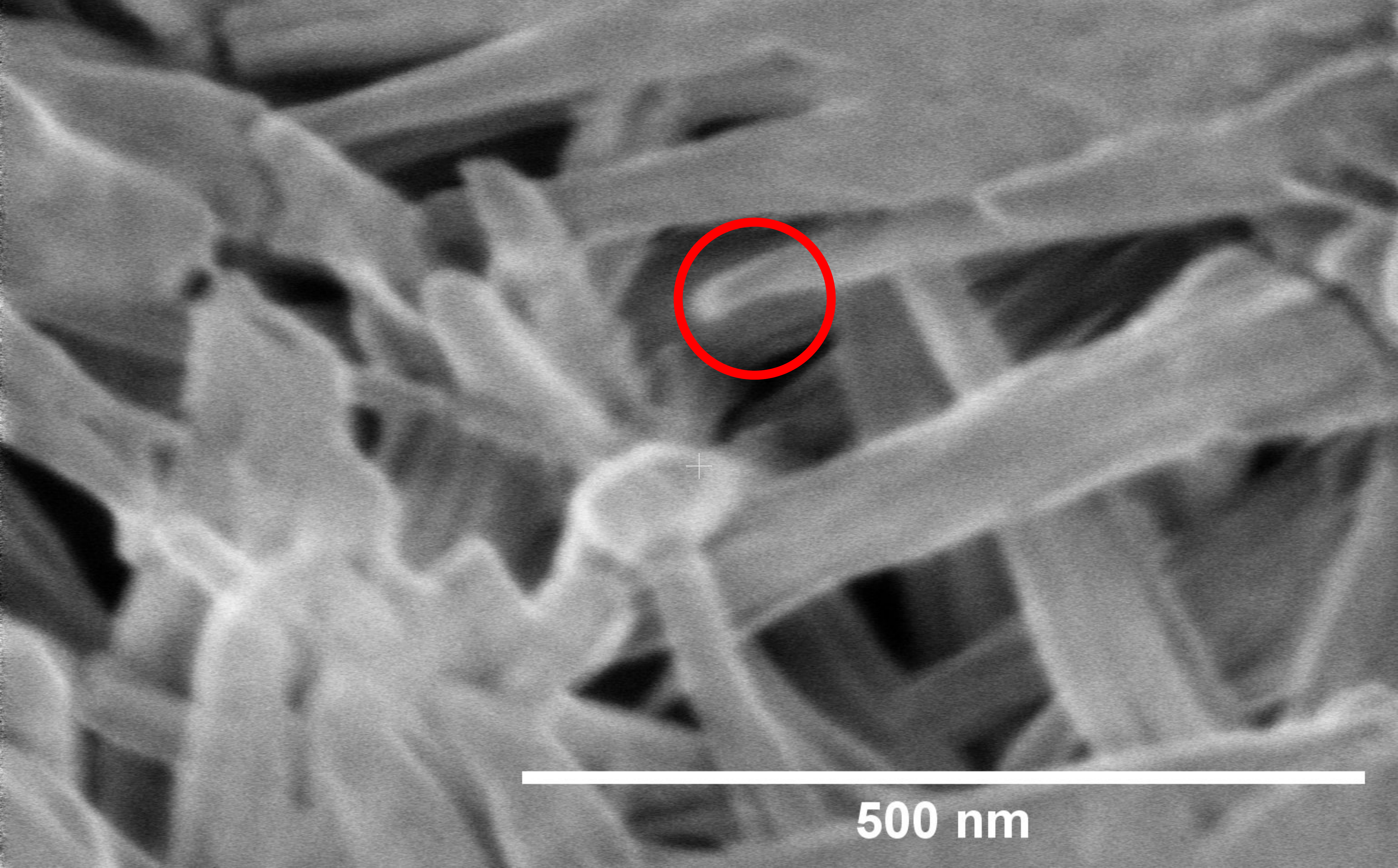}}\hspace{0.005\linewidth}
    \subfigure[Final focused image, 60\,s afterwards]{\includegraphics[width=0.49\linewidth]{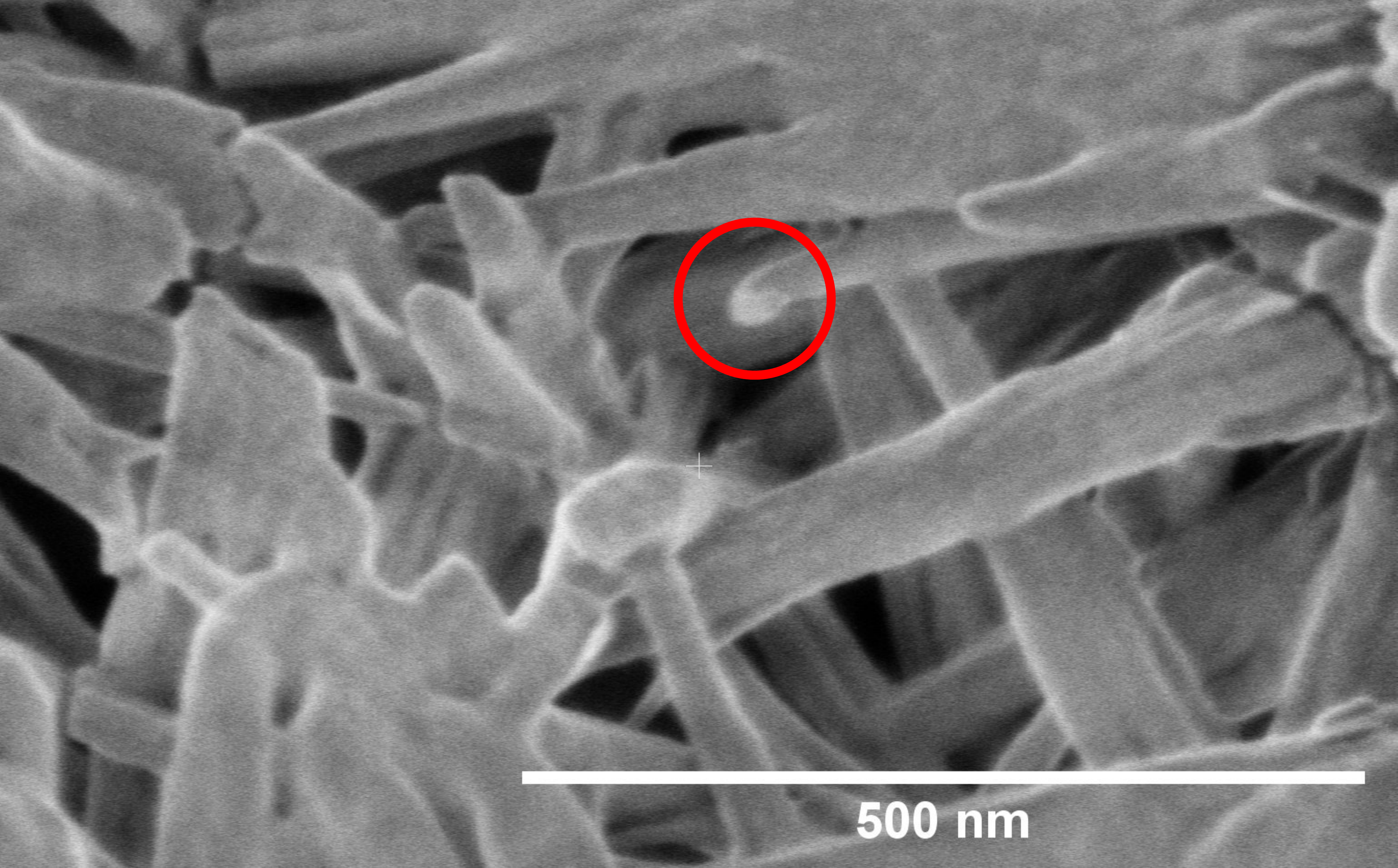}}
    \caption{Electron beam damaged \csh needle at 350\,V accelerating voltage and a stage bias of 200\,V. The left image was captured directly after moving to the position. The right image was captured 60\,s after the left image and a continuous electron beam exposure. A 5-frame-animation is also available \cite{Kleiner.2021d}.}
    \label{fig:electron_beam_damage}
\end{figure}

Due to the high water content, the electrical non-conductivity and the delicacy of the \csh needles, electron and ion beams can alter the shape of such components of the microstructure as needles and pores.
Even very low accelerating voltages and beam currents can induce damages to the needle-shaped structures of \csh in a manner of a few seconds.
As can be seen in Figure \ref{fig:electron_beam_damage}, the \csh needles start to bend or melt starting from the tip when they are exposed to an electron beam of 350\,V and 25\,pA for about 60\,s.
This is a direct consequence of the high energy density applied during the high resolution scanning.

\begin{figure}[!ht]
    \centering
    \includegraphics[width=\linewidth]{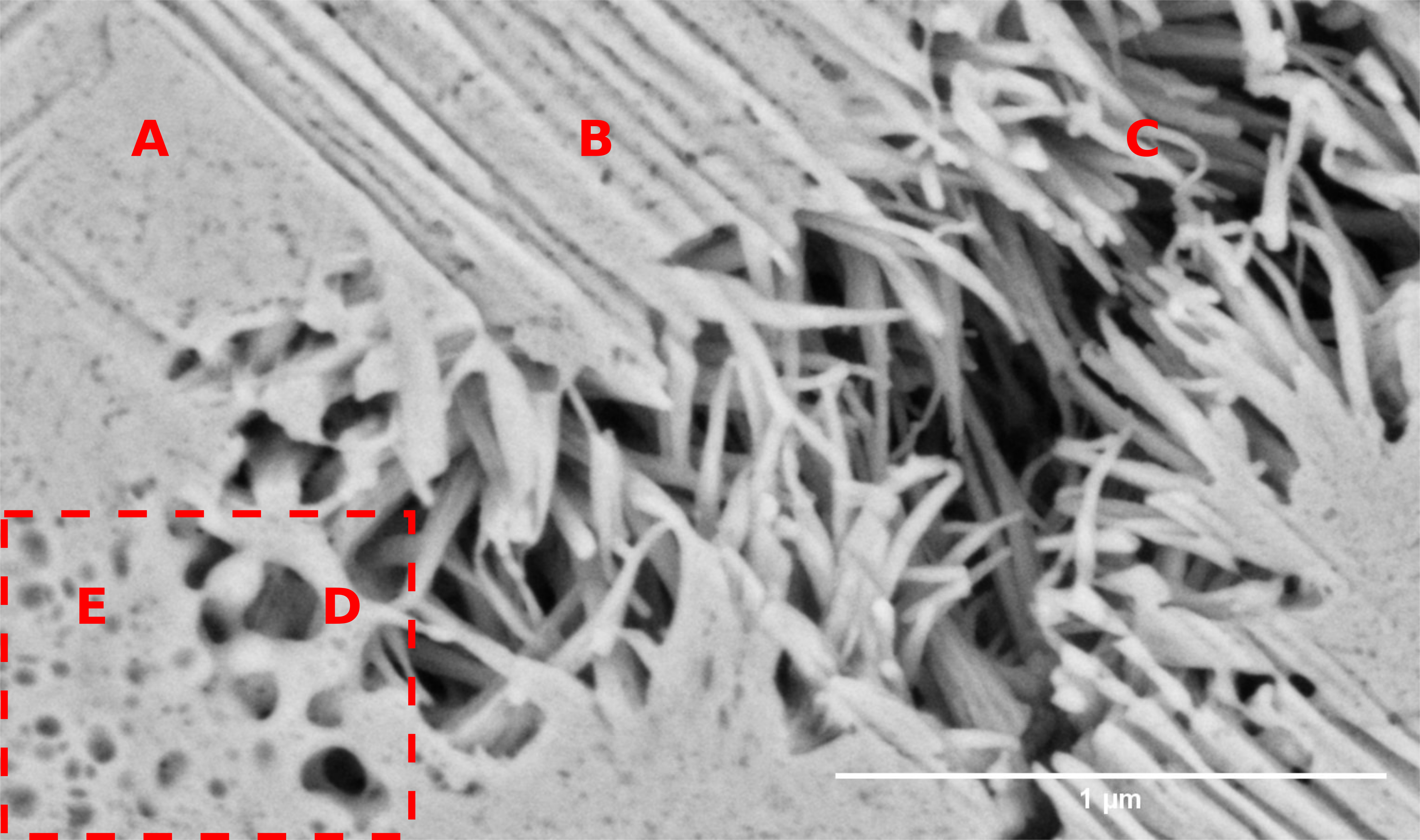}
    \caption{\csh with signs of various radiation damages.
    Undamaged pore structure (A), curtaining caused by BIB (B), deformed \csh-needles caused by BIB at 20\,$^\circ$C (C) and and by prolonged focusing of the electron beam (D), altered pore structure by focussing of the electron beam (E). (TLD-BSE, 2\,kV, 50\,pA)}
    \label{fig:BIB_damage}
\end{figure}

Furthermore, pores within the dense \csh-Phase seem to grow if the electron beam scans an area for a longer time period.
This can be seen in the marked area of Figure \ref{fig:BIB_damage}, where the electron beam stayed for 1\,-\,2 minutes to focus.
A comparable phenomenon has been observed by Rößler et al. \cite{Roler.2006} and Rossen \cite{Rossen.2014} using a TEM.

\subsection{Typical artefacts caused by argon BIB}

\begin{figure}[ht]
    \centering
    \includegraphics[width=\linewidth]{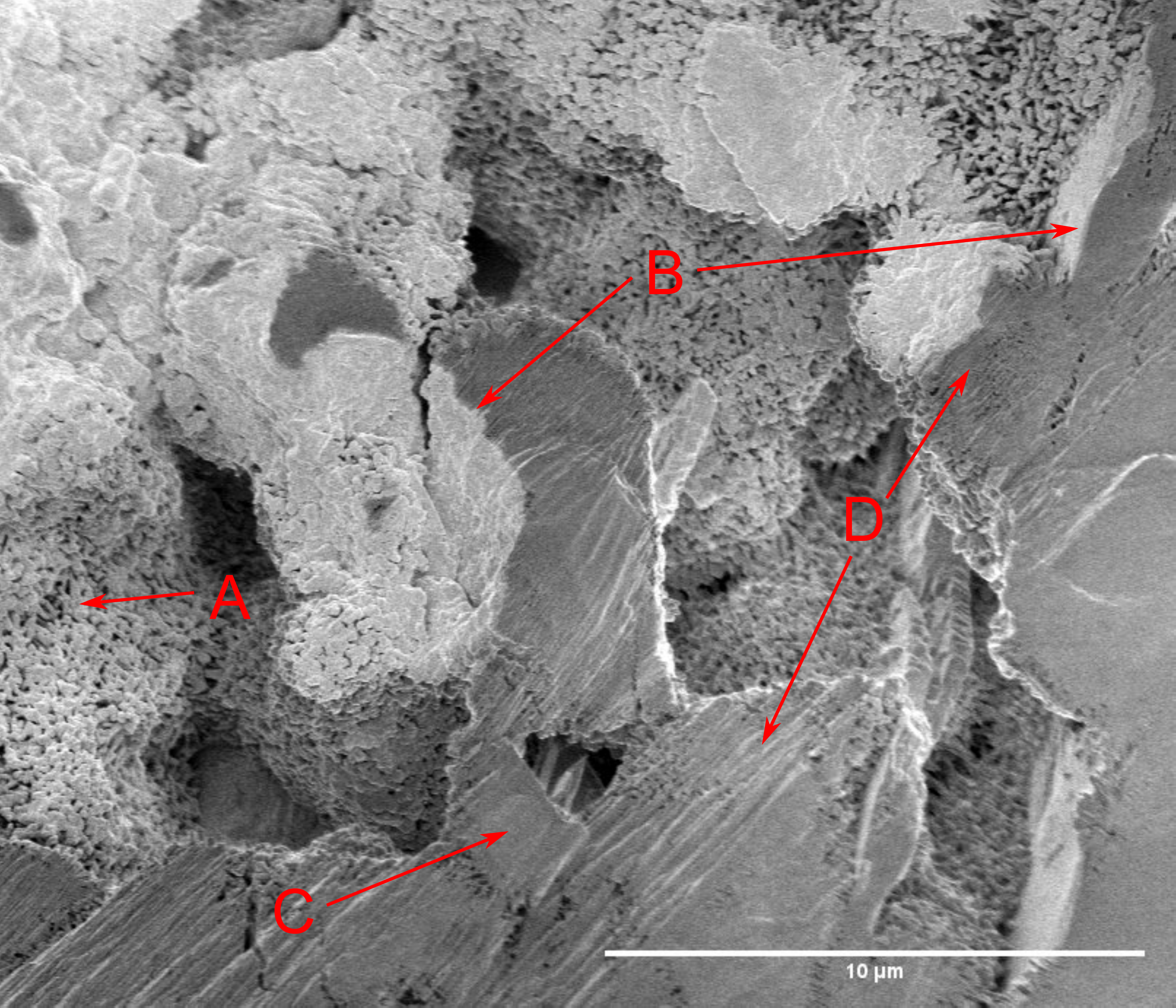}
    \caption{Material eroded by the BIB sectioning process deposits in a larger pore of a specimen. A: Exposed \csh needles seem to agglutinate due to coating with redeposited material.  B: Bright deposition layer containing tungsten (approximately 200\,nm thick). C: faintly visible alite grain, D: hydrates like \csh and CH. (TLD - beam deceleration mode, 500\,V, 6.3\,pA, stage bias 200\,V, immersion mode)}
    \label{fig:ablated_material}
\end{figure}

An ion beam is much more destructive than an electron beam and may induce a thermally damaged volume near the prepared surface \cite{Richardson.1993}.
Figure \ref{fig:BIB_damage} (Area C) visualizes the deformation of the \csh needles after the BIB preparation, especially in peripheral areas of the specimen or in very large pores.
As can be seen in Figure \ref{fig:ablated_material}, the \csh phases are damaged in larger pores after the BIB application.
This seems to be caused by the deposition of material eroded from the surface of the sample and the tungsten blade.

Another explanation for this appearance could be, that the tips of the \csh phases melt.
This seems unlikely, since the \csh needles directly at the surface or in smaller pores don't seem to be greatly affected after BIB sectioning.

\begin{figure}[htbp]
    \centering
    \subfigure[Curtaining artifacts are spread over the the surface.]{\label{fig:curtaining_a}\includegraphics[width=0.49\linewidth]{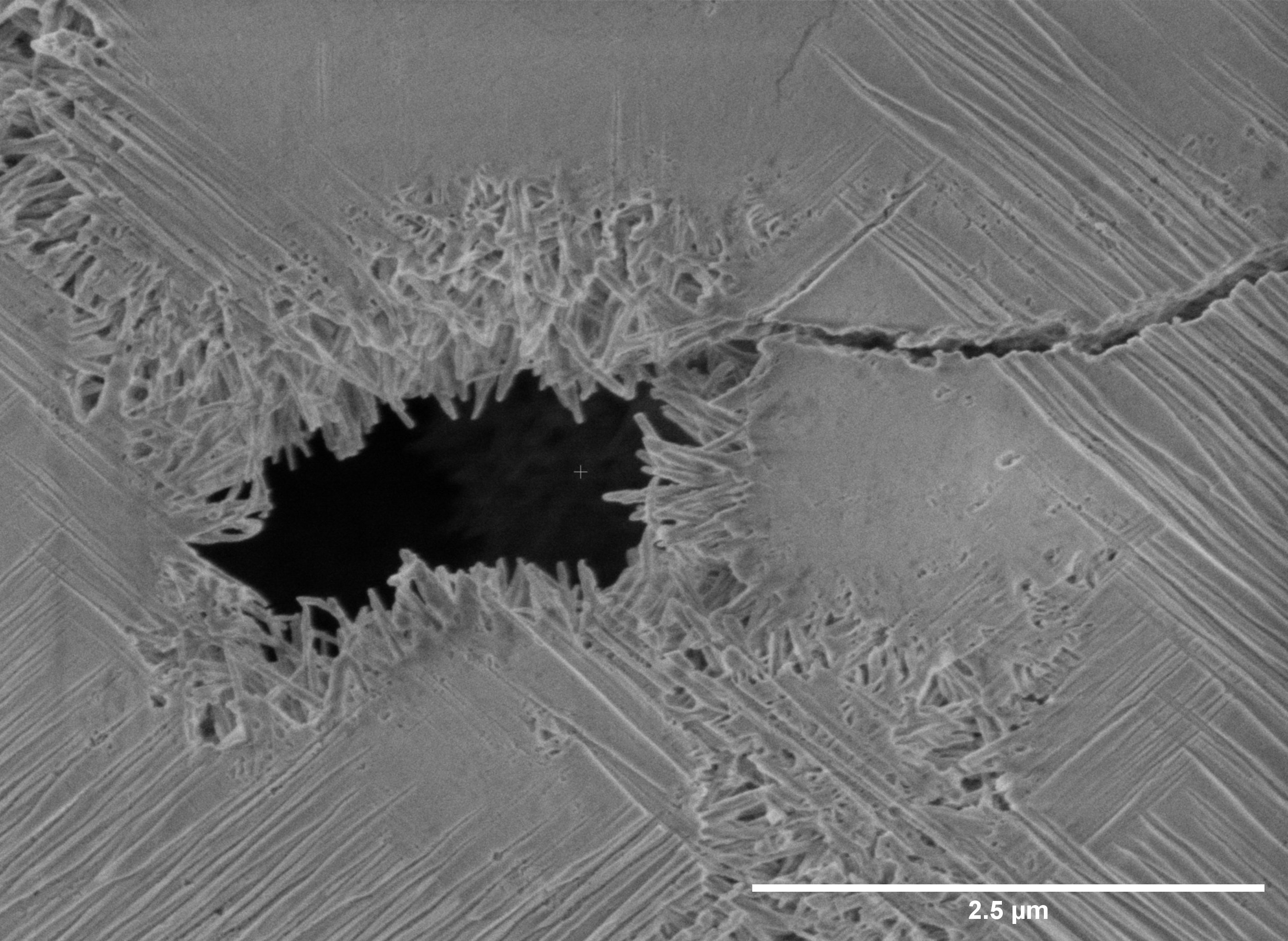}}\hspace{0.005\linewidth}
    \subfigure[Removed curtaining effect using FFT in both directions.]{\label{fig:curtaining_d}\includegraphics[width=0.49\linewidth]{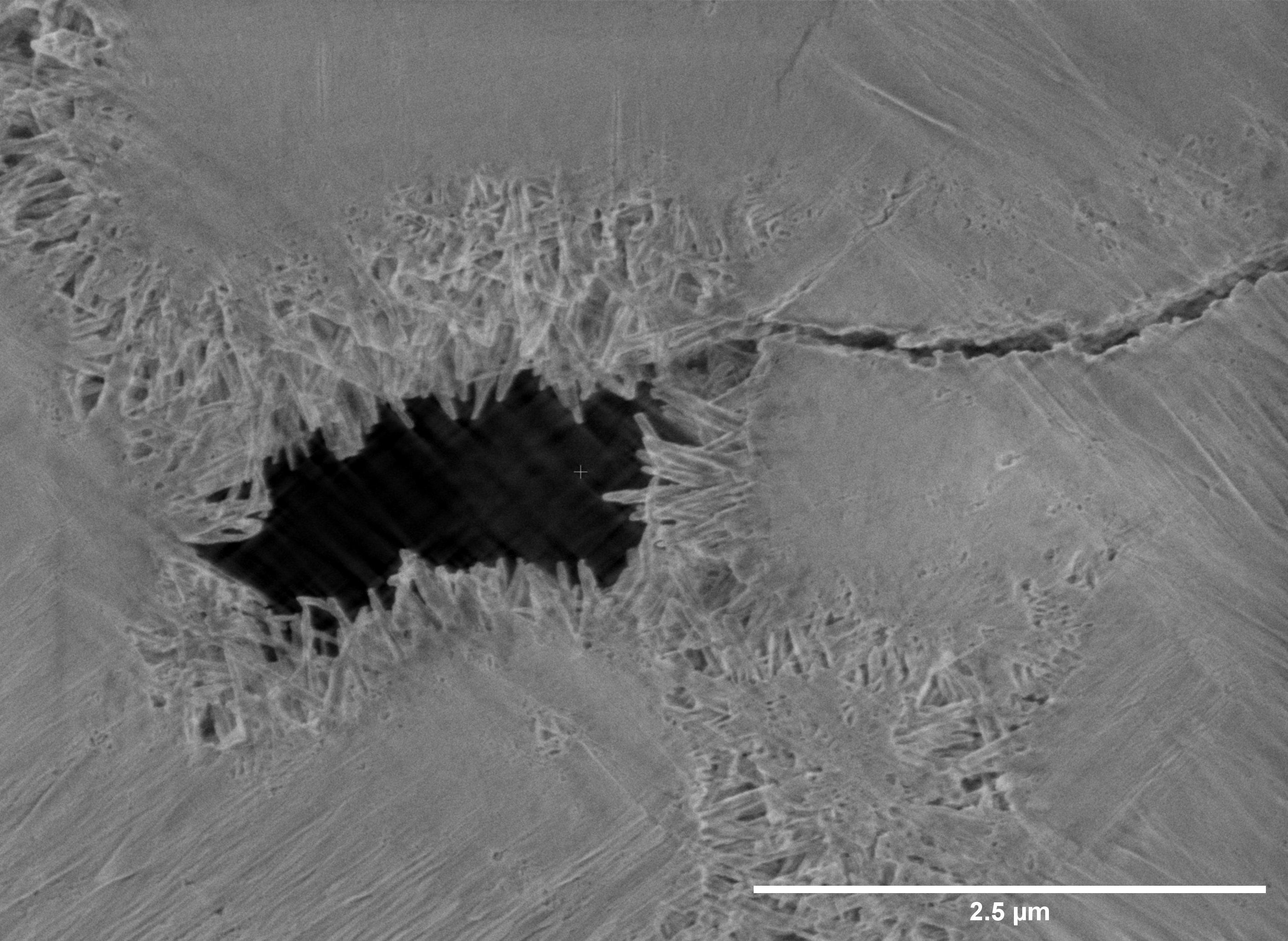}}
    \caption{Alite specimen (cured for 7 days) after cryo-BIB milling. (TLD-SE, 1\,kV, 3.1\,pA, stage bias 500\,V, immersion mode)}
    \label{fig:curtaining}
\end{figure}

Neither the specimen nor the ion sources are moveable in the used configuration of the ion beam device.
Therefore, the specimen shows some artefacts known as curtaining (also known as waterfall effect) \cite{Fitschen.2017,Mac.2018}.
As there were no corresponding damages to the tungsten blade, this effect has to be caused by different material densities and therefore varying resistance levels to the ion beam within the sample material.
Since three ion beams erode the surface, also multiple curtaining directions occur as can be seen in Figure \ref{fig:curtaining_a}.

\begin{figure}[!ht]
    \centering
    \includegraphics[width=0.7\linewidth]{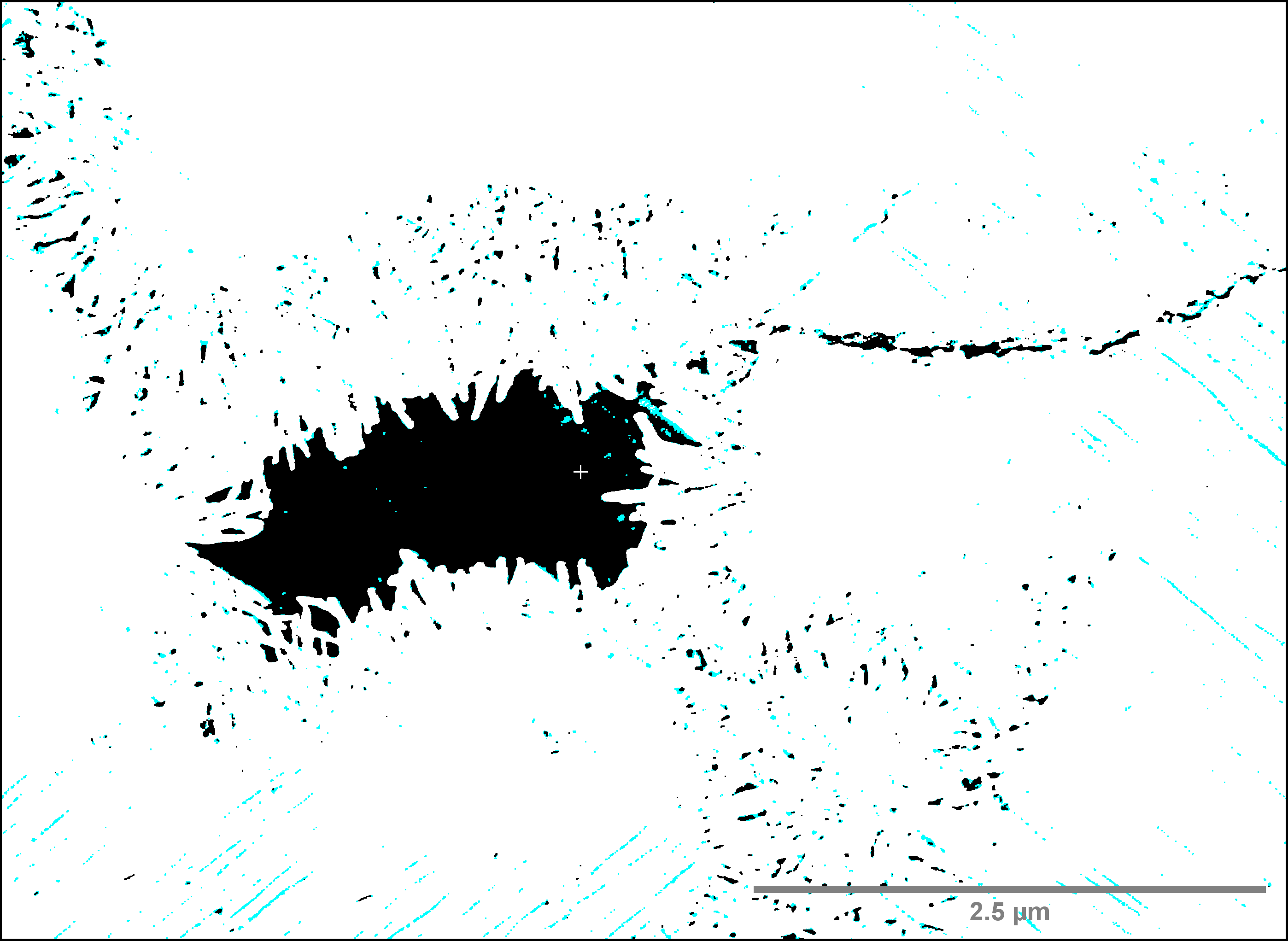}
    \caption{Thresholding results of Figure \ref{fig:curtaining} (Auto Local Threshold, method Phansalkar \cite{Phansalkar.10.02.201112.02.2011}). white: segmented pores of Figure \ref{fig:curtaining_a}.; blue: additional areas which were segmented due to curtaining in Figure \ref{fig:curtaining_a}.}
    \label{fig:curtaining_threshold}
\end{figure}

The effect can be diminished using Fast Fourier Transformation (FFT) based filtering provided in the open source software package Fiji \cite{Schindelin.2012}.
This process reduces the quality of the image, but makes it much easier to set a threshold for image binarization (compare Figure \ref{fig:curtaining_threshold}).
To limit the loss of image quality, in most cases only one direction of curtaining marks should be removed.
Nevertheless, imaging area with as little curtaining as possible  is preferable to any post-processing.

As can be seen in Figures \ref{fig:reference_images_a} and \ref{fig:curtaining_a} the \csh-phases are still intact after 6 hours of argon BIB application.
Also small pores and fibre structures are still visible. 

\subsection{Resin embedding}

\begin{figure}[htbp]
    \centering
    \subfigure[Raw image, TLD-SE, 2\,kV, 14\,pA]{\label{fig:oversegmentation_a}\includegraphics[width=0.48\linewidth]{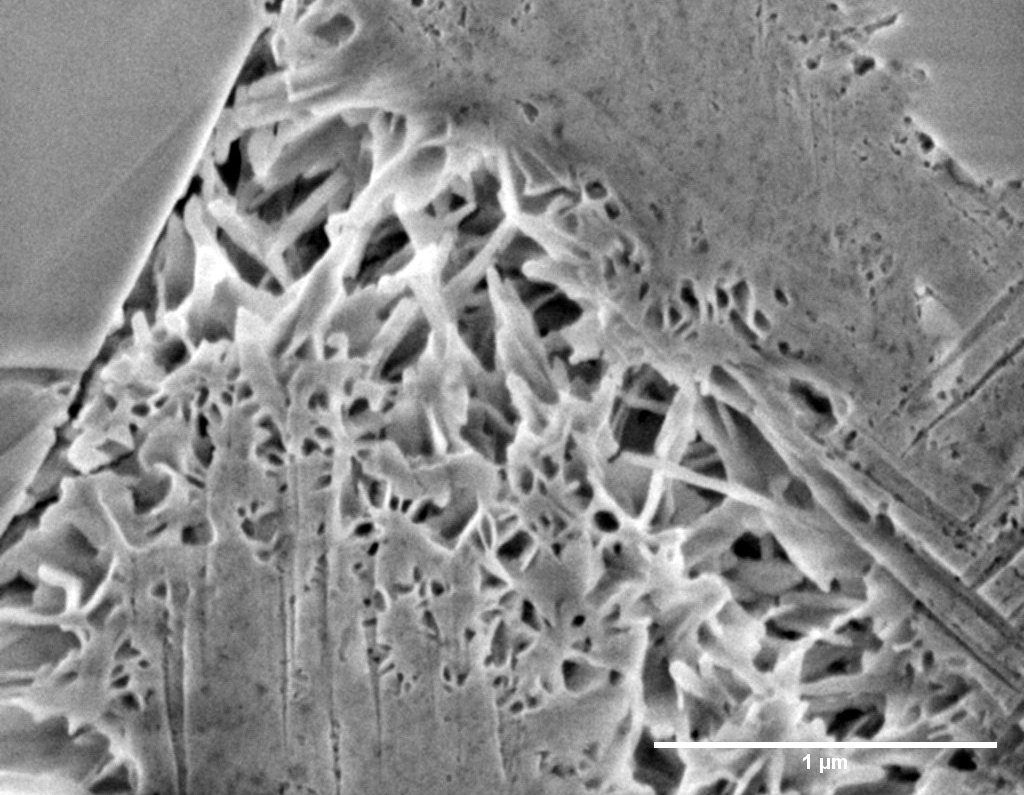}}\hspace{0.005\linewidth}
    \subfigure[Segmented image (manual thresholding)]{\label{fig:oversegmentation_b}\includegraphics[width=0.48\linewidth]{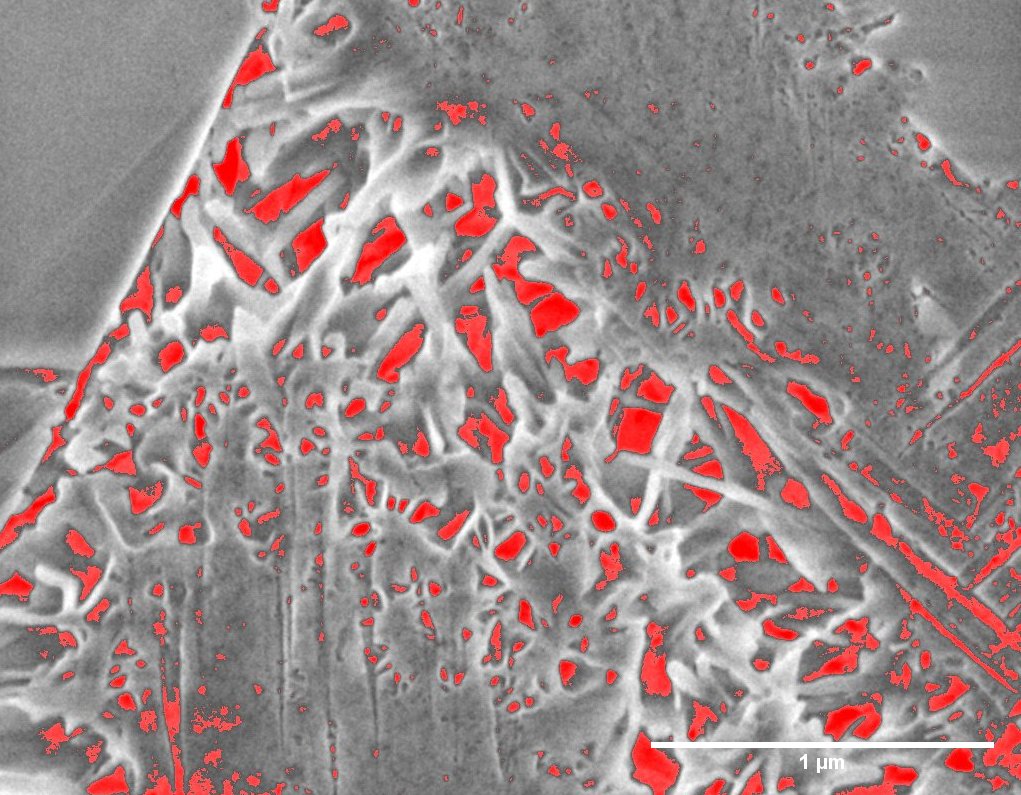}}
    \caption{7\,d hydrated \alite, segmented pores of porous outer \csh.}
    \label{fig:oversegmentation}
\end{figure}

Whereas BIB makes high quality surfaces possible while maintaining structures sensitive to higher temperatures or mechanical stress, it can be hard to correctly segment pores in an automated process.
It is not practical to prepare infinitely thin specimens for SEM investigations.
Therefore, the pore background of non-embedded samples is often visible (Figures \ref{fig:reference_images_a} and \ref{fig:oversegmentation_a}).

This leads to an incorrect segmentation of pores in the image analysis. 
In this case, the pore size is significantly underestimated and therefore, too many pores are recognized (Figure \ref{fig:oversegmentation}).

\begin{figure}[htbp]
    \centering
    \subfigure[Overview image, TLD-BSE, 1\,kV, 0.1\,nA]{\label{fig:resin_bse1}\includegraphics[width=0.7\linewidth]{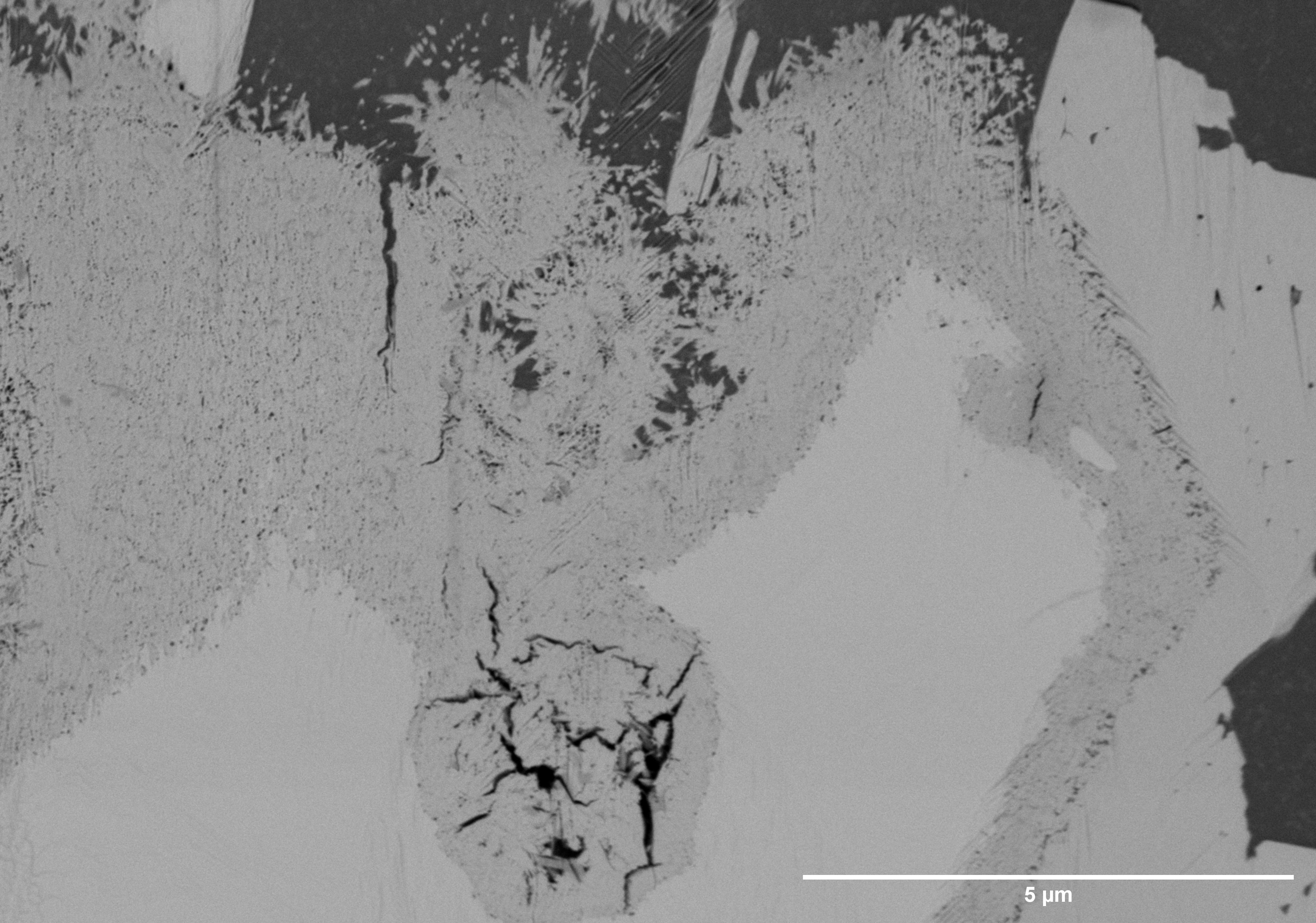}}
    \subfigure[Detail view of Figure \ref{fig:resin_bse1} showing very good contrast of larger pores and presence of nano-scaled pores in inner \csh]{\label{fig:resin_bse2}\includegraphics[width=0.7\linewidth]{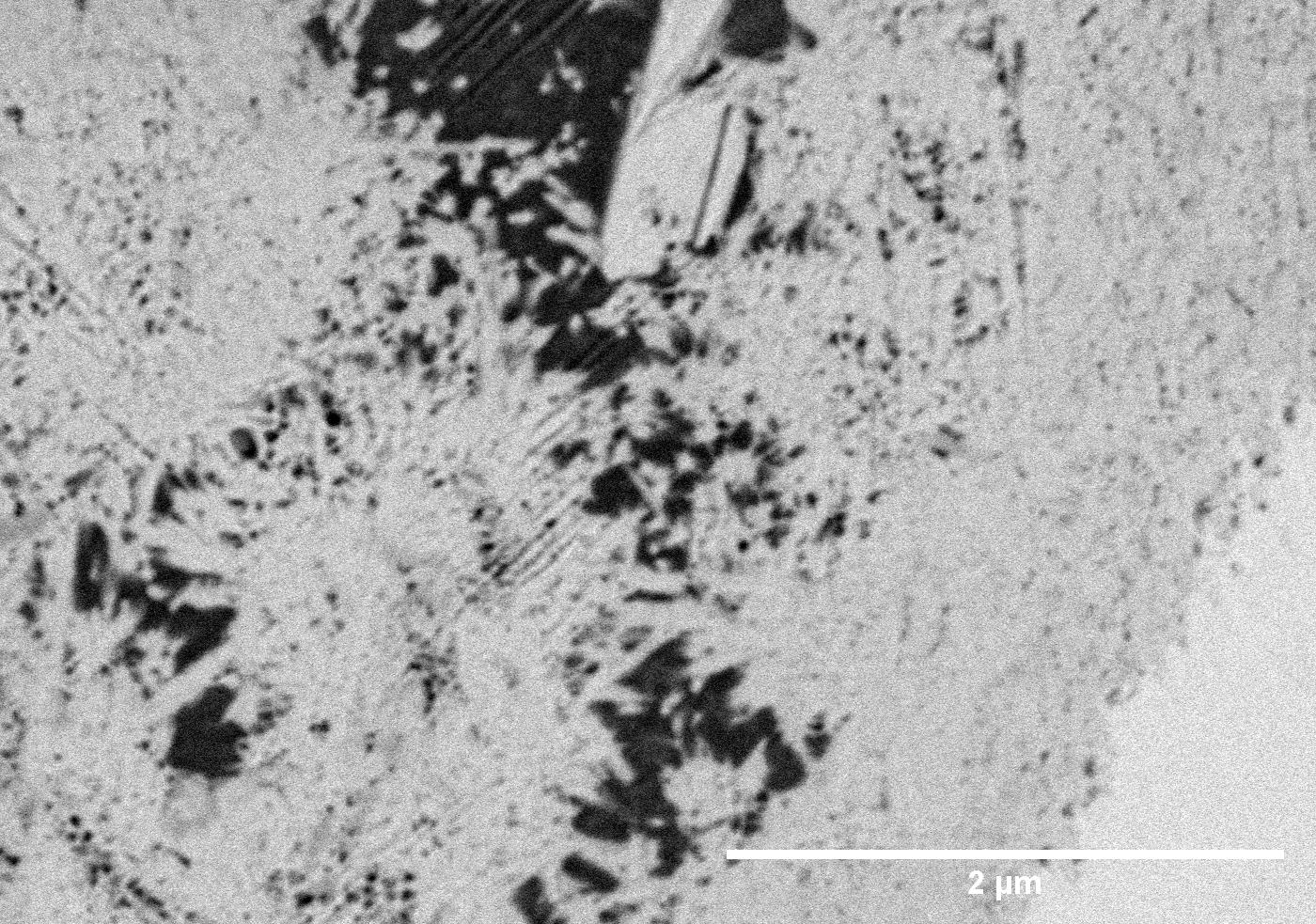}}
    \caption{BSE images of 7\,d hydrated \alite embedded in resin, polished using cryo-BIB. }
    \label{fig:resinembedding}
\end{figure}

A possibility to avoid this problem is to embed the the specimen in a low viscosity resin to fill the pores.
Resin embedded specimens are as easy to polish using argon BIB at room temperature as resin free specimens.
As shown in  Figure \ref{fig:resinembedding}, the visibility of the pore background of large pores is significantly reduced due to the embedding.

Due to the applied low acceleration voltage for imaging, mainly surface features are visible in SE mode, while the obtainable material contrast is very low.
While the trained human eye may be able to differentiate between resin filled pores and the \alite/\csh-phases, it is difficult to automatically segment those low contrast images.
In comparison, images made using BSE benefit from the high material contrast and the flat surface (Figure \ref{fig:resinembedding}).
Therefore, resin embedding is especially useful for the analysis of larger pores (i.e. $>$\,25\,nm diameter) using high resolution BSE-imaging.

\subsection{Pore analysis}

To determine the damage induced due to thermal stress during the BIB process at room temperature, it was compared against cryo-BIB.
Since subjectively no alteration of \csh structures in high resolution images after both preparation methods could be found, a pore analysis for the dense inner \csh (definition as proposed by Tennis and Jennings \cite{Tennis.2000}) was carried out.

\begin{figure}[htbp]
    \centering
    \subfigure[Raw image (TLD-SE, 2\,kV, 77\,pA)]{\label{fig:compare_cryo_a}\includegraphics[width=0.49\linewidth]{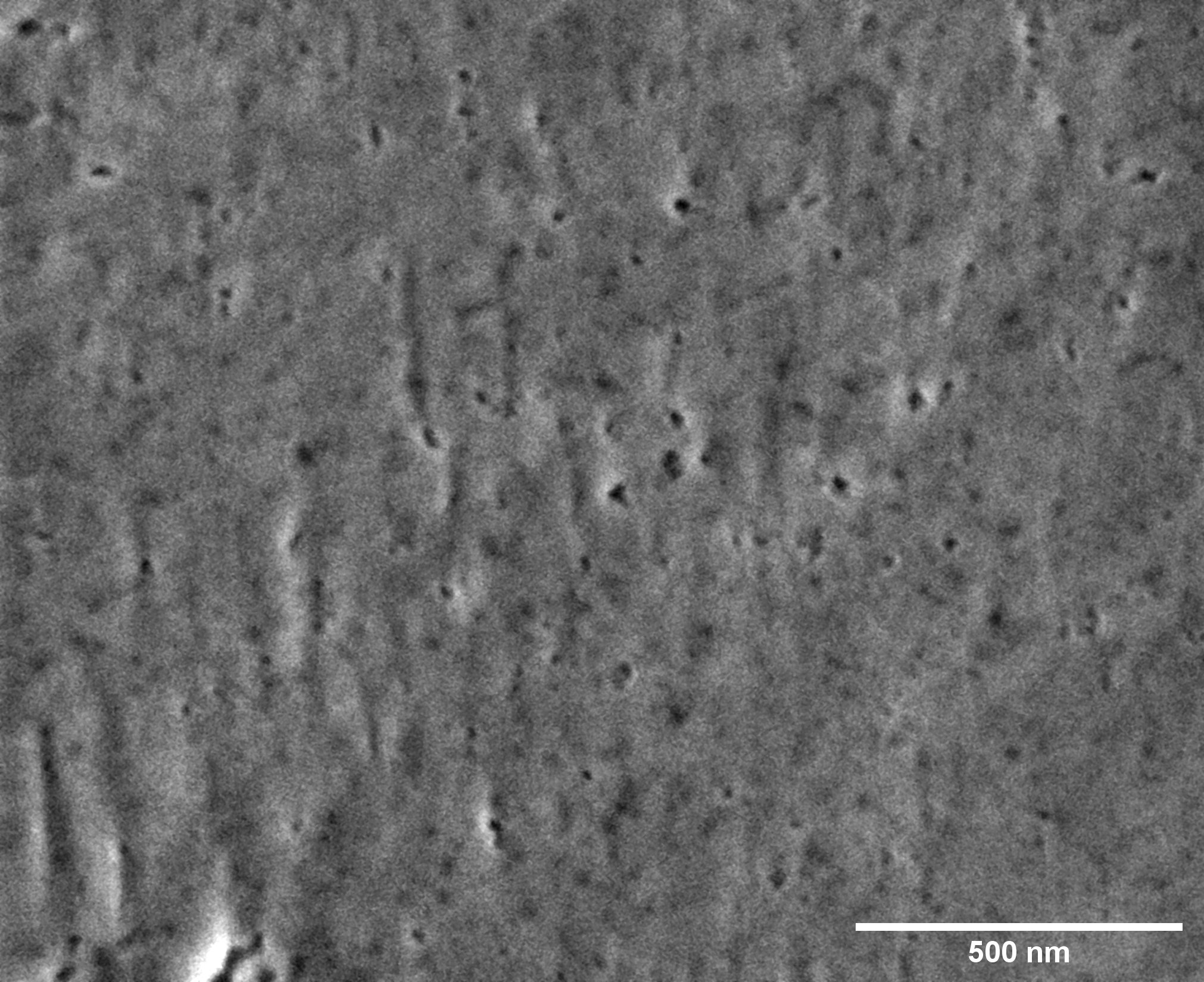}}\hspace{0.005\linewidth}
    \subfigure[Segmented pores]{\label{fig:compare_cryo_b}\includegraphics[width=0.49\linewidth]{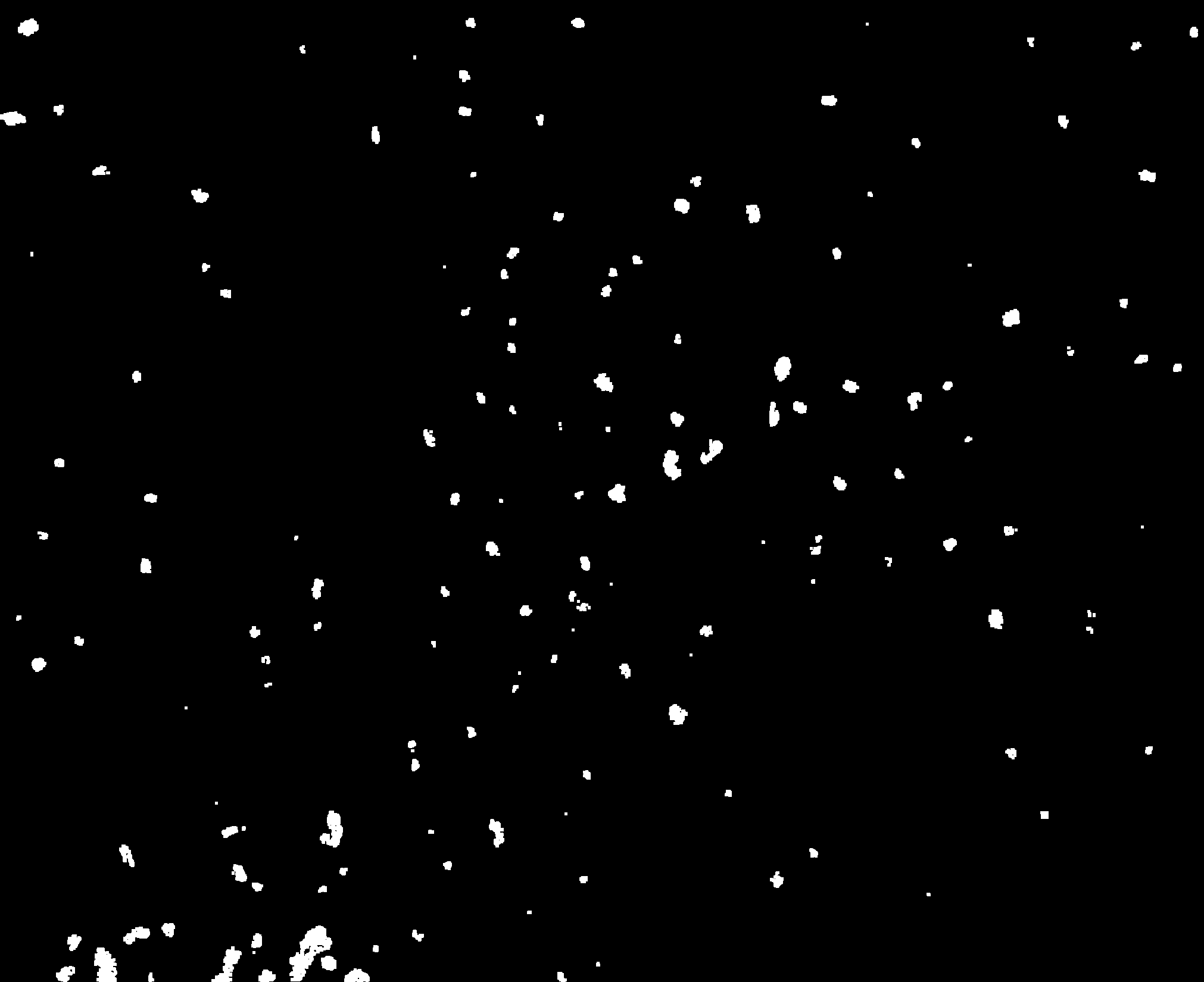}}
    \subfigure[CDF and PSD plot]{
   	 \label{fig:compare_cryo_c}
    \begin{tikzpicture}
        \begin{axis}[
            legend style={at={(1,1)},anchor=north east, font=\scriptsize},
            xmin=0, xmax=50,
            ymin=0, ymax=1.6,
            width=\linewidth,
            height=0.6\linewidth,
            grid=major,
            xlabel={pore diameter in nm},
            ylabel style={align=center},
            ylabel={normalized cumultative \\ pore diameter or area},
            xtick={0,10,...,50},
            ytick=\empty,
            grid style={line width=.1pt, draw=gray!10},
            ]
            \addplot[color=black] table[x=size,y=diameter,col sep=semicolon] {figure15c_CDF.csv};
            \addlegendentry{CDF (diameter)};
            \addplot[color=black, dashed] table[x=size,y=area,col sep=semicolon] {figure15c_CDF.csv};
            \addlegendentry{CDF (area)};
            \addplot[color=black, dashdotted] table[x=size,y=area,col sep=semicolon] {figure15c_PSD.csv};
            \addlegendentry{PSD (area)};
        \end{axis}
    \end{tikzpicture}
   }
    \caption{PSD of dense inner \csh (28\,d hydrated alite) obtained from SE image analysis (BIB at -140\,$^\circ$C).}
    \label{fig:compare_cryo}
\end{figure}
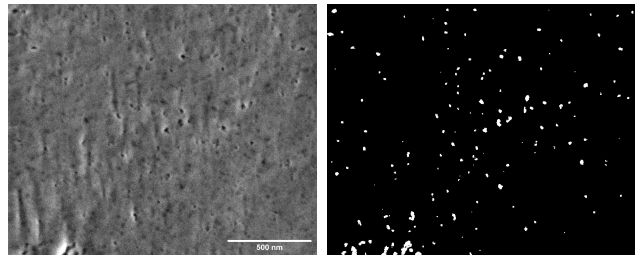
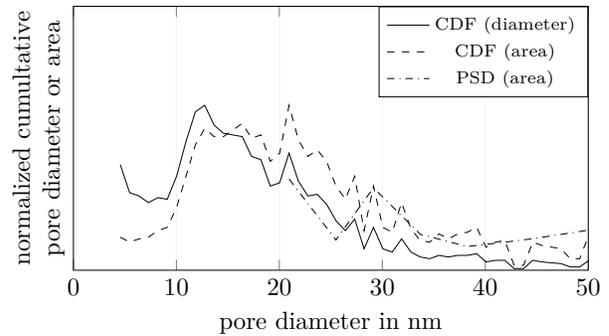

\begin{figure}[htbp]
    \centering
    \subfigure[Raw image (TLD-SE, 2\,kV, 89\,pA)]{\label{fig:compare_rt_a}\includegraphics[width=0.49\linewidth]{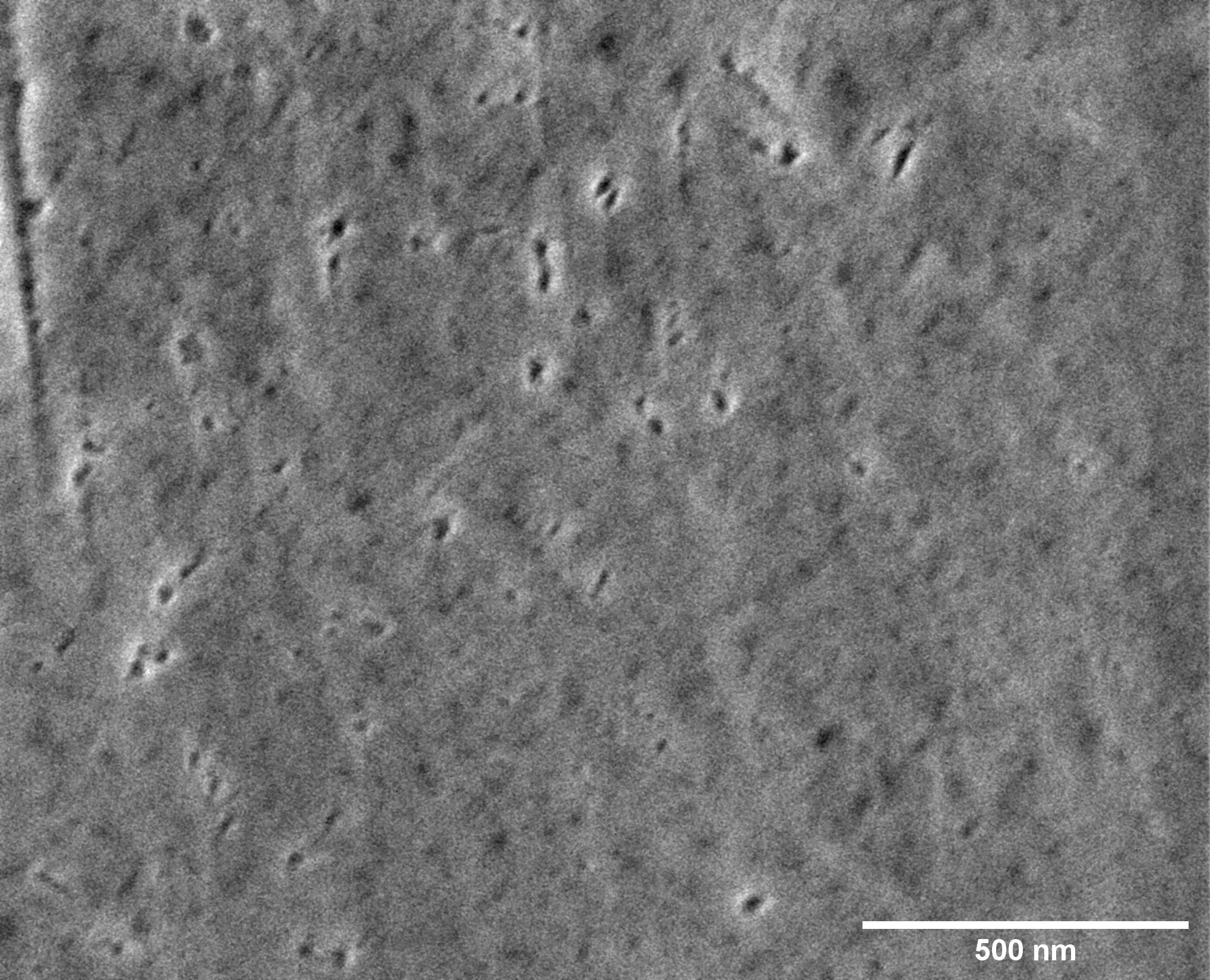}}\hspace{0.005\linewidth}
    \subfigure[Segmented pores]{\label{fig:compare_rt_b}\includegraphics[width=0.49\linewidth]{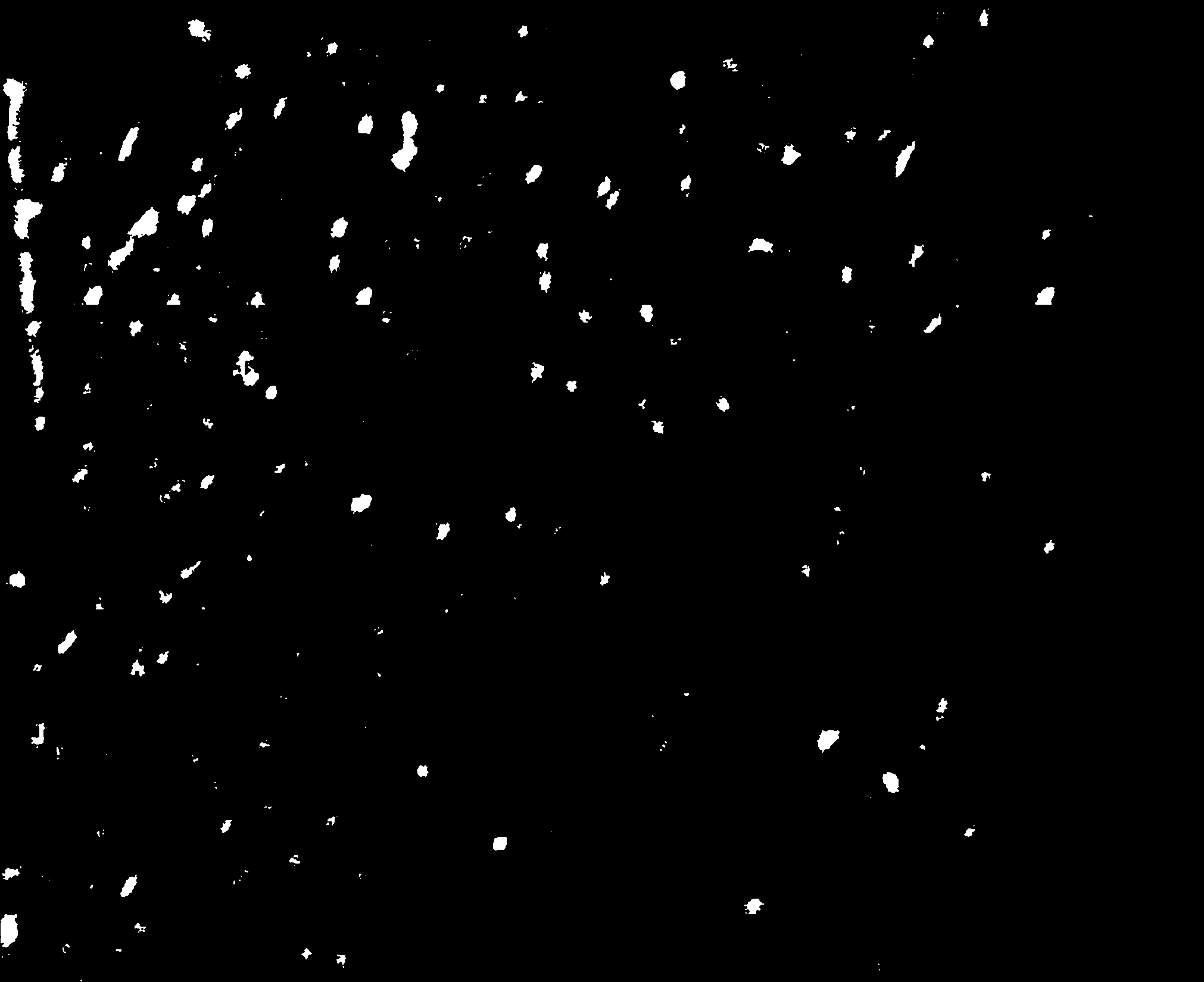}}
    \subfigure[CDF and PSD plot]{
   	 \label{fig:compare_rt_c}
    \begin{tikzpicture}
        \begin{axis}[
            legend style={at={(1,1)},anchor=north east, font=\scriptsize},
            xmin=0, xmax=50,
            ymin=0, ymax=1.6,
            width=\linewidth,
            height=0.6\linewidth,
            grid=major,
            xlabel={pore diameter in nm},
            ylabel style={align=center},
            ylabel={normalized cumultative \\ pore diameter or area},
            xtick={0,10,...,50},
            ytick=\empty,
            grid style={line width=.1pt, draw=gray!10},
            ]
            \addplot[color=black] table[x=size,y=diameter,col sep=semicolon] {figure16c_CDF.csv};
            \addlegendentry{CDF (diameter)};
            \addplot[color=black, dashed] table[x=size,y=area,col sep=semicolon] {figure16c_CDF.csv};
            \addlegendentry{CDF (area)};
            \addplot[color=black, dashdotted] table[x=size,y=area,col sep=semicolon] {figure16c_PSD.csv};
            \addlegendentry{PSD (area)};
        \end{axis}
    \end{tikzpicture}
   }
    \caption{PSD of dense inner \csh (28\,d hydrated alite) obtained from SE image analysis (BIB at 20\,$^\circ$C).}
    \label{fig:compare_rt}
\end{figure}
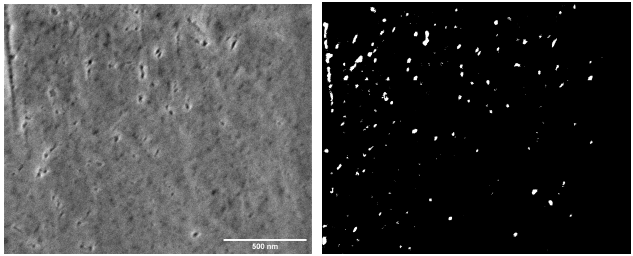
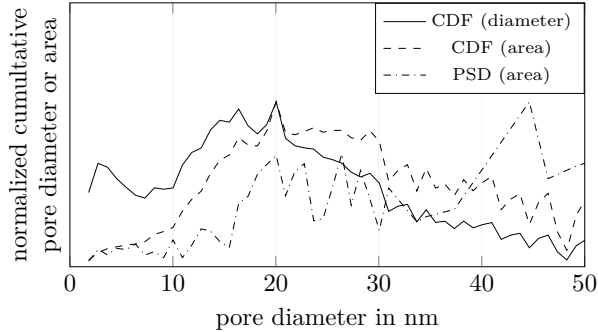

Figures \ref{fig:compare_cryo} and \ref{fig:compare_rt} show nano-size pores of a sample sectioned using BIB at -140 or 20\,$^\circ$C.
All images show dense inner \csh \ of 28\,d hydrated alite (as in Figure \ref{fig:resin_bse1}, upper left quadrant).
Pores with a diameter of 5\,nm can be clearly distinguished and analysed in the SE-images (resolution: 0.9\,nm per px, 2022\,$\times$\,1650).
As shown in Figures \ref{fig:compare_cryo_b} and \ref{fig:compare_rt_b}, the images are segmentable using the afore mentioned method.
Nevertheless, since it is still possible that the segmentation method may falsely detect some microstructural features, such as pores, the results should be checked manually and the areas with a erroneous segmentation should be excluded.

As can be seen in Figures \ref{fig:compare_cryo_c} and \ref{fig:compare_rt_c}, the CDF (converted from chord to area, dashed), shows a different distribution than the PSD, determined by the analysis of the pore areas (dash-dotted line).
Comparing these graphs, it becomes clear, that the PSD has very different gradients and is very noisy.
It is also dependent on the background noise of the image (compare Figures \ref{fig:compare_cryo_c} and  \ref{fig:compare_rt_c}), while the CDF provides more stable results.
The CDF analysis shows that the pores inside the \csh have a diameter of mainly 10 to 30\,nm.
However, these results base on single images and only illustrate the quality of the preparation method.

For each method (BIB at -140 or 20\,$^\circ$C), seven images were analysed using the presented process.
These images and the results can be found in the supplementary data. 
Since the pixel size and the image areas differed slightly, the images were analysed individually.
The position of the mean peak value of the CDF was used to compare both methods.
As presented in Table$~$1 in the supplementary attachment, the mean peak value for BIB at -140\,$^\circ$C is at a pore diameter of $17.5\,\pm\,2.8\,$nm and at a pore diameter of $22.7\,\pm\,5.4\,$nm for BIB at 20\,$^\circ$C.
Furthermore, Figures \ref{fig:compare_cryo_a} and \ref{fig:compare_rt_a} show no obvious difference in pore structure.
The detected variation in the CDF between both methods is negligible and in the opinion of the authors more dependent on the position and quality of the captured images.
This was confirmed by preparing \alite specimens with different BIB treatment times (3 or 6\,h) and voltages (3 or 6\,kV) at -\,140\,$^\circ$C and at 20\,$^\circ$C.
Instead, different surface qualities were found, at increasing distances from the crossover area of the three argon ion beams.
Nano pores, as shown in Figures \ref{fig:compare_cryo_a} and \ref{fig:compare_rt_a}, can be found especially in the green area marked in Figure \ref{fig:zoom_area}A, while in the yellow area these pores are rarely visible or covered by artefacts.
This was discovered in specimens treated at room temperature but also in cooled specimens.

\section{Conclusions}

Argon broad ion beam sectioning is widely used to prepare samples containing materials of varying hardness for SEM imaging.
This study has shown that it is also applicable for sensitive materials like 7 and 28\,d hydrated \alite.

In contrast to typically presented SEM images, specimens obtained using the BIB sectioning (at room temperature and -140\,$^\circ$C) show flat, high quality surfaces with minor alterations.
Similar to FIB serial sectioning, curtaining and redeposition of material can be found using argon BIB.
These features are not visible for every specimen or only in limited areas which can be excluded for image analysis.
Artefacts like curtaining are introduced using stationary sample holders, but are reducible by various image filters with a certain loss of image resolution.

\csh structures observed after cryo and room temperature BIB sectioning are very similar, which is in accordance to the findings of Groves et al. \cite{Groves.1986} based on sample preparation for TEM imaging.
The structural variation is masked by the damages caused by the electron beam of the SEM, which, depending on the acceleration voltage and beam current, can induce significant changes to the morphology of inner and outer \csh phases.
The results indicate, that on all investigated surfaces prepared using BIB sectioning, very small pores of about 5\,nm diameter can be visualized using the TLD-SE mode.
This was already proposed for BSE-images utilizing Monte Carlo simulations by Yio et al. \cite{Yio.2016}.

When performed at the temperature of -140\,$^\circ$C, the BIB method is more involved, and more time consuming than when it is performed at room temperature.
It can also be concluded that, regardless of the preparation temperature, the obtained pore size distribution of the \csh will be similar as long as the BIB accelerating voltage is kept within the range of 3\,kV to 6\,kV.
However, this conclusion may not hold for higher values of accelerating voltages.

The surfaces prepared using BIB sectioning can be used for image analysis to determine the pore size or the phase distribution.
Nevertheless, due to the small size of the area analysed, BIB sectioning as described in this paper is only applicable to statistically analyse a limited range of feature sizes.
For more heterogeneous samples like mortar or even concrete it can still provide a qualitative insight into the material, and quantitative-type observations for nano-meter sized features.

The good quality of the surface of BIB sections of hydrated alite allowed for the high resolution SEM images.
The imaging resolution is thereby depending mainly on the resolution of the microscope.
The highest resolution was obtained using the SEM immersion mode, modern in-lens BSE and SE detectors at low acceleration voltage (0.3 to 4.0\,kV).

Quantification of PSD on SEM images may be beneficial to improve the basic understanding of transport mechanisms in concrete.
In this study this was demonstrated using a limited number of images.

Furthermore, It must be noted that the PSD determined in this study is based on data derived from 2D-images.
A comparison with volumetric porosity measurements would require a transformation of the data, which is beyond the scope of the present work. 
Another approach could be serial sectioning and 3D-imaging using argon BIB as shown by Desbois et al. \cite{Desbois.2013}.
Nevertheless, the BIB section thickness achieved by Desbois et al. (>\,350\,nm) \cite{Desbois.2013} is much larger than the pores segmented in this study (5\,nm).
This leads to the conclusion that argon BIB is ideally combined with FIB-SEM to obtain high resolution volumetric data \cite{Hemes.2015, Kleiner.2021b}.

For future studies, considerably more images have to be segmented and analysed to be able to make statistically reliable statements considering the PSD in individual phases of hydrated alite.
Furthermore, the effects of this preparation method on hydrated Portland cement specimens should be studied.
For those types of specimens, using cooled-stage BIB might still be relevant as it will allow to preserve cement hydrates such as ettringite, which contain more structural water than \csh.

\section{Acknowledgment}

The research was supported by the Deutsche Forschungsgemeinschaft (DFG), grant number 344069666.

\bibliographystyle{elsarticle-num} 
\bibliography{BIB-Paper-CCR-arXiv}

\end{document}


\begin{frontmatter}

\title{Supplementary data: Argon Broad Ion beam sectioning and high resolution scanning electron microscopy imaging of hydrated alite}

\author{Florian Kleiner}
\author{Christian Matthes}
\author{Christiane Rößler}
\address{F. A. Finger-Institute for Building Materials Science, Bauhaus-University Weimar, 99423 Weimar, Germany}
 
\begin{abstract}
This document contains supplementary information (a selection of segmented images prepared using BIB at -140\,$^\circ$C and at 20\,$^\circ$C) for the article \textit{Argon Broad Ion beam sectioning and high resolution scanning electron microscopy imaging of hydrated alite}.
\end{abstract}

\end{frontmatter}



\section{Introduction}

This document contains the evaluation of 12 additional images as shown in the main article in Figures 16 and 17.
The document contains the images, the segmentation result, the evaluation of the pore size distribution (PSD) of every image and the chord density function (CDF) of the segmented pores.
All diagrams show the cumulative pore area of each pore diameter bin, since the pore count based histogram and a cumulative pore diameter overvalues the lower pore diameters.
For the PSD diagrams, the raw area measurements were used and for the CDF, the measured line lengths were calculated as diameters of a circle.

\section{Additional segmented images}

\begin{figure}[htbp]
    \centering
    \subfigure[raw image 1]{\label{fig:compare_cryo_1a}\includegraphics[width=0.43\textwidth]{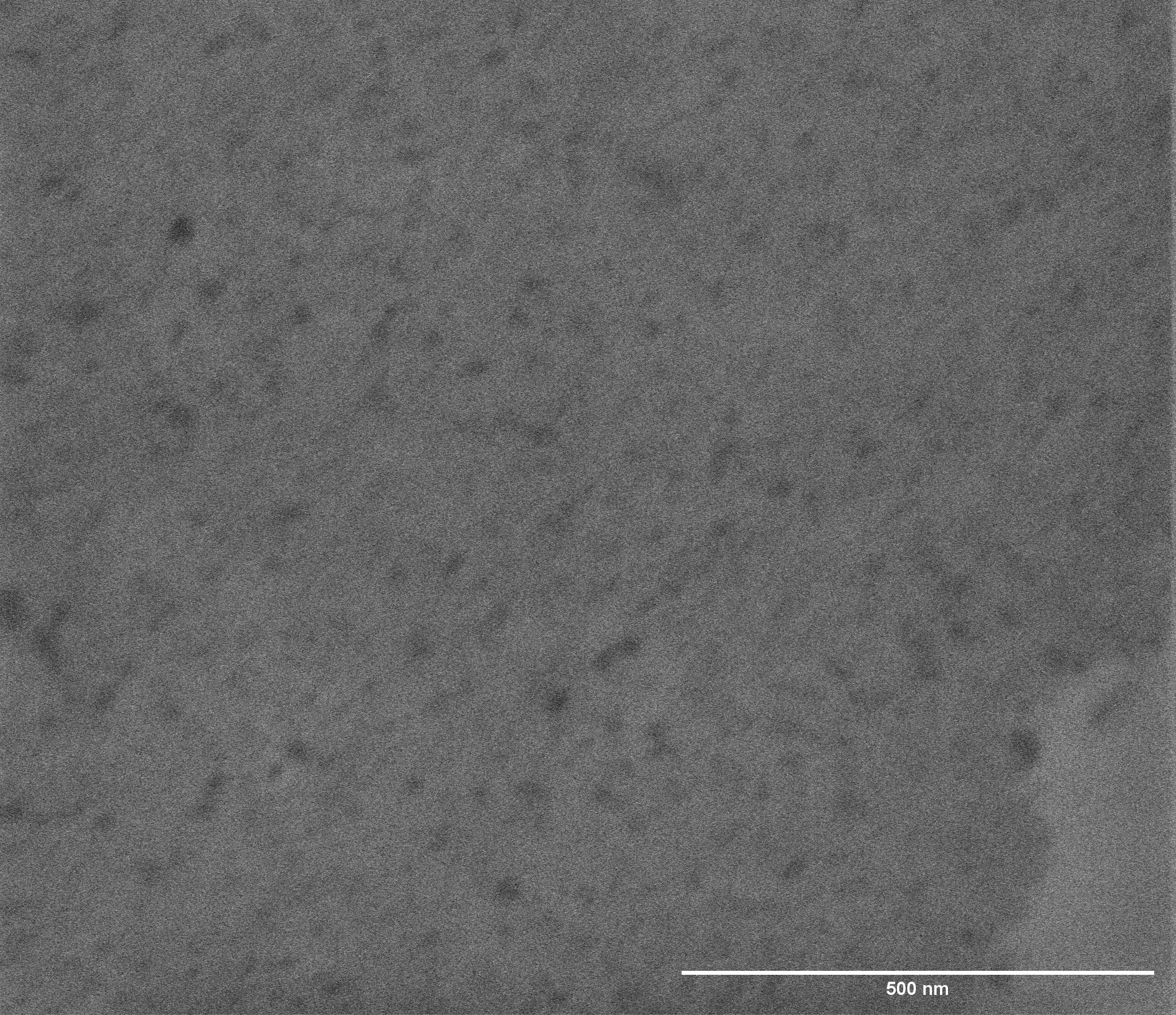}}\hspace{0.2cm}
    \subfigure[segmented image 1]{\label{fig:compare_cryo_1b}\includegraphics[width=0.43\textwidth]{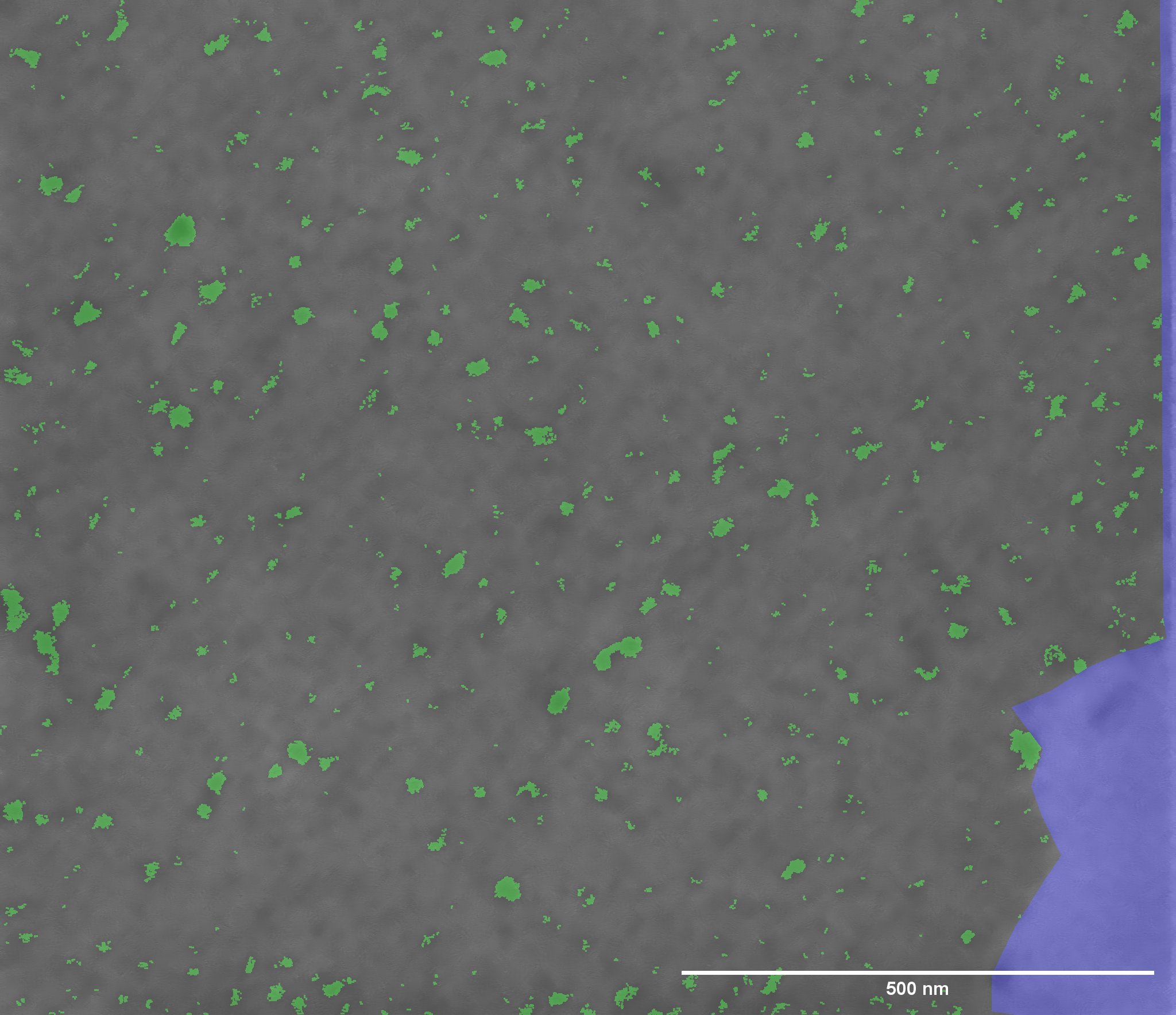}}
    \subfigure[raw image 2]{\label{fig:compare_cryo_1c}\includegraphics[width=0.43\textwidth]{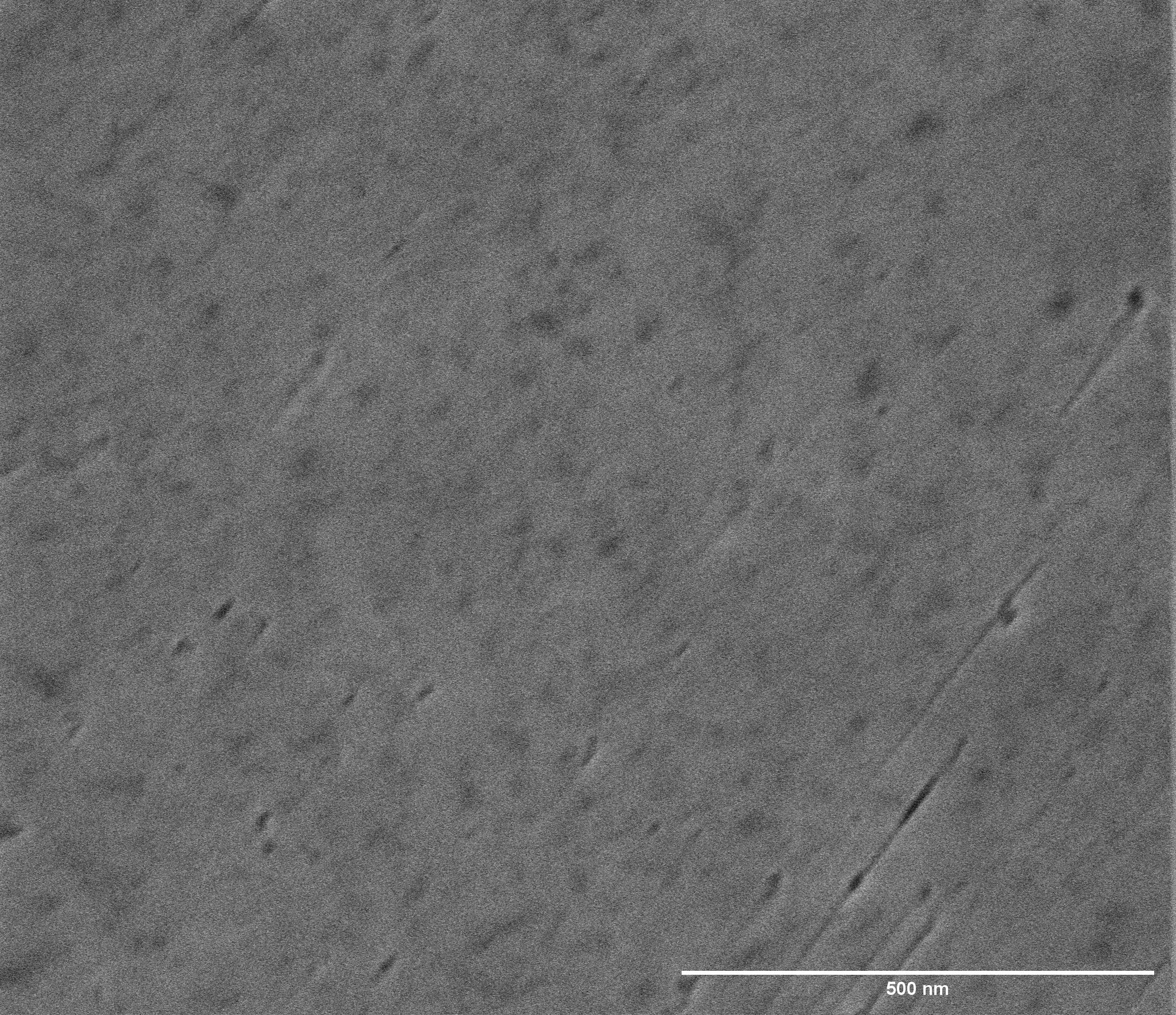}}\hspace{0.2cm}
    \subfigure[segmented image 2]{\label{fig:compare_cryo_1d}\includegraphics[width=0.43\textwidth]{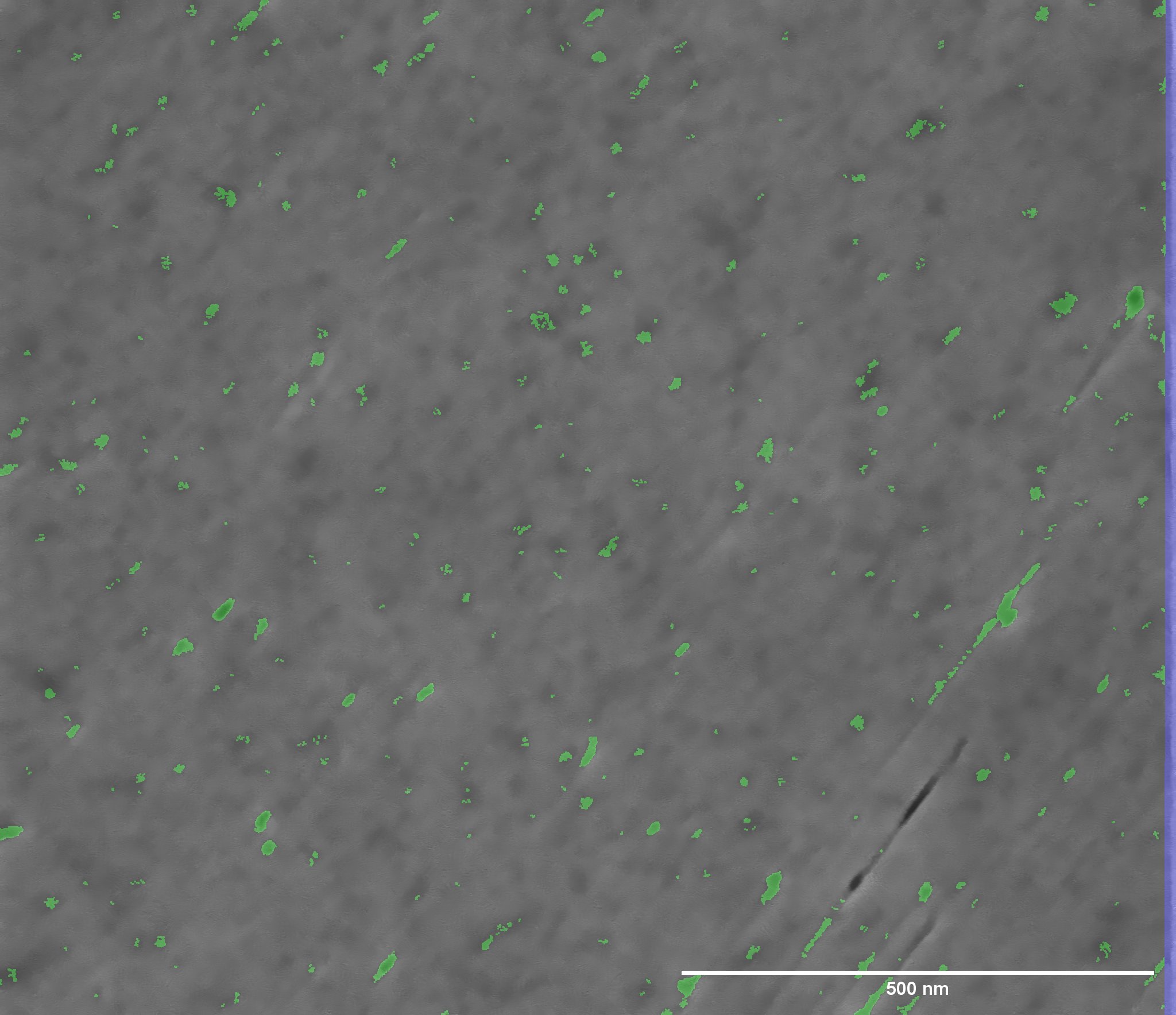}}
    \subfigure[raw image 3]{\label{fig:compare_cryo_1e}\includegraphics[width=0.43\textwidth]{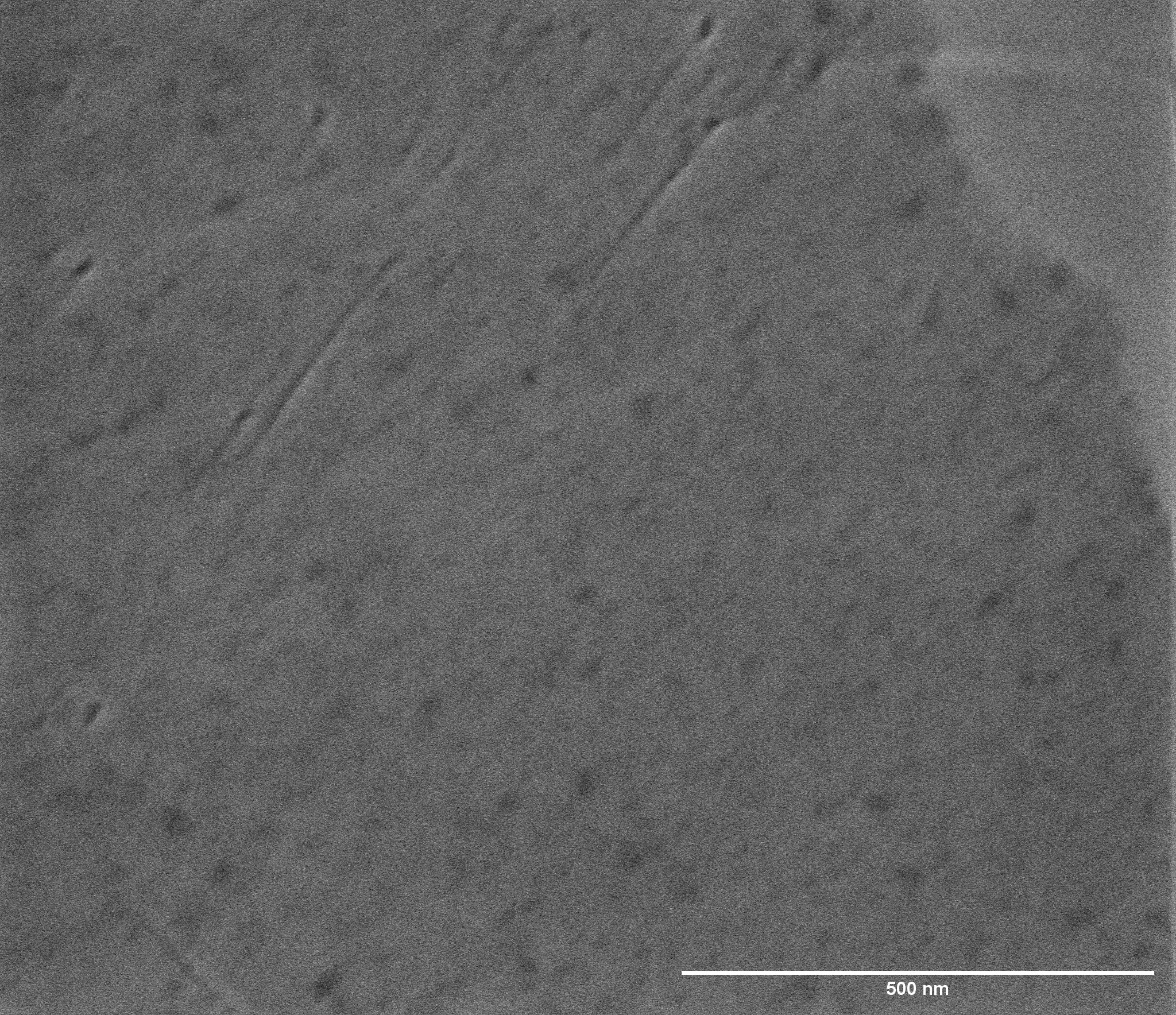}}\hspace{0.2cm}
    \subfigure[segmented image 3]{\label{fig:compare_cryo_1f}\includegraphics[width=0.43\textwidth]{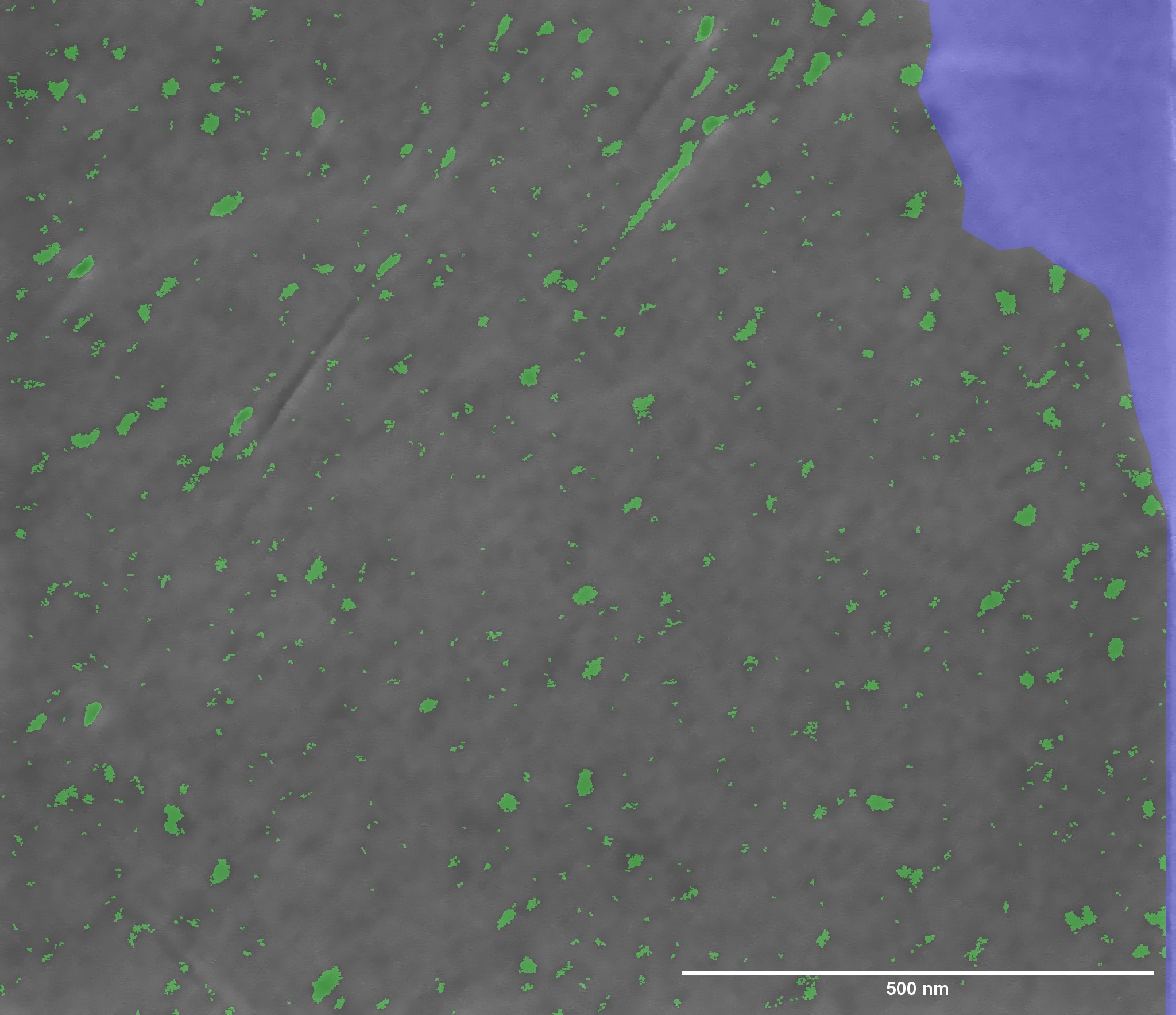}}
    \caption{Segmentation of pores within dense inner \csh, BIB at -140\,$^\circ$C, blue: ignored areas, green: segmented pores.}
    \label{fig:compare_cryo_1}
\end{figure}

\begin{figure}[htbp]
    \centering
    \subfigure[raw image 4]{\label{fig:compare_cryo_2a}\includegraphics[width=0.43\textwidth]{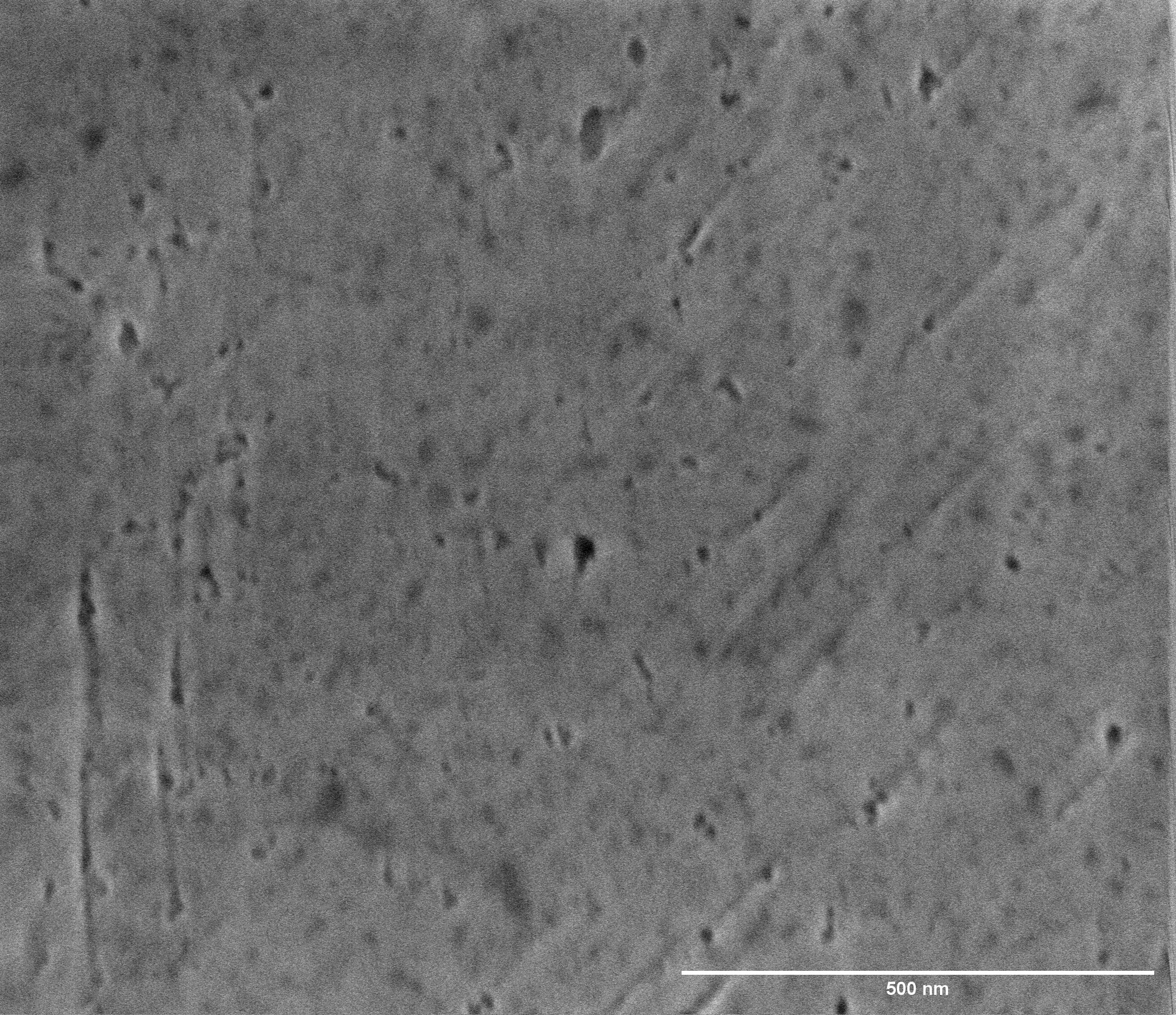}}\hspace{0.2cm}
    \subfigure[segmented image 4]{\label{fig:compare_cryo_2b}\includegraphics[width=0.43\textwidth]{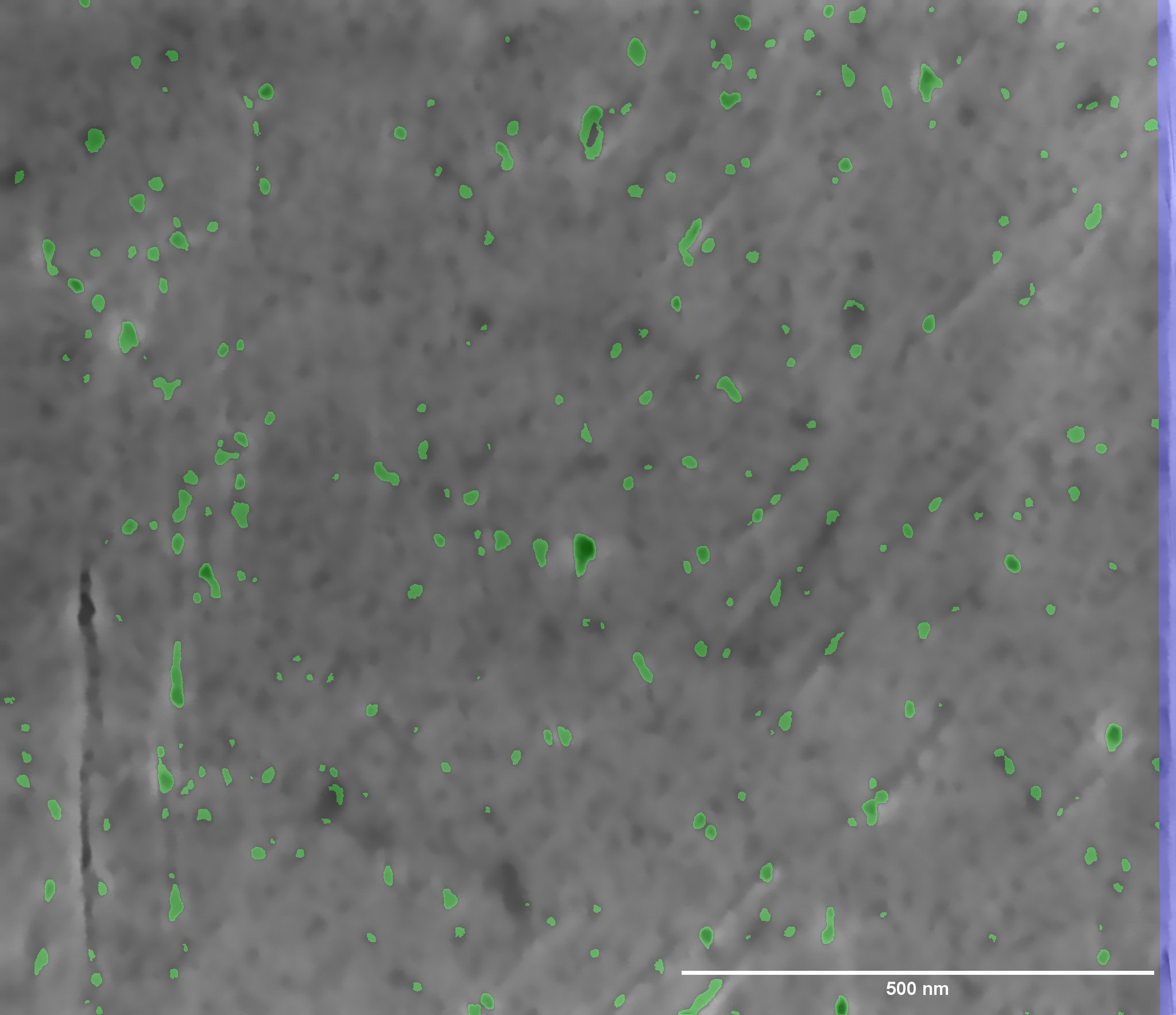}}
    \subfigure[raw image 5]{\label{fig:compare_cryo_2c}\includegraphics[width=0.43\textwidth]{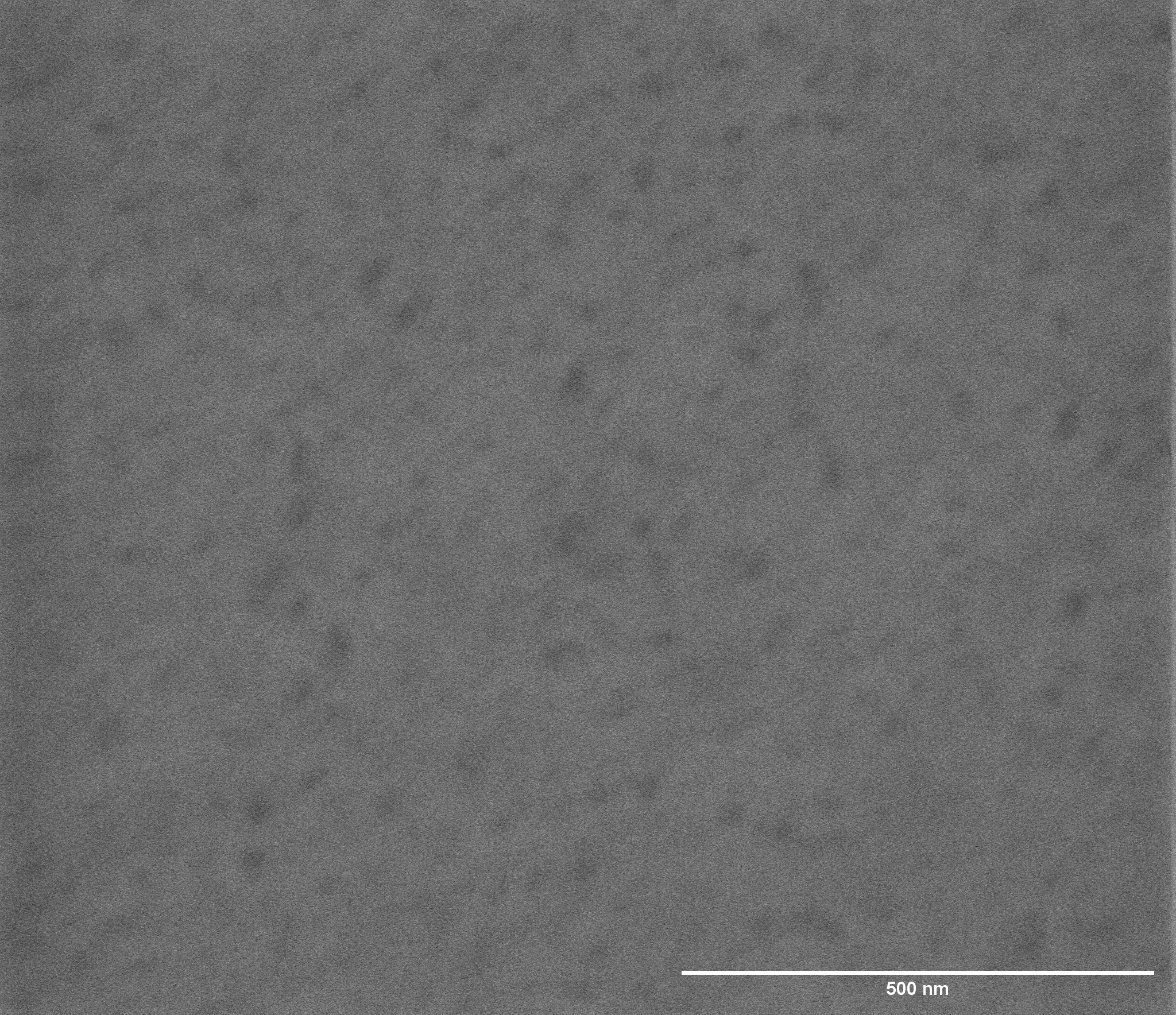}}\hspace{0.2cm}
    \subfigure[segmented image 5]{\label{fig:compare_cryo_2d}\includegraphics[width=0.43\textwidth]{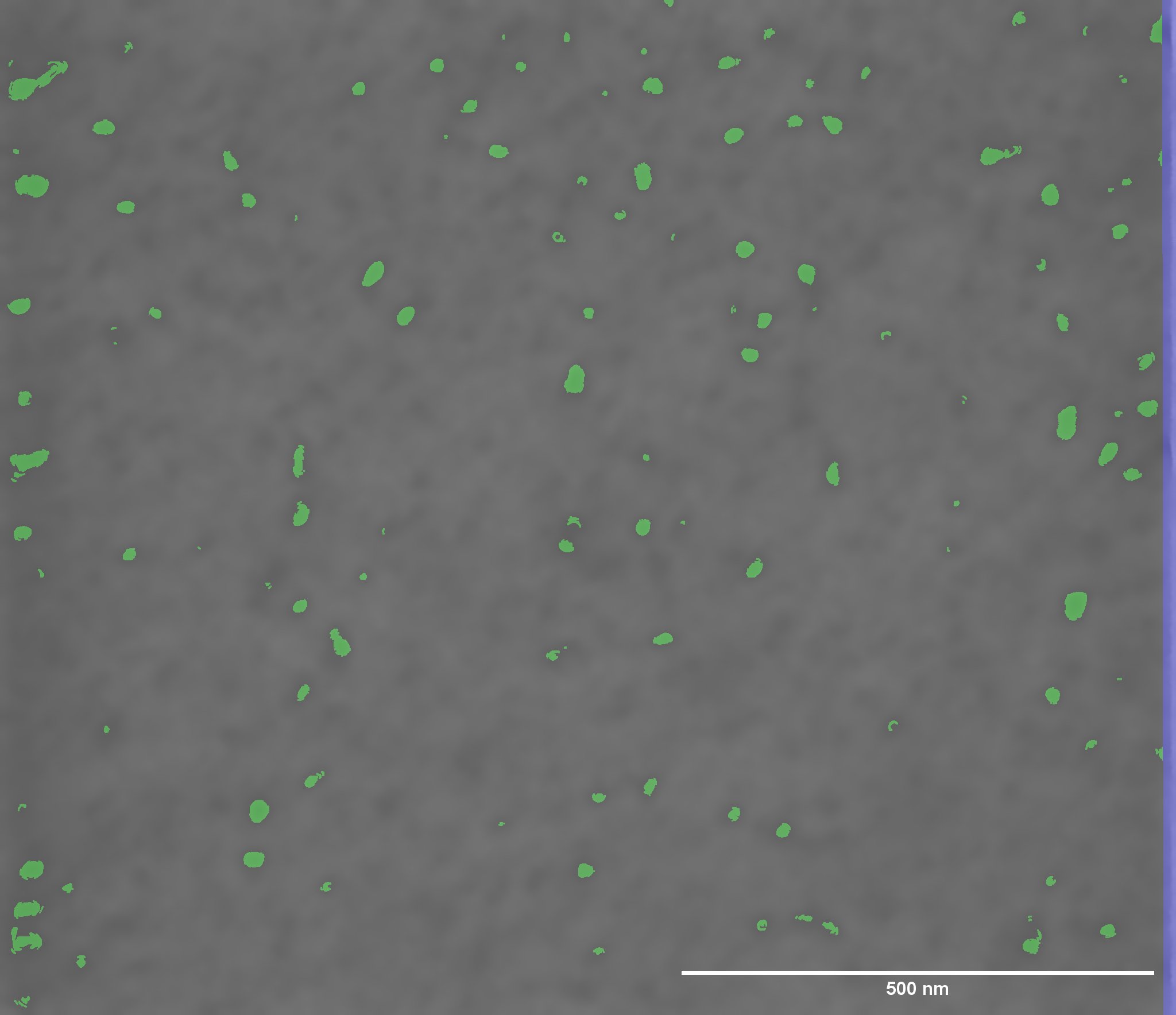}}
    \subfigure[raw image 6]{\label{fig:compare_cryo_2e}\includegraphics[width=0.43\textwidth]{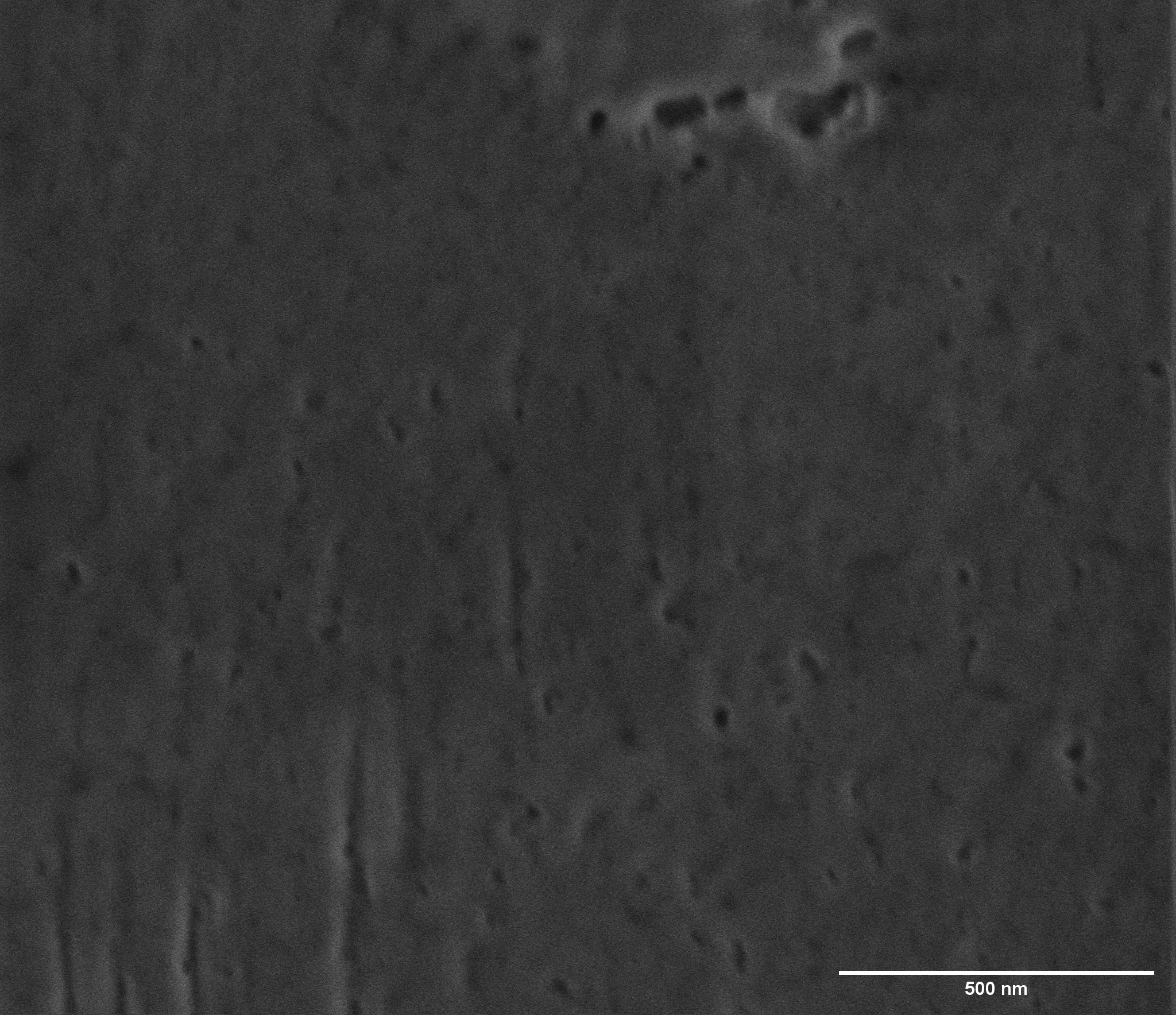}}\hspace{0.2cm}
    \subfigure[segmented image 6]{\label{fig:compare_cryo_2f}\includegraphics[width=0.43\textwidth]{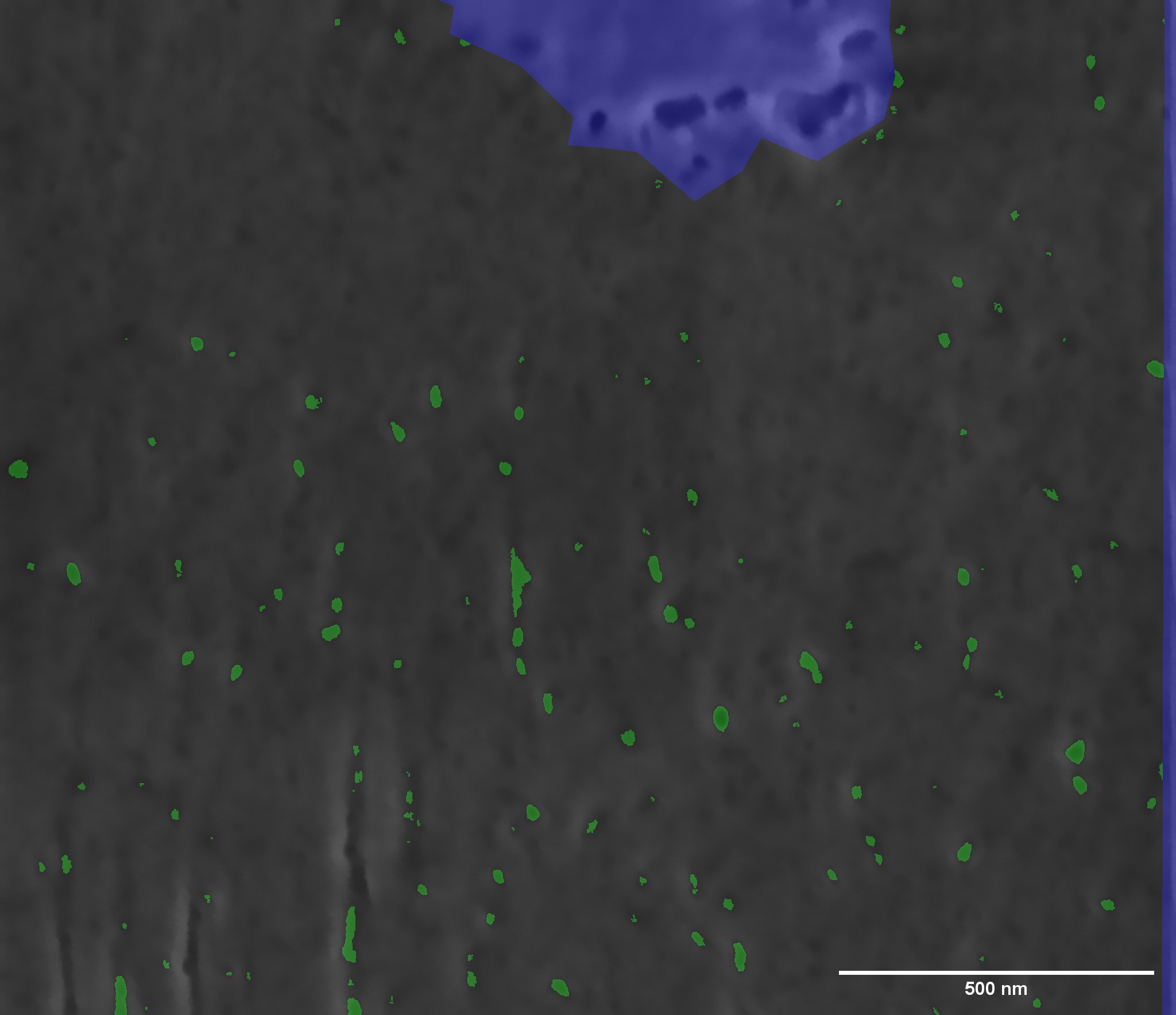}}
    \caption{Segmentation of pores within dense inner \csh, BIB at -140\,$^\circ$C, blue: ignored areas, green: segmented pores.}
    \label{fig:compare_cryo_2}
\end{figure}

\begin{figure}[htbp]
    \centering
    \subfigure[raw image 7]{\label{fig:compare_rt_1a}\includegraphics[width=0.43\textwidth]{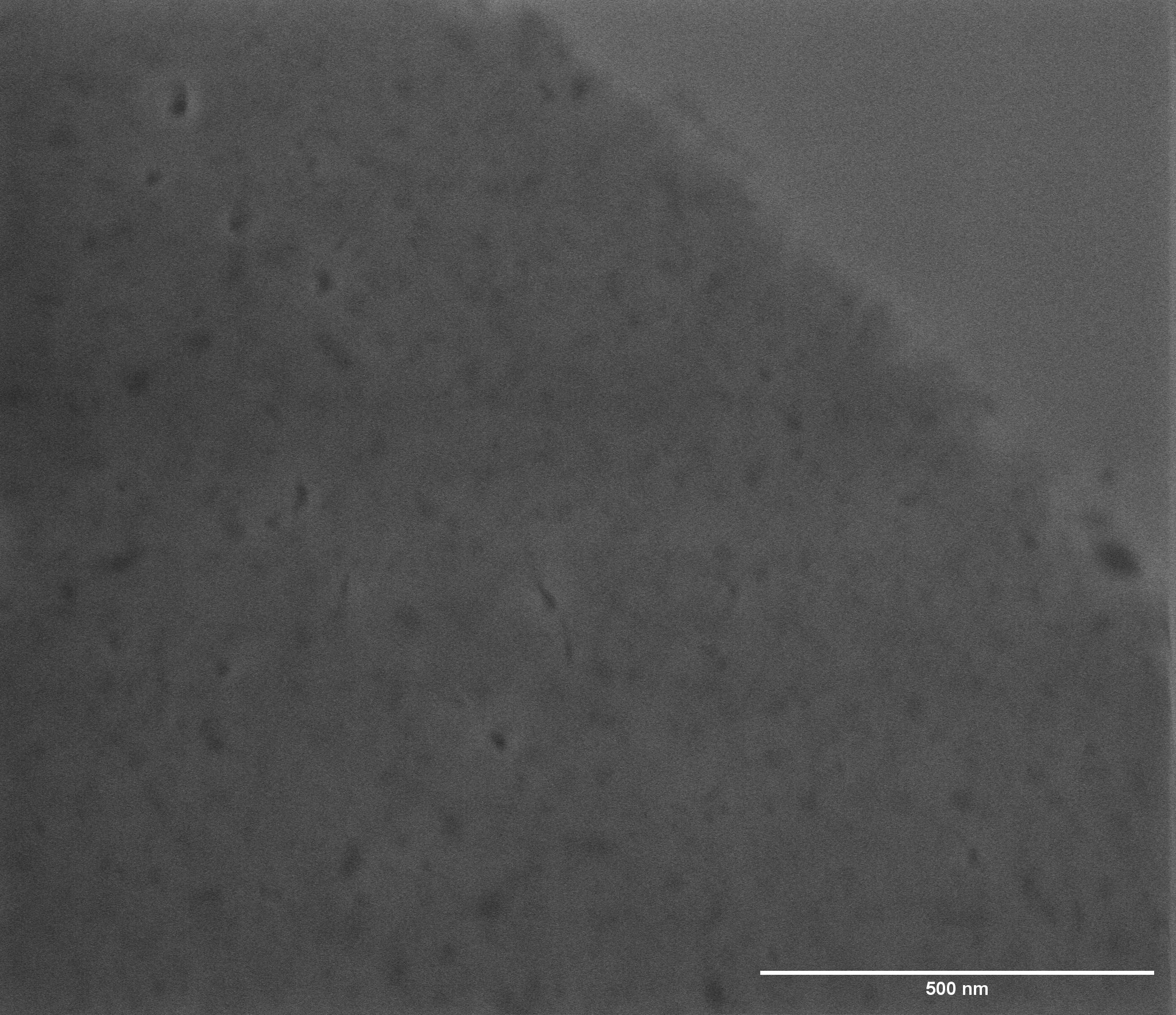}}\hspace{0.2cm}
    \subfigure[segmented image 7]{\label{fig:compare_rt_1b}\includegraphics[width=0.43\textwidth]{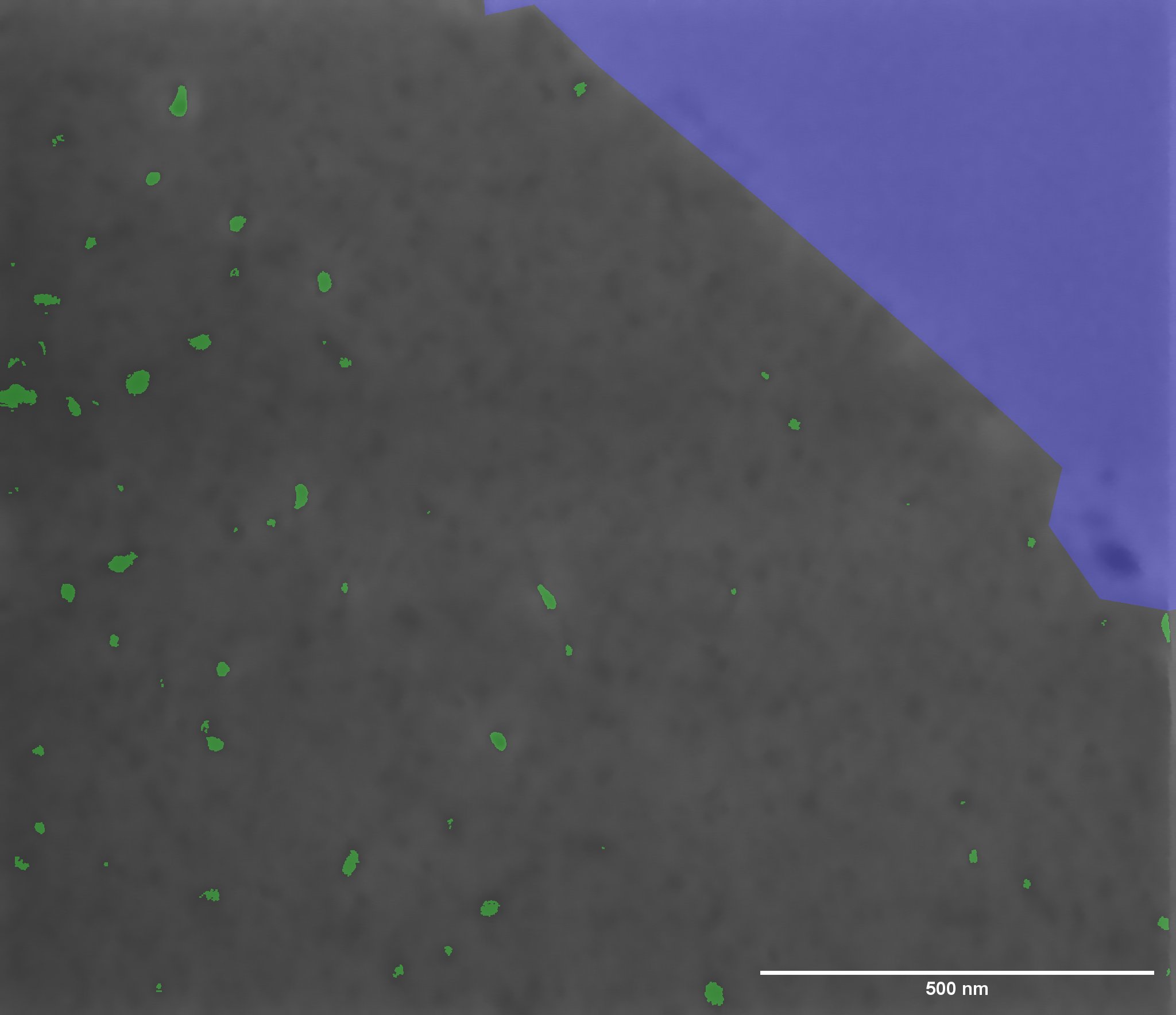}}
    \subfigure[raw image 8]{\label{fig:compare_rt_1c}\includegraphics[width=0.43\textwidth]{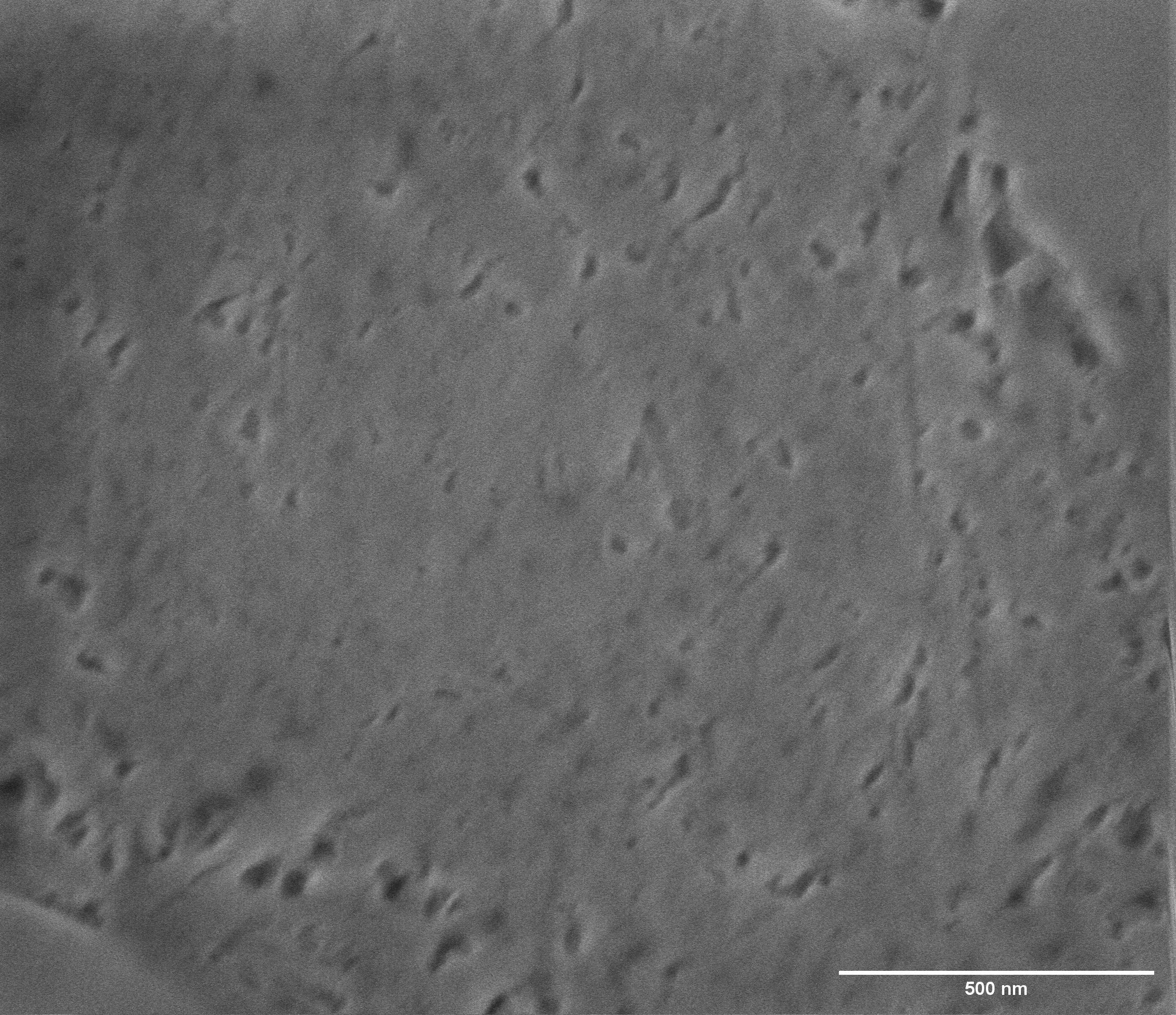}}\hspace{0.2cm}
    \subfigure[segmented image 8]{\label{fig:compare_rt_1d}\includegraphics[width=0.43\textwidth]{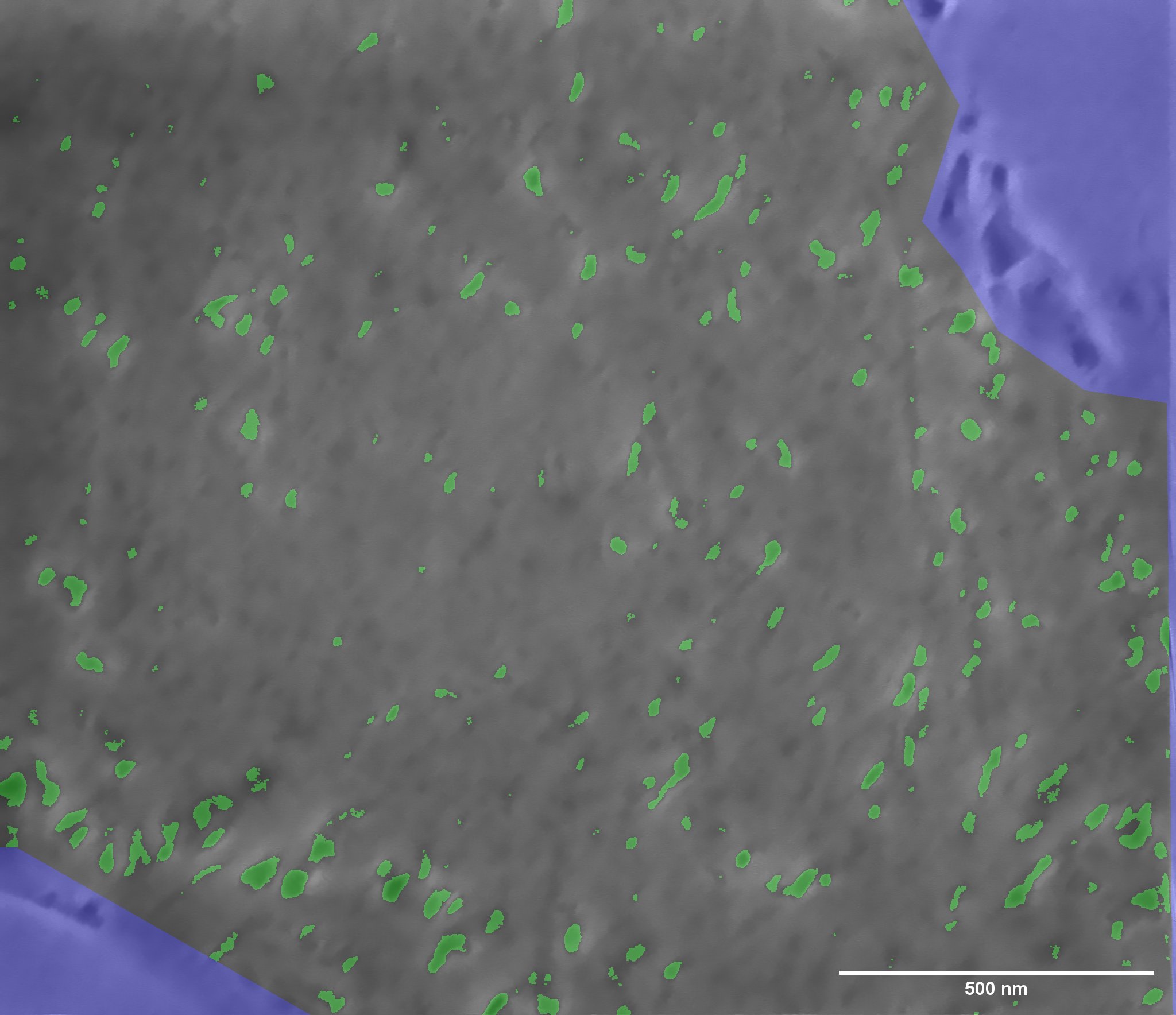}}
    \subfigure[raw image 9]{\label{fig:compare_rt_1e}\includegraphics[width=0.43\textwidth]{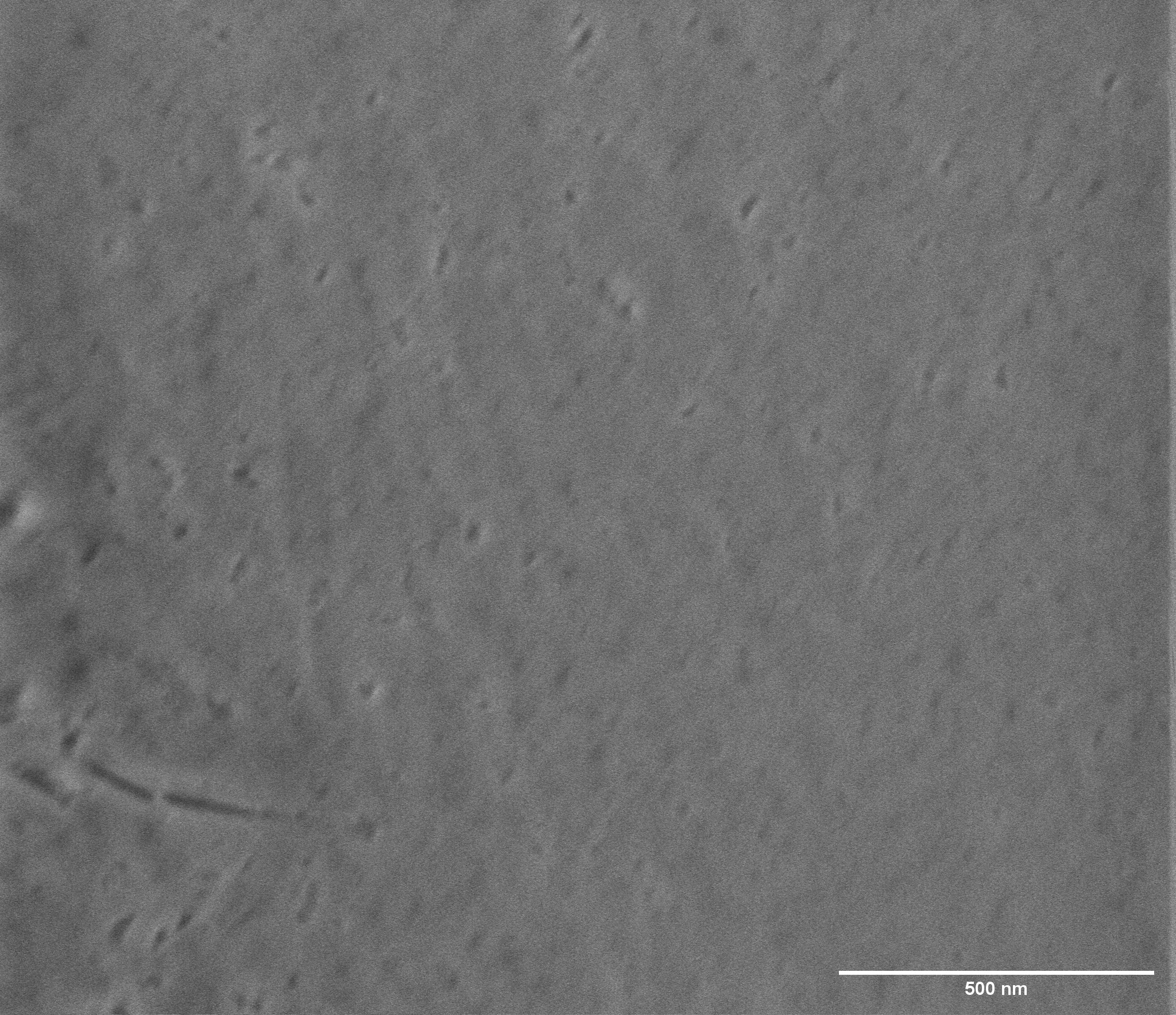}}\hspace{0.2cm}
    \subfigure[segmented image 9]{\label{fig:compare_rt_1f}\includegraphics[width=0.43\textwidth]{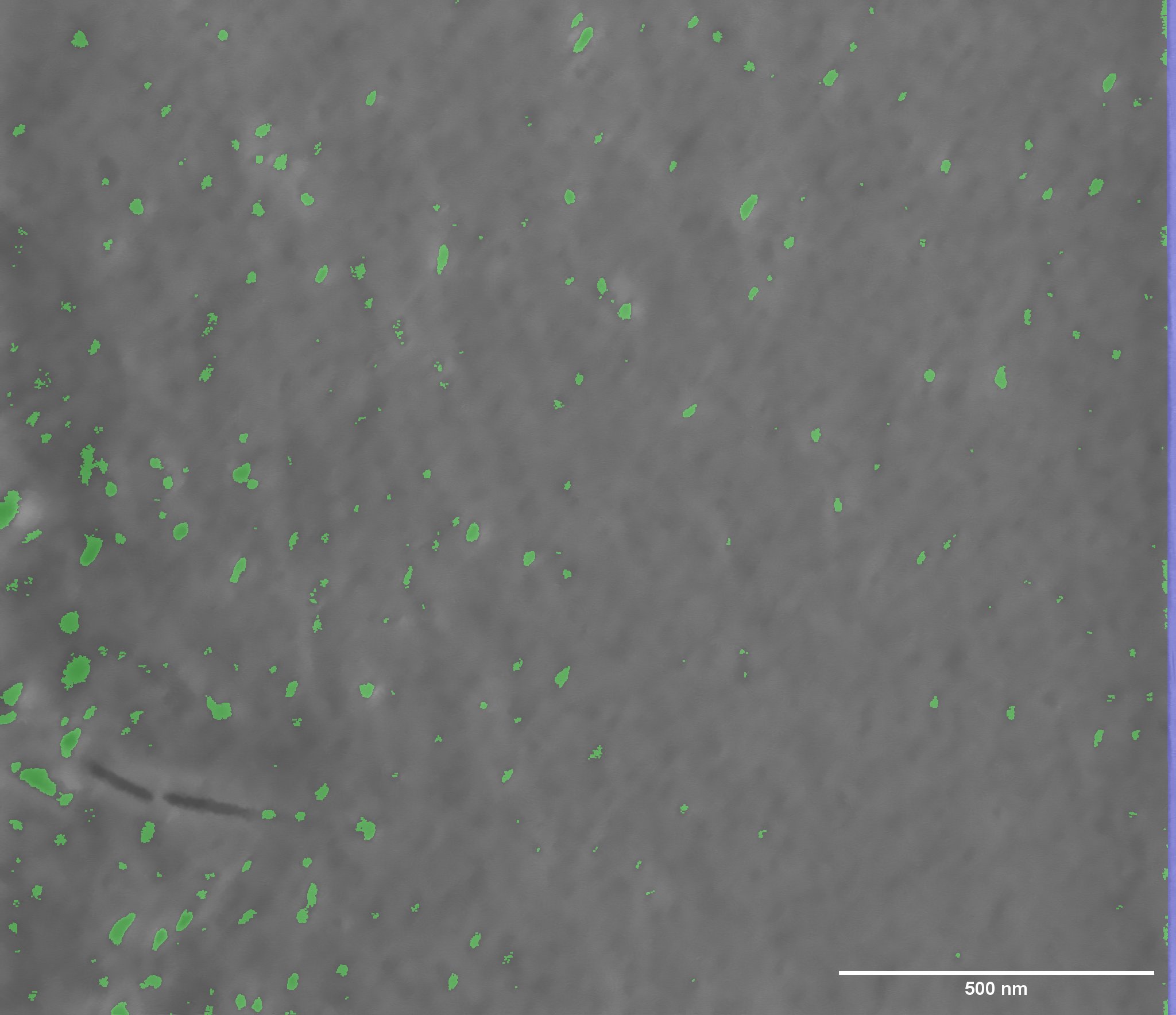}}
    \caption{Segmentation of pores within dense inner \csh, BIB at 20\,$^\circ$C, blue: ignored areas, green: segmented pores.}
    \label{fig:compare_rt_1}
\end{figure}

\begin{figure}[htbp]
    \centering
    \subfigure[raw image 10]{\label{fig:compare_rt_2a}\includegraphics[width=0.43\textwidth]{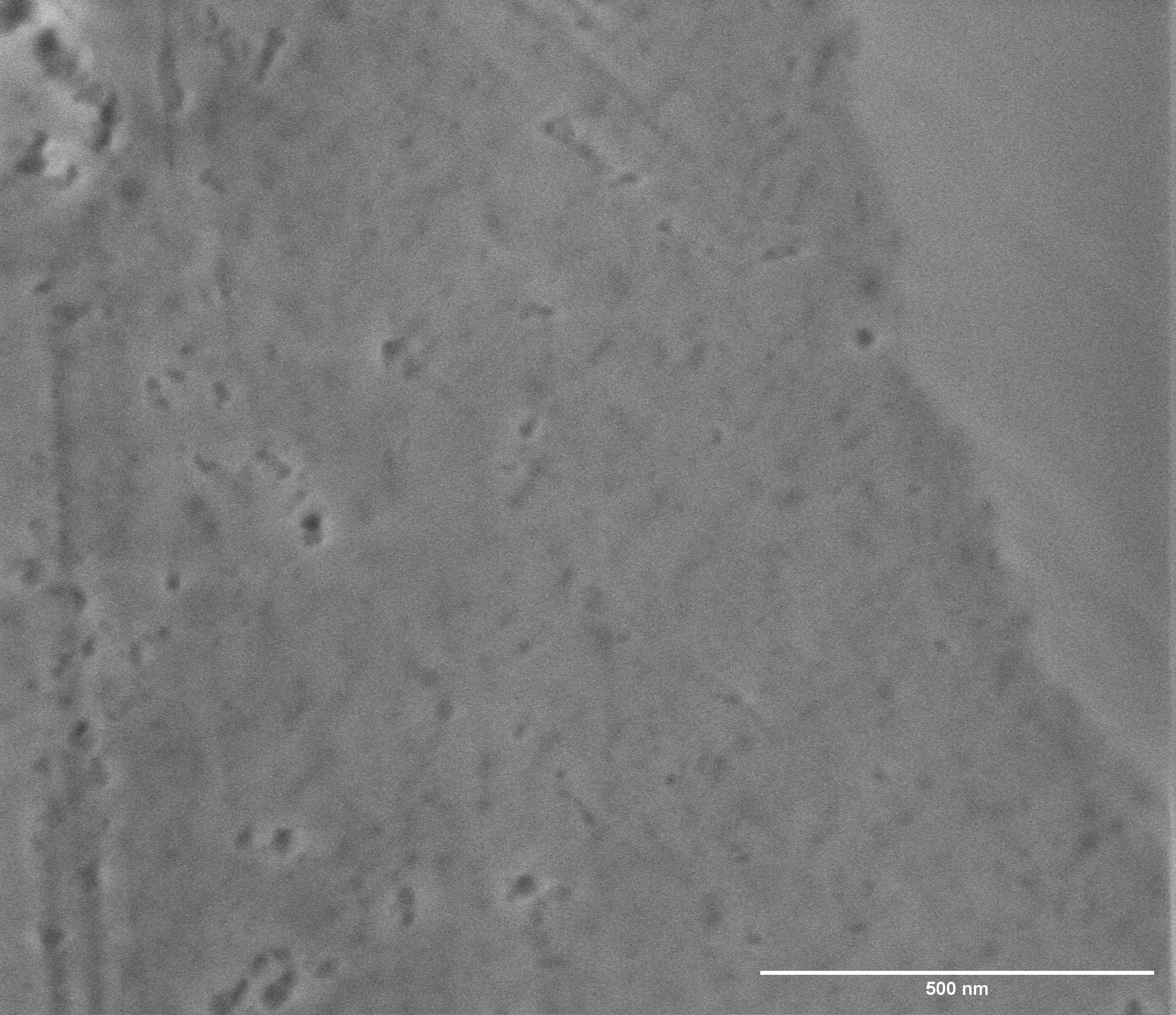}}\hspace{0.2cm}
    \subfigure[segmented image 10]{\label{fig:compare_rt_2b}\includegraphics[width=0.43\textwidth]{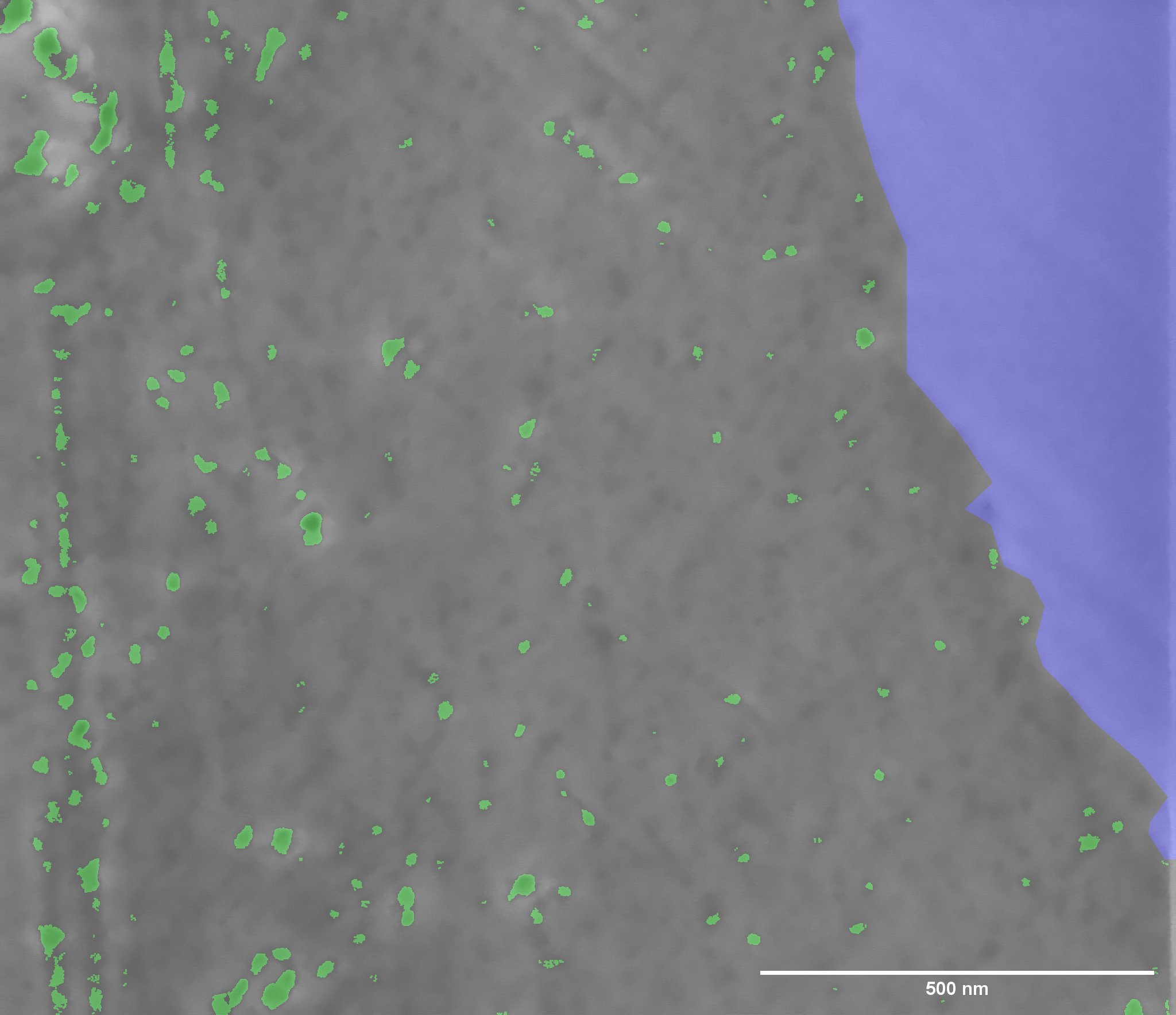}}
    \subfigure[raw image 11]{\label{fig:compare_rt_2c}\includegraphics[width=0.43\textwidth]{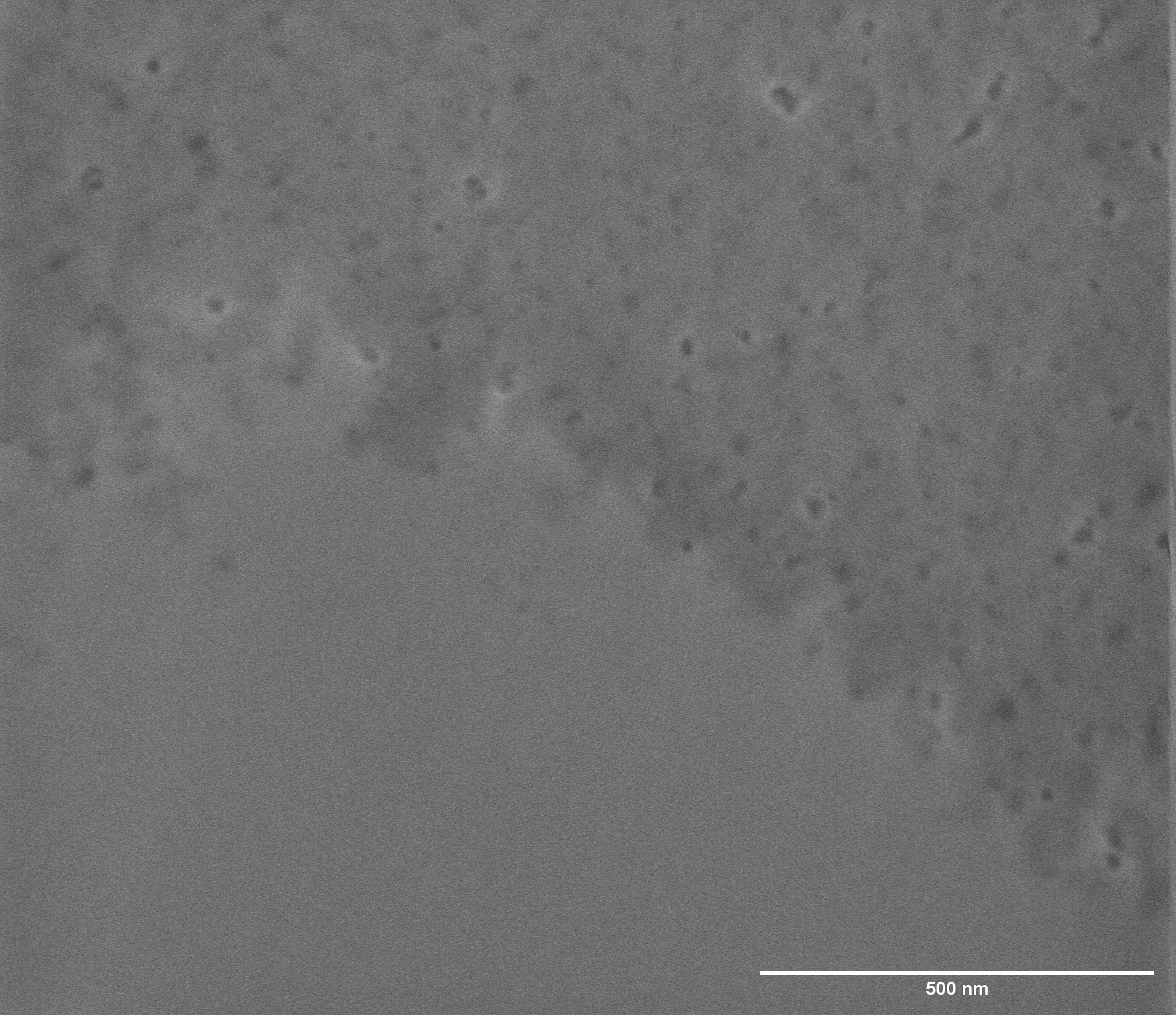}}\hspace{0.2cm}
    \subfigure[segmented image 11]{\label{fig:compare_rt_2d}\includegraphics[width=0.43\textwidth]{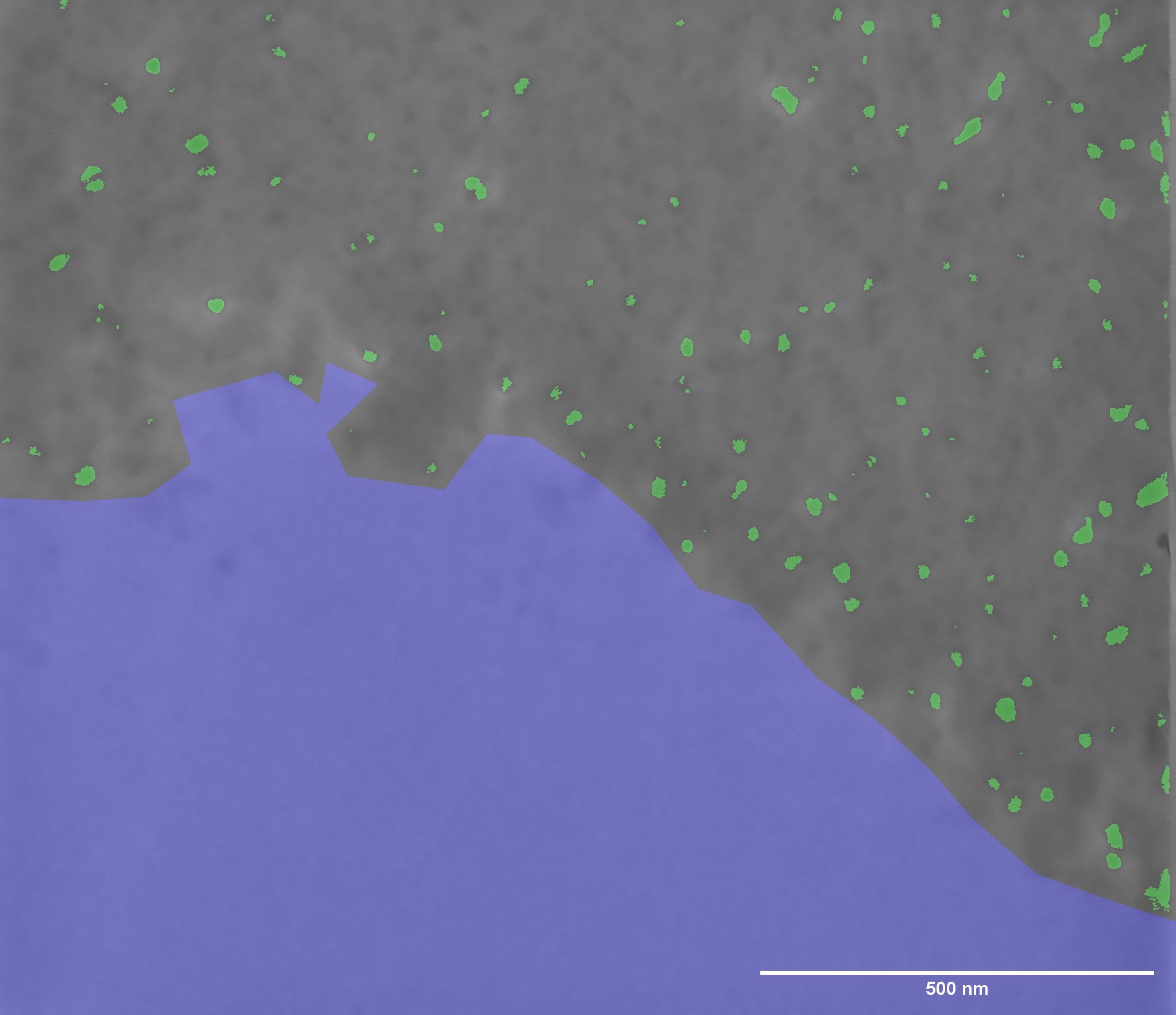}}
    \subfigure[raw image 12]{\label{fig:compare_rt_2e}\includegraphics[width=0.43\textwidth]{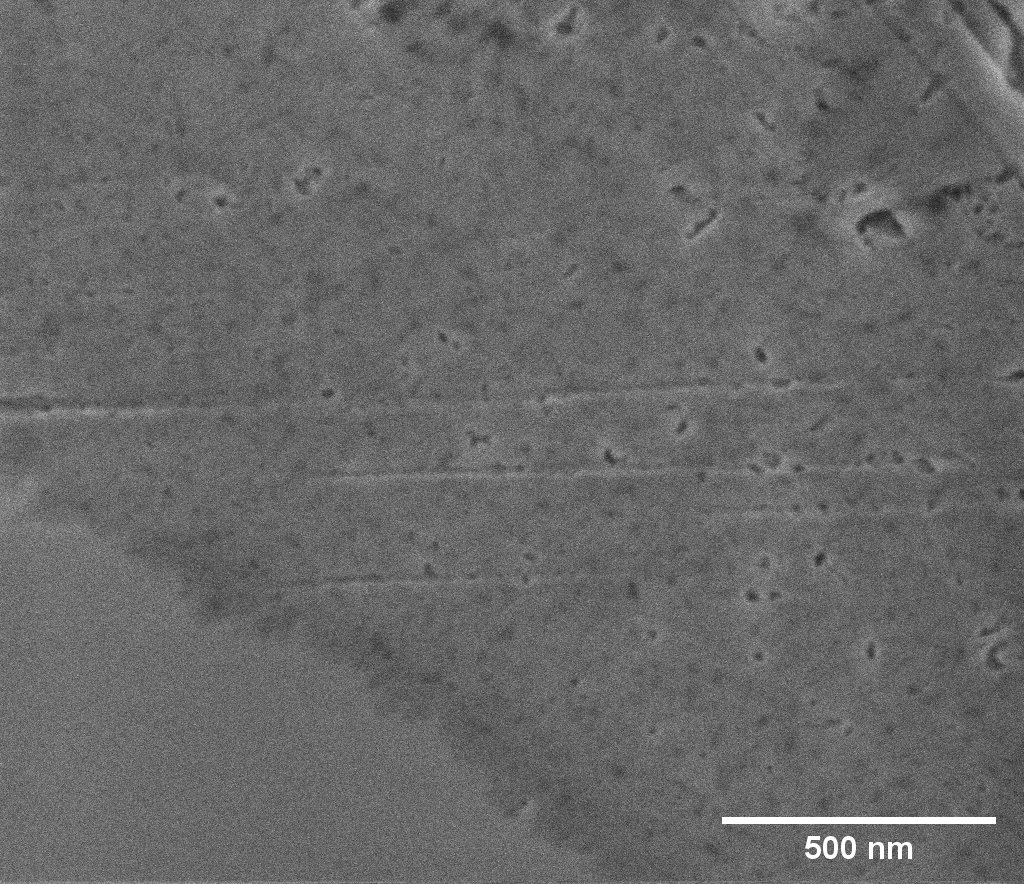}}\hspace{0.2cm}
    \subfigure[segmented image 12]{\label{fig:compare_rt_2f}\includegraphics[width=0.43\textwidth]{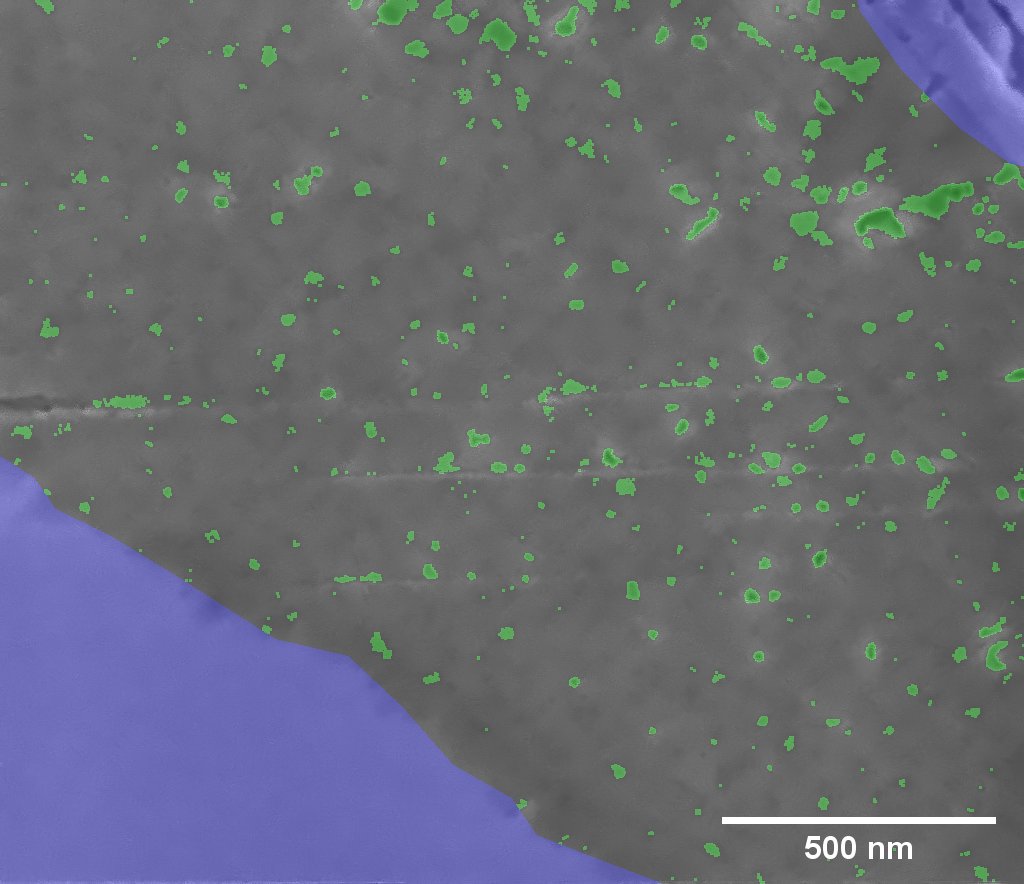}}
    \caption{Segmentation of pores within dense inner \csh, BIB at 20\,$^\circ$C, blue: ignored areas, green: segmented pores.}
    \label{fig:compare_rt_2}
\end{figure}

The following images were acquired as described in the main article.
Figures \ref{fig:compare_cryo_1} to \ref{fig:compare_rt_2} show images of dense inner \csh, which were prepared using Cryo-BIB.
The specimens shown in Figures \ref{fig:compare_cryo_1} and \ref{fig:compare_cryo_2} were prepared at -140\,$^\circ$C (raw images can be found at \cite{Kleiner.2021b}), while the specimens presented in Figure \ref{fig:compare_rt_1} and \ref{fig:compare_rt_2} were prepared at 20\,$^\circ$C (raw images can be found at \cite{Kleiner.2021a}).
On the left side the raw unedited images are shown, while on the right side, the denoised result image with the result of the semi-automatic segmentation \cite{Kleiner.2021} is presented.
The areas with a green overlay show the pores, while the blue overlay represents areas manually segmented. The latter were excluded from the pore segmentation since these areas contain large pores, alite or imaging artefacts, which were not subject to the analysis.

\FloatBarrier
\pagebreak
\section{PSD of the segmented images}
\begin{figure}[!h]
    \centering
    \subfigure[PSD plots for the images in Figure \ref{fig:compare_cryo_1}]{
   	 \label{fig:psd_compare_cryo_1a}
    \begin{tikzpicture}
        \begin{axis}[
            legend style={at={(1,1)},anchor=north east, font=\scriptsize},
            xmin=0, xmax=50,
            ymin=0, ymax=1,
            width=0.95\textwidth,
            height=7cm,
            grid=major,
            xlabel={pore diameter in nm},
            ylabel style={align=center},
            ylabel={normalized cumultative \\ pore area},
            xtick={0,5,...,50},
            ytick=\empty,
            grid style={line width=.1pt, draw=gray!10},
            ]
            \addplot[color=black] table[x=bins,y=area,col sep=comma] {CryoBIB_002_pores_psd_sum_hist.csv};
            \addlegendentry{image row 1};
            \addplot[color=black, dashed] table[x=bins,y=area,col sep=comma] {CryoBIB_009_pores_psd_sum_hist.csv};
            \addlegendentry{image row 2};
            \addplot[color=black, dashdotted] table[x=bins,y=area,col sep=comma] {CryoBIB_011_pores_psd_sum_hist.csv};
            \addlegendentry{image row 3};
        \end{axis}
    \end{tikzpicture}
   }
    \subfigure[PSD plots for the images in Figure \ref{fig:compare_cryo_2}]{
   	 \label{fig:psd_compare_cryo_1b}
    \begin{tikzpicture}
        \begin{axis}[
            legend style={at={(1,1)},anchor=north east, font=\scriptsize},
            xmin=0, xmax=50,
            ymin=0, ymax=1,
            width=0.95\textwidth,
            height=7cm,
            grid=major,
            xlabel={pore diameter in nm},
            ylabel style={align=center},
            ylabel={normalized cumultative \\ pore area},
            xtick={0,5,...,50},
            ytick=\empty,
            grid style={line width=.1pt, draw=gray!10},
            ]
            \addplot[color=black] table[x=bins,y=area,col sep=comma] {CryoBIB_041_pores_psd_sum_hist.csv};
            \addlegendentry{image row 4};
            \addplot[color=black, dashed] table[x=bins,y=area,col sep=comma] {CryoBIB_020_pores_psd_sum_hist.csv};
            \addlegendentry{image row 5};
            \addplot[color=black, dashdotted] table[x=bins,y=area,col sep=comma] {CryoBIB_6h_018_pores_psd_sum_hist.csv};
            \addlegendentry{image row 6};
        \end{axis}
    \end{tikzpicture}
   }
    \caption{PSD (area) of the images in Figures \ref{fig:compare_cryo_1} and \ref{fig:compare_cryo_2}, (BIB at -140\,$^\circ$C).}
    \label{fig:psd_compare_cryo_1}
\end{figure}

\FloatBarrier
\pagebreak

\begin{figure}[!h]
    \centering
    \subfigure[PSD plots for the images in Figure \ref{fig:compare_rt_1}]{
   	 \label{fig:psd_compare_cryo_2a}
    \begin{tikzpicture}
        \begin{axis}[
            legend style={at={(1,1)},anchor=north east, font=\scriptsize},
            xmin=0, xmax=50,
            ymin=0, ymax=1,
            width=0.95\textwidth,
            height=7cm,
            grid=major,
            xlabel={pore diameter in nm},
            ylabel style={align=center},
            ylabel={normalized cumultative \\ pore area},
            xtick={0,5,...,50},
            ytick=\empty,
            grid style={line width=.1pt, draw=gray!10},
            ]
            \addplot[color=black] table[x=bins,y=area,col sep=comma] {BIB_6KV_6h_004_pores_psd_sum_hist.csv};
            \addlegendentry{image row 7};
            \addplot[color=black, dashed] table[x=bins,y=area,col sep=comma] {BIB_6KV_6h_015_pores_psd_sum_hist.csv};
            \addlegendentry{image row 8};
            \addplot[color=black, dashdotted] table[x=bins,y=area,col sep=comma] {BIB_6KV_6h_016_pores_psd_sum_hist.csv};
            \addlegendentry{image row 9};
        \end{axis}
    \end{tikzpicture}
   }
    \subfigure[PSD plots for the images in Figure \ref{fig:compare_rt_2}]{
   	 \label{fig:psd_compare_cryo_2b}
    \begin{tikzpicture}
        \begin{axis}[
            legend style={at={(1,1)},anchor=north east, font=\scriptsize},
            xmin=0, xmax=50,
            ymin=0, ymax=1,
            width=0.95\textwidth,
            height=7cm,
            grid=major,
            xlabel={pore diameter in nm},
            ylabel style={align=center},
            ylabel={normalized cumultative \\ pore area},
            xtick={0,5,...,50},
            ytick=\empty,
            grid style={line width=.1pt, draw=gray!10},
            ]
            \addplot[color=black] table[x=bins,y=area,col sep=comma] {BIB_6KV_6h_pores_psd_sum_hist.csv};
            \addlegendentry{image row 10};
            \addplot[color=black, dashed] table[x=bins,y=area,col sep=comma] {BIB_6KV_6h_003_pores_psd_sum_hist.csv};
            \addlegendentry{image row 11};
            \addplot[color=black, dashdotted] table[x=bins,y=area,col sep=comma] {BIB_6KV_6h_041_pores_psd_sum_hist.csv};
            \addlegendentry{image row 12};
        \end{axis}
    \end{tikzpicture}
   }
    \caption{PSD (area) of the images in Figures \ref{fig:compare_rt_1} and \ref{fig:compare_rt_2}, (BIB at 20\,$^\circ$C).}
    \label{fig:psd_compare_cryo_2}
\end{figure}

\FloatBarrier
\pagebreak
\section{CDF of the segmented images}

\begin{figure}[!h]
    \centering
    \subfigure[CDF plots for the images in Figure \ref{fig:compare_cryo_1}]{
   	 \label{fig:cdf_compare_cryo_1a}
    \begin{tikzpicture}
        \begin{axis}[
            legend style={at={(1,1)},anchor=north east, font=\scriptsize},
            xmin=0, xmax=50,
            ymin=0, ymax=1,
            width=0.95\textwidth,
            height=7cm,
            grid=major,
            xlabel={pore diameter in nm},
            ylabel style={align=center},
            ylabel={normalized cumultative \\ pore area},
            xtick={0,5,...,50},
            ytick=\empty,
            grid style={line width=.1pt, draw=gray!10},
            ]
            \addplot[color=black] table[x=bins,y=area,col sep=comma] {CryoBIB_002_pores_cld_sum_hist.csv};
            \addlegendentry{image row 1};
            \addplot[color=black, dashed] table[x=bins,y=area,col sep=comma] {CryoBIB_009_pores_cld_sum_hist.csv};
            \addlegendentry{image row 2};
            \addplot[color=black, dashdotted] table[x=bins,y=area,col sep=comma] {CryoBIB_011_pores_cld_sum_hist.csv};
            \addlegendentry{image row 3};
        \end{axis}
    \end{tikzpicture}
   }
    \subfigure[CDF plots for the images in Figure \ref{fig:compare_cryo_2}]{
   	 \label{fig:cdf_compare_cryo_1b}
    \begin{tikzpicture}
        \begin{axis}[
            legend style={at={(1,1)},anchor=north east, font=\scriptsize},
            xmin=0, xmax=50,
            ymin=0, ymax=1,
            width=0.95\textwidth,
            height=7cm,
            grid=major,
            xlabel={pore diameter in nm},
            ylabel style={align=center},
            ylabel={normalized cumultative \\ pore area},
            xtick={0,5,...,50},
            ytick=\empty,
            grid style={line width=.1pt, draw=gray!10},
            ]
            \addplot[color=black] table[x=bins,y=area,col sep=comma] {CryoBIB_041_pores_cld_sum_hist.csv};
            \addlegendentry{image row 4};
            \addplot[color=black, dashed] table[x=bins,y=area,col sep=comma] {CryoBIB_020_pores_cld_sum_hist.csv};
            \addlegendentry{image row 5};
            \addplot[color=black, dashdotted] table[x=bins,y=area,col sep=comma] {CryoBIB_6h_018_pores_cld_sum_hist.csv};
            \addlegendentry{image row 6};
        \end{axis}
    \end{tikzpicture}
   }
    \caption{CDF (area) of the images in Figures \ref{fig:compare_cryo_1} and \ref{fig:compare_cryo_2}, (BIB at -140\,$^\circ$C).}
    \label{fig:cdf_compare_cryo_1}
\end{figure}

\FloatBarrier
\pagebreak
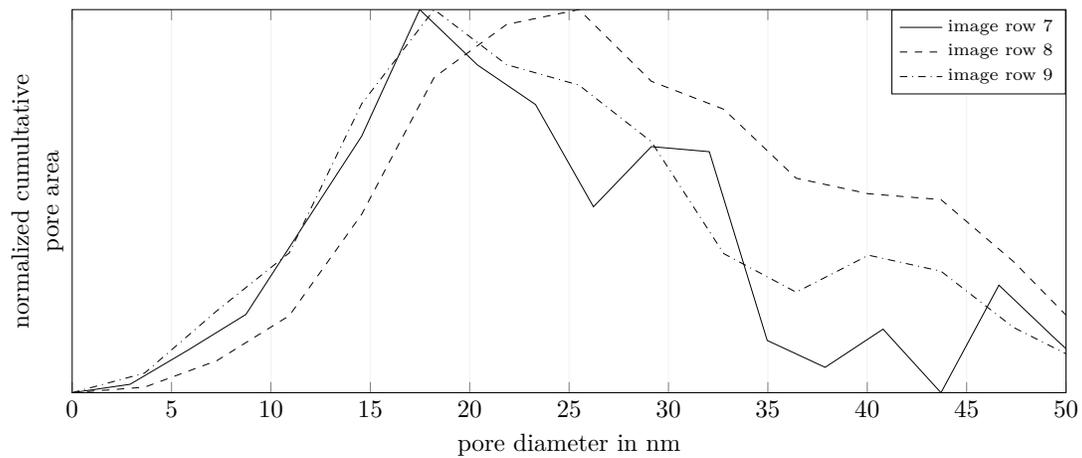
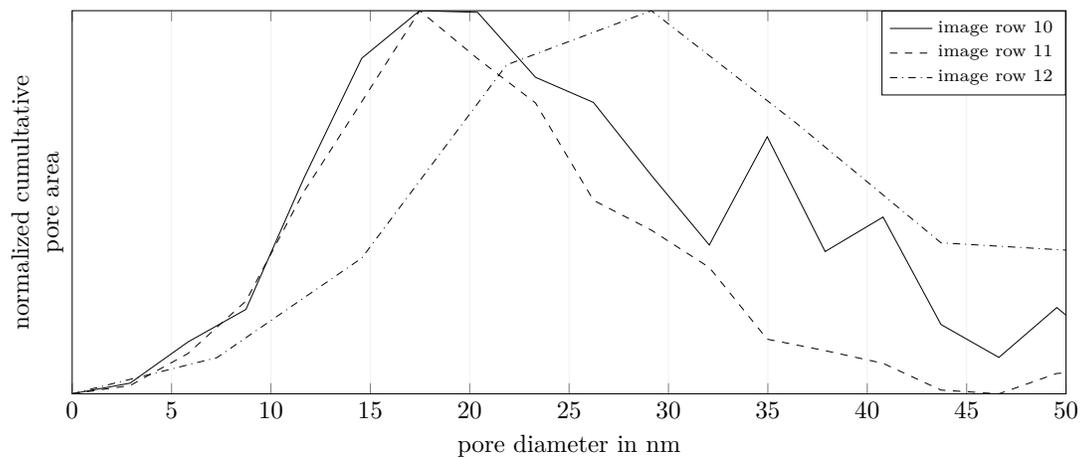
\begin{figure}[!h]
    \centering
    \subfigure[CDF plots for the images in Figure \ref{fig:compare_rt_1}]{
   	 \label{fig:cdf_compare_cryo_2a}
    \begin{tikzpicture}
        \begin{axis}[
            legend style={at={(1,1)},anchor=north east, font=\scriptsize},
            xmin=0, xmax=50,
            ymin=0, ymax=1,
            width=0.95\textwidth,
            height=7cm,
            grid=major,
            xlabel={pore diameter in nm},
            ylabel style={align=center},
            ylabel={normalized cumultative \\ pore area},
            xtick={0,5,...,50},
            ytick=\empty,
            grid style={line width=.1pt, draw=gray!10},
            ]
            \addplot[color=black] table[x=bins,y=area,col sep=comma] {BIB_6KV_6h_004_pores_cld_sum_hist.csv};
            \addlegendentry{image row 7};
            \addplot[color=black, dashed] table[x=bins,y=area,col sep=comma] {BIB_6KV_6h_015_pores_cld_sum_hist.csv};
            \addlegendentry{image row 8};
            \addplot[color=black, dashdotted] table[x=bins,y=area,col sep=comma] {BIB_6KV_6h_016_pores_cld_sum_hist.csv};
            \addlegendentry{image row 9};
        \end{axis}
    \end{tikzpicture}
   }
    \subfigure[CDF plots for the images in Figure \ref{fig:compare_rt_2}]{
   	 \label{fig:cdf_compare_cryo_2b}
    \begin{tikzpicture}
        \begin{axis}[
            legend style={at={(1,1)},anchor=north east, font=\scriptsize},
            xmin=0, xmax=50,
            ymin=0, ymax=1,
            width=0.95\textwidth,
            height=7cm,
            grid=major,
            xlabel={pore diameter in nm},
            ylabel style={align=center},
            ylabel={normalized cumultative \\ pore area},
            xtick={0,5,...,50},
            ytick=\empty,
            grid style={line width=.1pt, draw=gray!10},
            ]
            \addplot[color=black] table[x=bins,y=area,col sep=comma] {BIB_6KV_6h_pores_cld_sum_hist.csv};
            \addlegendentry{image row 10};
            \addplot[color=black, dashed] table[x=bins,y=area,col sep=comma] {BIB_6KV_6h_003_pores_cld_sum_hist.csv};
            \addlegendentry{image row 11};
            \addplot[color=black, dashdotted] table[x=bins,y=area,col sep=comma] {BIB_6KV_6h_041_pores_cld_sum_hist.csv};
            \addlegendentry{image row 12};
        \end{axis}
    \end{tikzpicture}
   }
    \caption{CDF (area) of the images in Figures \ref{fig:compare_rt_1} and \ref{fig:compare_rt_2}, (BIB at 20\,$^\circ$C).}
    \label{fig:cdf_compare_cryo_2}
\end{figure}

\FloatBarrier
\pagebreak
\section{Basic statistics}

Table \ref{tab:mean_max_CDF} shows the peak position (pore diameter) $d_\text{max}$ of the CDF-histograms shown in Figures \ref{fig:cdf_compare_cryo_1} and \ref{fig:cdf_compare_cryo_2}.
Since the histograms are composed of bins $b$ (bin width equals 4 times the pixel size of the respective image), the bin center $b_\text{center}$ was used as the peak position.

\begin{equation}
d_\text{max} = b_\text{center} = (b_\text{lower} + b_\text{upper})/2
\end{equation}

\begin{table}[ht]
	\centering
	\caption{Peak position $d_\text{max}$ of the CDF as shown in Figures \ref{fig:cdf_compare_cryo_1} and \ref{fig:cdf_compare_cryo_2}}
	\begin{tabular}{cccc}
		\toprule
\multicolumn{2}{c}{BIB at -140\,$^\circ$C} & \multicolumn{2}{c}{BIB at 20\,$^\circ$C} \\
image reference & $d_\text{max}$ & image reference & $d_\text{max}$\\
		\midrule
main document, Fig. 16 & 20.5\,nm & main document, Fig. 17 & 20.0\,nm \\
image row 1, Fig. \ref{fig:compare_cryo_1a} & 13.4\,nm & image row 7, Fig. \ref{fig:compare_rt_1a} & 16.0\,nm \\
image row 2, Fig. \ref{fig:compare_cryo_1c} & 10.9\,nm & image row 8, Fig. \ref{fig:compare_rt_1c} & 23.7\,nm \\
image row 3, Fig. \ref{fig:compare_cryo_1e} & 15.8\,nm & image row 9, Fig. \ref{fig:compare_rt_1e} & 16.4\,nm \\
image row 4, Fig. \ref{fig:compare_cryo_2a} & 13.4\,nm & image row 10, Fig. \ref{fig:compare_rt_2a} & 16.0\,nm \\
image row 5, Fig. \ref{fig:compare_cryo_2c} & 15.8\,nm & image row 11, Fig. \ref{fig:compare_rt_2c} & 16.0\,nm \\
image row 6, Fig. \ref{fig:compare_cryo_2e} & 16.4\,nm & image row 12, Fig. \ref{fig:compare_rt_2e} & 25.5\,nm \\
		\midrule
mean & $17.5\,\pm\,2.8\,$nm & mean & $22.7\,\pm\,5.4\,$nm \\
		\bottomrule
	\end{tabular}
	\label{tab:mean_max_CDF}
\end{table}

\bibliographystyle{elsarticle-num} 
\bibliography{supplementary_bibliography}